\documentclass[reqno,11pt]{amsart}

\usepackage{amssymb,amsfonts}
\usepackage[mathscr]{eucal}
\usepackage{amsmath}

\usepackage{graphics}
\usepackage{graphicx}

\allowdisplaybreaks[1]
\numberwithin{equation}{section}

\theoremstyle{plain}
\newtheorem{theorem}{Theorem}[section]
\newtheorem{proposition}[theorem]{Proposition}

\theoremstyle{remark}
\newtheorem{remark}{Remark}[section]

\newtheorem{observation}{Observation}

\theoremstyle{definition}

\newcommand{\DP}{DP}

\newcommand{\cl}{{\mathrm h}}

\newcommand{\orth}{\perp}

\newcommand{\id}{{ \rm id}}

\newcommand{\bN}{{\mathbb N}}
\newcommand{\bR}{{\mathbb R}}
\newcommand{\bC}{{\mathbb C}}
\newcommand{\bZ}{{\mathbb Z}}

\newcommand{\cA}{{\mathcal A}}

\newcommand{\cB}{{\mathcal B}}

\newcommand{\cN}{{\mathcal N}}

\newcommand{\ct}{{\mathfrak t}}
\newcommand{\cM}{{\mathcal M}}
\newcommand{\cG}{{\mathfrak g}}

\newcommand{\G}{{\mathcal G}}
\newcommand{\cH}{{\mathcal H}}

\newcommand{\cP}{{\mathcal P}}
\newcommand{\cC}{{\mathcal C}}

\newcommand{\eps}{\epsilon}

\DeclareMathOperator{\Image}{Image}
\DeclareMathOperator{\arc}{arc}
 \DeclareMathOperator{\Tr}{Tr}
\DeclareMathOperator{\GL}{GL}
\DeclareMathOperator{\gl}{gl}
 \DeclareMathOperator{\Mat}{Mat}
\DeclareMathOperator{\supp}{supp}

 \DeclareMathOperator{\End}{End}
\DeclareMathOperator{\Hom}{Hom}
\DeclareMathOperator{\Ad}{Ad}
\DeclareMathOperator{\ad}{ad}

\DeclareMathOperator{\Pairs}{Pairs}

\DeclareMathOperator{\WLF}{WLF}
 \DeclareMathOperator{\WLO}{WLO}

\DeclareMathOperator{\LK}{LK^*}

\DeclareMathOperator{\Link}{LK}

\DeclareMathOperator{\wind}{wind}

\DeclareMathOperator{\ind}{ind}
\DeclareMathOperator{\Ind}{ind}

\DeclareMathOperator{\sgn}{sgn}

\DeclareMathOperator{\rank}{rank}
\DeclareMathOperator{\Ker}{ker}

\DeclareMathOperator{\symb}{symb}

\begin{document}

\title{An analytic Approach to Turaev's Shadow Invariant}

\maketitle

\begin{center} \large
Atle Hahn\footnote{Email: atle.hahn@gmx.de}
\end{center}

\begin{center} \em   Grupo de F{\'i}sica Matem{\'a}tica da Universidade de Lisboa\\
Av. Prof. Gama Pinto, 2\\
PT-1649-003 Lisboa, Portugal
  \end{center}

\begin{abstract}
In the present paper we  extend the ``torus gauge fixing'' approach
by Blau and Thompson, which  was developed in  \cite{BlTh1} for the
study of Chern-Simons models with base manifolds $M$ of the form $M=
\Sigma \times S^1$, in a suitable way. We  arrive at a heuristic
path integral formula for the Wilson loop observables associated to
general links in $M$.
 We then show that the right-hand side of this formula can be
 evaluated explicitly in a non-perturbative way and that this
 evaluation naturally leads to the face models  in terms of which Turaev's
shadow invariant is defined.
\end{abstract}

\medskip


\medskip


\medskip

\section{Introduction}
\label{intro}

The study of the heuristic Chern-Simons path integral functional
in \cite{Wi} inspired the following
 two general  approaches to quantum topology:

\begin{itemize}
\item[(A1)] The perturbative approach
based on the Chern-Simons path integral in the Landau gauge (cf. \cite{GMM,Bar,Bar2,AxSi1,BoTa,AxSi2,AlFr}).

\item[(A2)] The  non-perturbative  ``quantum group approach''
that comes in two different versions:
the  ``surgery'' version   (cf. \cite{ReTu2,ReTu1} and the first part of \cite{Tu3})
and the  ``state sum'' or ``shadow'' version (cf.  \cite{ReKi,TuVi,Tu2} and the second part
of \cite{Tu3}).
\end{itemize}
While for the first approach
the relationship to the Chern-Simons path integral is obvious
it is not fully understood yet how the expressions that appear
in the second approach are related to the Chern-Simons path integral.
In other words the following problem
has so far remained open (cf., e.g.,  \cite{Freed1}):
\begin{enumerate}
\item[(P1)]
Derive the  algebraic expressions (in particular,
the $R$-matrices resp. the quantum 6j-symbols)
 that  appear in approach (A2)
directly from the Chern-Simons path integral.
\end{enumerate}

Approach (A2) is considerably less complicated than approach (A1).
Thus it is reasonable to expect that if one can solve problem (P1)
then  the  corresponding path integral derivation
will be less complicated than the  path integral derivation\footnote{we remark that
while the final perturbation series appearing in approach (A1)
is rigorous  (cf. \cite{AlFr}) the path integral expressions that are used for its derivation are not}
given in \cite{GMM,Bar,Bar2,AxSi1,BoTa,AxSi2,AlFr}.
One could therefore hope that after having solved  problem (P1)
one can make progress  towards the solution of
the following problem, which can be considered to be one of
the major open problems in quantum topology (cf. \cite{Kup}):
\begin{enumerate}
\item[(P2)] Make rigorous sense of the  heuristic path integral expressions
for the Wilson loop observables  (WLOs) that were studied in \cite{Wi} (cf. formula \eqref{eq_WLO}
below).
\end{enumerate}
As a first step towards the solution of problem (P2)
one can try to solve the following weakened version:
\begin{enumerate}
\item[(P2)'] Make rigorous sense either of  the original
 path integral expressions for the Wilson loop observables
 or, alternatively, of those path integral expressions
 that arise from the original ones after one has fixed a suitable gauge.
\end{enumerate}

The aim of the present paper  is to give a partial solution
of problems (P1) and (P2)'.
In order to do so we  will concentrate on the special situation where
the base manifold $M$ of the Chern-Simons model is of the form $M = \Sigma \times S^1$
and then apply the  so-called ``torus gauge fixing'' procedure
which was successfully used in  \cite{BlTh1} for
 the computation of the partition function of Chern-Simons models on
 such manifolds (cf. eq. (7.1) in  \cite{BlTh1})
 and for the  computation of the Wilson loop observables of a special type of links
 in $M$, namely links $L$ that consist of ``vertical'' loops (cf. eq. (7.24) in   \cite{BlTh1},
 see also our  Subsec. \ref{subsec6.2}).
 The first question which we study in the present paper is
 the question whether is is possible to generalize
  the formulae (7.1) and (7.24)  in   \cite{BlTh1} to general links $L$ in $M$.
  The answer to this questions turns out to be ``yes'', cf. Eq. \eqref{eq_WLO_end} below.\par
  Next we study the question whether it is possible
  to give a rigorous meaning to the heuristic path integral expressions
  on the right-hand side of Eq. \eqref{eq_WLO_end}.
  Fortunately, it is very likely that also this question has a positive answer
  (cf. \cite{Ha6} and point (4) in Subsec. \ref{subsec7.2}).
  In fact, due to the remarkable property of Eq. \eqref{eq_WLO_end} that
  all the heuristic measures that appear there are of ``Gaussian type''
  we can apply  similar techniques as
  in the axial gauge approach
  to Chern-Simons models on $\bR^3$  developed in \cite{FK,ASen,Ha1,Ha2}.
  In particular, we can make use of white noise analysis and
  of the two regularization techniques ``loop smearing'' and
  ``framing''.\par
  Finally,  we study  the question
  if and how the right-hand side of Eq. \eqref{eq_WLO_end} can be evaluated explicitly and if,
   by doing this, one arrives at the same algebraic
   expressions for the corresponding quantum invariants as in the shadow version of approach (A2).
   It turns out that
  also this question has a positive answer, at least
   in all the special cases that we will study in detail.\par
The present paper is organized as follows. In Sec. \ref{sec2} we
recall and extend the relevant definitions and results from
\cite{BlTh1,BlTh3,Ha3b,Ha3c} on Blau and Thompson's torus gauge fixing procedure.
 In Secs \ref{subsec3.1}--\ref{subsec3.3}, we then
apply the torus gauge fixing procedure to Chern-Simons models with
compact base manifolds of the form  $M=\Sigma \times S^1$.
 After introducing  a suitable decomposition  $\cA^{\orth} = \hat{\cA}^{\orth} \oplus \cA_c^{\orth}$
in Subsec.~\ref{subsec3.4} we finally arrive in
Subsec.~\ref{subsec3.5} at the aforementioned heuristic path
integral formula \eqref{eq_WLO_end} for the WLOs. \par The rest of
the paper is concerned with the question how one can make rigorous
sense of (the inner integral in) the heuristic formula
\eqref{eq_WLO_end} and how one can evaluate its right-hand side
explicitly.  We proceed in three steps. In Sec. \ref{sec4} (Step 1)
we briefly summarize the rigorous realization of the integral
functional $\Phi^{\orth}_B$ found in \cite{Ha3b} and we then show in
Sec. \ref{sec5} how the whole inner integral can be realized
rigorously and evaluated explicitly (Step 2). In Sec. \ref{sec6} we
then evaluate
 the whole right-hand side of formula \eqref{eq_WLO_end} (Step 3) in several
  special cases.
First we consider the  case where the group $G$ is Abelian (cf. Subsec. \ref{subsec6.1}).
Next we  consider the  case where
$G=SU(2)$ and where the link $L$ consists exclusively of ``vertical'' loops
(this case  was already studied successfully  in Sec. 7.6 in \cite{BlTh1}).
In Subsec. \ref{subsec6.3} we then study the
case where $G=SU(2)$ and where the link $L$ has no double points,
and demonstrate how in this situation the
  face models by which the shadow invariant is defined arise
 naturally.
Finally, in Sec. \ref{sec7} we  comment on the results to be
expected when completing the computations for the WLOs of general
links, which we plan to carry out in the near future, and  we then
give a list of suggestions for additional generalizations/extensions
of the results of the present paper.

\medskip

{\em Convention:} In the present paper, the symbol ``$\sim$'' will denote
``equality up to a multiplicative constant''. Sometimes we allow this multiplicative ``constant''
to depend on the ``charge'' $k$ of the model, but it will never depend on the
link $L$ which we will fix in Subsec. \ref{subsec3.1} below.

\section{Torus gauge fixing for manifolds $M=\Sigma \times S^1$}
\label{sec2}

Let $M$ be a smooth manifold of the form
 $M = \Sigma \times S$ where $S \in \{S^1,\bR\}$
 and let  $G$ be a   compact connected
Lie group with Lie algebra $\cG$.
For $X \in \{M,\Sigma\}$ we will denote by $\cA_X$
the space of all smooth $\cG$-valued 1-forms on $X$
and by $\G_X$ the group of all smooth $G$-valued functions on $X$.
In the special case $X=M$ we will often write $\cA$ instead of $\cA_X$ and
$\G$ instead of $\G_X$.\par

We now fix a point $\sigma_0 \in \Sigma$
and a point $t_0 \in S^1$.
In \cite{Ha3b,Ha3c} we consider only the special case
 $t_0=1 \in S^1$ ($\cong \{ z
\in \bC \mid \|z\| = 1 \}$).
In the present paper we will not assume this anymore.

\subsection{Quasi-axial and torus gauge fixing: the basic idea}
\label{subsec2.1}

In order to motivate the definition of
quasi-axial gauge fixing for  manifolds
of the form  $M = \Sigma \times S^1$
we first recall the definition of  axial gauge fixing for manifolds of the
form $M = \Sigma \times \bR$.\par

Let $M = \Sigma \times \bR$ and let $\tfrac{\partial}{\partial t}$ (resp. $dt$)
denote the vector field (resp. 1-form) on $\bR$ which is
induced by $\id_{\bR}: \bR \to \bR$.
By lifting  $\tfrac{\partial}{\partial t}$ and $dt$
to $M = \Sigma \times \bR$ in the obvious way
we obtain a vector field and a 1-form on $M$
which will also be denoted by $\tfrac{\partial}{\partial t}$ resp. $dt$.
 Clearly,  every $A \in \cA = \cA_M$ can be written uniquely in the form
 $A = A^{\orth} + A_0 dt $
with $A_0 \in C^{\infty}(M,\cG)$
and $A^{\orth} \in \cA^{\orth} := \{A \in \cA \mid A(\tfrac{\partial}{\partial t}) =0\}$.\par

Let us now consider manifolds
 $M$ of the form $M=\Sigma \times S^1$. In this situation
  $\tfrac{\partial}{\partial t}$ will denote
 the vector field on $S^1$ which is induced by the curve
 \begin{equation} i_{S^1}: [0,1] \ni s \mapsto \exp(2 \pi i s) \in \{ z
\in \bC \mid \|z\| = 1 \} \cong S^1
\end{equation}
 and  $dt$  the  1-form
 on $S^1$ which is dual to   $\tfrac{\partial}{\partial t}$. Again we can lift $\tfrac{\partial}{\partial t}$ and $dt$
 to a vector field resp. a 1-form on $M$,
 which will again be denoted by $\tfrac{\partial}{\partial t}$ resp. $dt$.
  As before every   $A \in \cA$ can be written uniquely in the form
  $A = A^{\orth} + A_0 dt $ with $A^{\orth} \in \cA^{\orth}$ and $A_0 \in C^{\infty}(M,\cG)$
  where $\cA^{\orth}$ is defined in total analogy to the $\Sigma \times \bR$ case by
 \begin{equation}
 \cA^{\orth} := \{A \in \cA \mid A(\tfrac{\partial}{\partial t}) =0 \}
 \end{equation}

However, there is a crucial difference
between the case $M = \Sigma \times \bR$
and the case $M = \Sigma \times S^1$.
For $M = \Sigma \times \bR$
the condition $A_0 = 0$ (which is equivalent
to the condition  $A \in \cA^{\orth}$) defines a gauge.
More precisely:
Every 1-form $A \in \cA$
is gauge equivalent to a 1-form in $\cA^{\orth}$.
By contrast for $M = \Sigma \times S^1$
  the condition $A_0 = 0$  does \underline{not} define a gauge.
 There are 1-forms $A$ which are not gauge equivalent to any
 1-form in $\cA^{\orth}$.
 For example this is the case for any 1-form $A$ with the property that
  the holonomy $\cP \exp(\int_{l_{\sigma}} A) $ is not equal to $1$ for some $\sigma \in \Sigma$.
  Here $l_{\sigma}$ denotes
  the ``vertical'' loop $[0,1] \ni s \mapsto (\sigma,i_{S^1}(s)) \in M$ ``above'' the
  fixed point  $\sigma \in \Sigma$. This follows immediately
   from the two observations that, firstly,
$\cP \exp(\int_{l_{\sigma}} A^{\orth}) =1$
 for every $A^{\orth} \in \cA^{\orth}$
 and, secondly,   for two gauge equivalent 1-forms $A_1 \in \cA$ and
$A_2 \in \cA$  the corresponding holonomies $\cP
\exp(\int_{l_{\sigma}} A_1)$ and $\cP \exp(\int_{l_{\sigma}} A_2)$
are conjugated to each other.\par
  Thus, in order to obtain a proper gauge we have to weaken the condition $A_0 = 0$.
 There are two natural candidates for such a weakened condition.\\

  \noindent {\bf 1. Option:} Instead of demanding  $A_0(\sigma,t) = 0$ for all $\sigma \in \Sigma, t \in S^1$
  we just demand that $A_0(\sigma,t)$ is independent of the second variable $t$, i.e.
  we demand that
  $A_0 = B$ holds where $B \in C^{\infty}(\Sigma,\cG) \subset C^{\infty}(M,\cG)$ (``quasi-axial gauge fixing'').\\

  \noindent {\bf 2. Option:} We  demand, firstly, that $A_0(\sigma,t)$ is independent of the variable $t$ and, secondly, that it takes values in the  Lie algebra $\ct$ of a fixed  maximal torus $T \subset G$
 (``torus gauge fixing'').\par

Accordingly, let us  introduce the spaces
\begin{align}
 \cA^{qax} & := \cA^{\orth} \oplus  \{ B dt \mid B \in C^{\infty}(\Sigma,\cG)\} \\
  \cA^{qax}(T) & := \cA^{\orth} \oplus  \{ B dt \mid B \in C^{\infty}(\Sigma,\ct)\}
\end{align}

\subsection{Some technical details for quasi-axial gauge fixing}
\label{subsec2.2}

Let us first analyze when/if quasi-axial gauge fixing really is a ``proper'' gauge fixing
in the sense that  every gauge field is gauge-equivalent to a ``quasi-axial''
gauge field. In order to answer this question we start with a fixed gauge field
$A \in \cA$ and  try to find a
$A^q = \cA^{\orth} + B dt \in \cA^{qax}$, $\cA^{\orth} \in \cA$, $B \in C^{\infty}(\Sigma,\cG)$,
 and an element  $\tilde{\Omega}$ of the subgroup\footnote{it is
 more convenient to work with the subgroup  $\tilde{\G}$
 of $\G$ here. We remark that $\G$
is the semi-direct product of $\tilde{\G}$ and $\G_{\Sigma} \subset
\G$. The  factor $\G_{\Sigma}$ will be taken care of in Sec.
\ref{subsec2.4} below.} $\tilde{\G}:= \{ \Omega \in \G \mid
\Omega(\sigma,t_0) = 1
 \text{ for all } \sigma \in \Sigma \}$ of $\G$
 such that
\begin{equation} \label{eq_gauging}
A = A^q \cdot \tilde{\Omega} = (A^{\orth} + B dt) \cdot
\tilde{\Omega}
\end{equation}
holds. Here ``$\cdot$'' denotes the standard right operation of $\G$
on $\cA$ (the ``gauge operation''; cf. Appendix C below). Taking into account that Eq.
\eqref{eq_gauging} implies
\begin{equation}
g_A(\sigma):= \cP \exp(\int_{l_{\sigma}} A) =
 \cP \exp(\int_{l_{\sigma}} A^{\orth} + B dt) = \exp(B(\sigma)) \quad \forall \sigma \in \Sigma
\end{equation}
where $l_{\sigma}$  denotes again the ``vertical'' loop  above the point
$\sigma$ it is clear that in order to find such a $A^q \in \cA^{qax}$
 one first has to find a lift $B:\Sigma \to \cG$
of $g_A: \Sigma \to G$ w.r.t. the projection $\exp:\cG \to G$.
 In order to find such a  lift $B$ it is tempting to
 apply the standard theory of coverings, see e.g. \cite{Hu}.
 What  complicates matters
 somewhat is that $\exp:\cG \to G$ is not a covering
 if $G$ is Non-Abelian.
 On the other hand $\exp:S^* \to G_{reg}$ where $G_{reg}$  denotes the set of all
 ``regular''\footnote{i.e. the
set of all $g \in G$ such that $g$ is contained in a unique maximal torus of $G$}
   elements of $G$  and where $S^*$ is any fixed connected component of $\exp^{-1}(G_{reg})$
 is a (connected) covering.
 So if $g_A: \Sigma \to G$ takes only values in $G_{reg}$ then we can apply the standard  theory of coverings
 and conclude  that at least in the following two situations
 there is a  (smooth) lift
 $B:\Sigma \to S^*$ of $g_A$:
 \begin{enumerate}
 \item  $\Sigma$ is simply-connected. In this case the existence of the lift $B$ follows
  from the well-known ``Lifting Theorem''.
 \item $G$  is simply-connected. In this case the existence of the lift $B$ follows
  from the fact that then also $G_{reg}$ is simply-connected  (cf. \cite{Br_tD}, Chap. V, Sec. 7) and, consequently,
  the covering $\exp:S^* \to G_{reg}$ is just a bijection.
\end{enumerate}

Accordingly, let us assume for the rest of this paper that $G$ or $\Sigma$ is simply-connected.\par

 Once such a lift $B$ is found it is not difficult to find also a $\tilde{\Omega} \in \tilde{\G}$ and a $A^{\orth}$
 such that \eqref{eq_gauging} is fulfilled with $A^q:= A^{\orth} + B dt$.
 Thus   if $\Sigma$ or $G$ is simply-connected then
 $\cA_{reg} \subset \cA^{qax} \cdot \tilde{\G}  $
where
\begin{equation}
\cA_{reg} := \{ A \in \cA \mid g_A:\Sigma \to G \text{ takes values in } G_{reg} \}
\end{equation}
It can be shown that the codimension
of the  subset $G \backslash G_{reg}$ of $G$
is at least $3$.
So in the special case when $\dim(\Sigma)=2$
it is intuitively clear that for ``almost all'' $A \in \cA$
the function $g_A$ will take values in $G_{reg}$.
In other words: the set $\cA  \backslash \cA_{reg}$
 is then ``negligible''.
Accordingly, let us assume for the rest of this paper that $\Sigma$
is 2-dimensional.

\subsection{The Faddeev-Popov determinant for quasi-axial gauge fixing}
\label{subsec2.3}

Let us begin with some general remarks.
Assume that $\cA_{gf}$ is a ``gauge fixing'' linear\footnote{in view of the discussion
 after Eq. \eqref{eq_cAgf_tilde} below
let us mention that the  condition that $\cA_{gf}$ is a linear subspace can be weakened,
cf. Remark \ref{rm_appC} and Remark  \ref{rm_final_appC} in Appendix C} subspace of $\cA$,
i.e. a linear subspace with the property that
$$q:\cA_{gf} \times \G \ni (A,\Omega) \mapsto A \cdot \Omega \in \cA$$
is a bijection. Let us fix a linear bijection\footnote{here and later, when we introduce $DA$,
we assume implicitly that we have fixed a scalar product on $\cA$} $(\cA_{gf})^{\orth} \to   C^{\infty}(M,\cG) $ or,
equivalently,   a linear surjective map
$F:\cA \to C^{\infty}(M,\cG)$ with $\ker(F) = \cA_{gf}$.\par

Using the standard  physicists procedure (cf., e.g., \cite{Pok}) we obtain, informally,   for every gauge-invariant (i.e. $\G$-invariant) function $\chi:\cA \to \bC$
 \begin{align} \label{eq_FadPop00}
 \int_{\cA} \chi(A) DA  = \int_{\cA} \chi(A) \triangle_{FP}[A] \delta(F(A)) DA
\end{align}
where is $DA$  is the informal   ``Lebesgue measure'' on $\cA$
and  $\triangle_{FP}$ is  the so-called ``Faddeev-Popov-determinant associated to $F$''
 (cf. Sec. 3.1 in \cite{Pok}).
Since $F$ is linear, the expression on the r.h.s. of Eq. \eqref{eq_FadPop00} can be simplified
and we obtain\footnote{this step (and the meaning of the notation $\delta(F(A))$ above)
 should become clear by considering a simple finite-dimensional
analogue: for $\phi \in C^{\infty}(\bR^2,\bR)$
and  $F:\bR^2 \to \bR$ given by $F(x)=x_2$
the finite dimensional analogue of the argument above
is $\int_{\bR^2} \phi(x) \delta(F(x)) dx = \int_{\bR}\int_{\bR} \phi(x_1,x_2) \delta(x_2) dx_2 dx_1
= \int_{\bR} \phi(x_1,0)  dx_1 = \int_{\ker(F)} \phi(x) dx$
where the last $dx$ is the normalized Lebesgue measure on $\ker(F) = \bR \times \{0\}$
equipped with the scalar product inherited from $\bR^2$.
For a more general linear  form $F: \bR^2 \to \bR$ equality will hold only up to a multiplicative constant}
\begin{subequations} \label{eq_genFadPop}
 \begin{align} \label{eq_genFadPop1}
 \int_{\cA} \chi(A) DA  \sim \int_{\cA_{gf}} \chi(A) \triangle_{FP}[A]  DA
\end{align}
 For   $A \in \cA_{gf}$ the value  $\triangle_{FP}[A]$
 is given explicitly by
\begin{equation} \label{eq_genFadPop2}
\triangle_{FP}[A] = \det\bigl( \tfrac{\delta F(A \cdot \exp(\eta))}{\delta \eta}_{| \eta=0}\bigr)
\end{equation}
\end{subequations}
 where $\frac{\delta F(A \cdot \exp(\eta))}{\delta \eta}_{| \eta=0}$
 is a physicist type of notation for the (informal) total derivative of the map
 $C^{\infty}(M,\cG) \ni \eta \mapsto F(A \cdot \exp(\eta)) \in C^{\infty}(M,\cG)$
 in the point $0 \in C^{\infty}(M,\cG)$.

 \medskip

 For the convenience of the reader  we will give in Appendix C a direct derivation of
  Eqs. \eqref{eq_genFadPop1}--\eqref{eq_genFadPop2}, which may be more accessible for  readers with a mathematics background than   the standard derivation of  Eqs. \eqref{eq_genFadPop1}--\eqref{eq_genFadPop2} in the physics literature.

\medskip

Let us now come back to the situation relevant for us.
As in the previous subsection fix  an arbitrary  connected component  $S^*$  of $\exp^{-1}(G_{reg})  \subset \cG$.
Moreover, set
\begin{equation} \label{eq_cAgf_tilde}
\tilde{\cA}_{gf} := \cA^{\orth} \oplus C^{\infty}(\Sigma,S^*)dt \subset \cA^{qax} \cap \cA_{reg}
\end{equation}
Observe that  $\tilde{\cA}_{gf}$ is not a gauge fixing subspace of $\cA$.
However, it has the  following analogous property: the   map
$$\tilde{q}: \tilde{\cA}_{gf} \times \tilde{\G} \ni (A,\tilde{\Omega})
\mapsto A \cdot \tilde{\Omega} \in  \cA_{reg}$$

 is a bijection, cf.  Proposition 3.1 in \cite{Ha3b} and
Proposition 3.1 in \cite{Ha3c}.
We can therefore hope to be able
to derive an analogue of Eqs. \eqref{eq_genFadPop1}--\eqref{eq_genFadPop2}
where  the role of $\G= C^{\infty}(M,G) = C^{\infty}(\Sigma \times S^1,G)$ is played by the group
$\tilde{\G} = \{ \Omega \in \G \mid
\Omega(\sigma,t_0) = 1  \text{ for all } \sigma \in \Sigma \}$
and the role  of the space $C^{\infty}(M,\cG) = C^{\infty}(\Sigma \times S^1,\cG)$,
which arises naturally as the  informal
(infinite-dimensional) ``Lie algebra'' of $\G$,
will be played by the space
\begin{equation}
\tilde{C}^{\infty}(\Sigma \times S^1,\cG)  := \{f \in  C^{\infty}(\Sigma \times S^1,\cG) \mid
f(\sigma,t_0) = 0 \quad \forall \sigma \in \Sigma \}\\
\end{equation}
which can be considered as the informal ``Lie algebra'' of $\tilde{\G}$.
Beside the space $\tilde{C}^{\infty}(\Sigma \times S^1,\cG)$
let us also introduce the space
\begin{equation}
\Check{C}^{\infty}(\Sigma \times S^1,\cG)  := \{f \in  C^{\infty}(\Sigma \times S^1,\cG) \mid
\int f(\sigma,t) dt = 0 \quad \forall \sigma \in \Sigma \}
\end{equation}
It is not difficult to see that  the map
\begin{equation} \label{eq_diff_bijection} \tfrac{\partial}{\partial t}: \tilde{C}^{\infty}(\Sigma \times S^1,\cG) \to \Check{C}^{\infty}(\Sigma \times S^1,\cG)
\end{equation}
is a well-defined bijection and that  the linear map
$F:\cA \to \Check{C}^{\infty}(\Sigma \times S^1,\cG)$  given by
$F(A) = \frac{\partial}{\partial t} A_0$
is a well-defined surjection with $\ker(F) = \cA^{qax}$.

\smallskip

 Taking into account that the set $\cA  \backslash \cA_{reg}$
 is ``negligible'' when $\dim(\Sigma)=2$
 and modifying the derivation of Eqs. \eqref{eq_genFadPop1}--\eqref{eq_genFadPop2} in Appendix C
 in an appropriate way\footnote{Clearly, $\tilde{\cA}_{gf}$ and $\cA_{reg}$
 are not linear spaces but they can be considered as ``open'' subspaces of the vector spaces
 $\cA^{qax}$ and $\cA$ in the sense of  Remark \ref{rm_final_appC} in Appendix C.
 Thus, informally, we have the identifications
 $T_{A} \cA_{reg} \cong \cA$ for each $A \in \cA_{reg} \subset \cA$ and
 $T_{A_0} \tilde{\cA}_{gf} \cong \cA^{qax}$ for each $A_0 \in \tilde{\cA}_{gf} \subset \cA^{qax}$
 and we can therefore consider  each $d\tilde{q}_{| (A_0,\Omega)}$,  $A_0 \in \tilde{\cA}_{gf}$, $\Omega \in \tilde{\G}$ as a linear map
 $\cA^{qax} \oplus \tilde{C}^{\infty}(\Sigma \times S^1,\cG) \to \cA$.
 We now choose $\Psi: \cA \to \cA^{qax} \oplus \tilde{C}^{\infty}(\Sigma \times S^1,\cG)$ to be a linear isomorphism with the extra property $\Psi_2 = \tfrac{\partial}{\partial t}^{-1}  \cdot F$
 and define $\triangle_{FP}[A_0]$ for $A_0 \in \tilde{\cA}_{gf}$ by the analogues of Eqs. \eqref{eq_def_triangle}  and \eqref{eq_def_triangle_FP} in Appendix C.
 Setting  $H:= \Psi \circ \tilde{q} \circ (\id_{\cA^{qax}},\tilde{\exp})$
we obtain    $\triangle_{FP}[A_0] = \det\bigl(\frac{\delta H_2(A_0, \tilde{\eta})}{\delta \tilde{\eta}}_{| \tilde{\eta} = 0 }\bigr) = \det\bigl(\tfrac{\partial}{\partial t}^{-1}  \cdot \frac{ \delta F(A_0 \cdot \exp(\tilde{\eta}))}{\delta \tilde{\eta}}_{| \tilde{\eta} = 0 }\bigr)$
for each $A_0 \in \tilde{\cA}_{gf}$}
  we obtain,  informally,   for every  $\tilde{\G}$-invariant function $\chi:\cA \to \bC$
\begin{subequations} \label{eq_FadPop0_ges}
 \begin{align} \label{eq_FadPop0}
 \int_{\cA} \chi(A) DA  = \int_{\cA_{reg}} \chi(A) DA  \sim \int_{\tilde{\cA}_{gf}} \chi(A) {\triangle}_{FP}[A] DA
\end{align}
where  for each $A \in \tilde{\cA}_{gf}$ we have set
 \begin{equation} \label{eq_FadPop0'}
 {\triangle}_{FP}[A] := \det\bigl(\tfrac{\partial}{\partial t}^{-1} \cdot \tfrac{\delta F(A \cdot \tilde{\exp}(\tilde{\eta}))}{\delta \tilde{\eta}}_{|\tilde{\eta}=0}\bigr)
\end{equation}
\end{subequations}
Here  $\tilde{\exp}: \tilde{C}^{\infty}(\Sigma \times S^1,\cG)  \to \tilde{\G}$ is
 given by $(\tilde{\exp}(f))(x) = \exp_G (f(x))$ for $f \in  \tilde{C}^{\infty}(\Sigma \times S^1,\cG)$
  and $x \in M = \Sigma \times S^1$ and
  the zero element $0$ refers to the space $ \tilde{C}^{\infty}(\Sigma \times S^1,\cG)$.
  Moreover,      $\tfrac{\partial}{\partial t}^{-1}$ is the inverse of the bijection \eqref{eq_diff_bijection} above.

\smallskip

   After a short computation  we  obtain the following explicit  expression
\begin{multline}
{\triangle}_{FP}(A^{\orth}+Bdt) = \det\bigl( \tfrac{\partial}{\partial t}^{-1} \cdot \tfrac{\delta F((A^{\orth}+Bdt) \cdot \tilde{\exp}(\tilde{\eta}))}{\delta\tilde{\eta}}_{| \tilde{\eta}=0}\bigr) \\
=  \det \bigl(  \tfrac{\partial}{\partial t}^{-1} \cdot \bigl(\tfrac{\partial}{\partial t} + \ad(B) \bigl) \cdot   \tfrac{\partial}{\partial t}\bigr) = \det\bigl( \tfrac{\partial}{\partial t} + \ad(B)\bigl)
\end{multline}
where $\tfrac{\partial}{\partial t} + \ad(B)$ is the obvious operator
on $\Check{C}^{\infty}(\Sigma \times S^1,\cG)$.

\smallskip

 The last determinant can be evaluated explicitly in several ways,
for example by using a suitable $\zeta$-function regularization. By doing so we obtain\footnote{
for step $(*)$ observe that the operator $\tfrac{\partial}{\partial t} + \ad(B)$
on $\Check{C}^{\infty}(\Sigma \times S^1,\cG) \cong \Check{C}^{\infty}(\Sigma \times S^1,\cG_0) \oplus \Check{C}^{\infty}(\Sigma \times S^1,\ct)$
coincides with $\tfrac{\partial}{\partial t}$ on the subspace $\Check{C}^{\infty}(\Sigma \times S^1,\ct)$}
\begin{equation} \label{eq_det_eval}
 \det\bigl( \tfrac{\partial}{\partial t} + \ad(B)\bigl) \overset{(*)}{\sim}
 \det\bigl( (\tfrac{\partial}{\partial t} + \ad(B))_{|\Check{C}^{\infty}(\Sigma \times S^1,\cG_0)})  \bigl)
 \sim   \det\bigl( \sum\nolimits_{n=0}^{\infty} \tfrac{(\ad(B)_{| \cG_0})^n}{(n+1)!}\bigr)
\end{equation}
where the  space $\Check{C}^{\infty}(\Sigma \times S^1,\cG_0)$ is defined
in an analogous way as the space $\Check{C}^{\infty}(\Sigma \times S^1,\cG)$
and where on the right hand side $\ad(B)_{| \cG_0}$ is a short notation
for the obvious linear operator $\ad(B)_{| C^{\infty}(\Sigma,\cG_0)}: C^{\infty}(\Sigma,\cG_0)
\to C^{\infty}(\Sigma,\cG_0)$.

\smallskip

Combining Eqs. \eqref{eq_FadPop0_ges}--\eqref{eq_det_eval}
we now arrive at\footnote{observe that  in the published version of this paper
(and also in \cite{Ha3b,Ha3c}) we used a wrong variant of this formula
where instead of the factor  $\det\bigl( \sum\nolimits_{n=0}^{\infty} \tfrac{(\ad(B))^n}{(n+1)!}\bigr)$
 the factor $ \det\bigl( \tfrac{\partial}{\partial t} + \ad(B)\bigl)
\det\bigl( \sum\nolimits_{n=0}^{\infty} \tfrac{(\ad(B))^n}{(n+1)!}\bigr)
\sim \det\bigl( \sum\nolimits_{n=0}^{\infty} \tfrac{(\ad(B))^n}{(n+1)!}\bigr)^2$ appeared
(in this footnote we write simply $\ad(B)$ instead of $\ad(B)_{| \cG_0}$).
 However, this did not affect the final formulas at the end of the paper
since after the replacements in Sec. \ref{subsec3.5} below this error was implicitly corrected}
 \begin{equation} \label{eq2.8}
  \int_{\cA} \chi(A) DA
   \sim  \int_{C^{\infty}(\Sigma,S^*)} \biggl[ \int_{\cA^{\orth}} \chi(A^{\orth} + B dt)
   DA^{\orth}  \biggr]   \det\bigl( \sum\nolimits_{n=0}^{\infty} \tfrac{(\ad(B)_{| \cG_0})^n}{(n+1)!}\bigr) DB
  \end{equation}
 where $DA^{\orth}$ resp. $DB$ is the informal Lebesgue measure
 on $\cA^{\orth}$ resp. $C^{\infty}(\Sigma,\cG)$.

\subsection{From quasi-axial to torus gauge fixing}
\label{subsec2.4}

Let us fix an $\Ad$-invariant scalar product
 $(\cdot,\cdot)_{\cG}$ on $\cG$ and,
 once and for all, a maximal torus  $T$ of $G$.
 The  Lie algebra of $T$ will be denoted by $\ct$.
We set
\begin{equation}
\cG_0:= \ct^{\orth}
\end{equation}
where $\ct^{\orth}$ denotes the
$(\cdot,\cdot)_{\cG}$-orthogonal complement of $\ct$ in
$\cG$.
Moreover, let us  fix   an open ``alcove'' (or ``affine
Weyl chamber'')  $P \subset \ct$
and set  $S^* := P \cdot G$
where ``$\cdot$'' denotes the right operation of $G$
on $\cG$ given by $B \cdot g  = g^{-1}B g$.
(Note that $P \cdot G$ is indeed a connected component of $\exp^{-1}(G_{reg})$).
We then have
\begin{align} \label{eq2.11}
 P & \cong S^*/G \\
  \label{eq2.12}
\pi_*(dx) & = \det(-\ad(x)_{| \cG_0}) dx
\end{align}
 Here  $\pi: S^* \to S^*/G \cong P$ is the canonical projection,
 $dx$ denotes both the restriction of Lebesgue measure on $\ct$ onto
$P$  and  the restriction of Lebesgue measure on $\cG$ onto $S^*$,
  and, finally,  $\pi_*(dx)$ is the image of the measure $dx$ on  $S^*$ under the projection
  $\pi$.
  In view of $\G_{\Sigma}=C^{\infty}(\Sigma,G)$
 and   Eqs. \eqref{eq2.11} and  \eqref{eq2.12}
  one could expect
 naively that
\begin{align} \label{eq_wrong}
C^{\infty}(\Sigma,P) & \cong C^{\infty}(\Sigma,S^*)/\G_{\Sigma}\\
\label{eq2.14}
 \pi_*(DB) & = \det(-\ad(B)_{| \cG_0}) DB
\end{align}
holds where   $\pi: C^{\infty}(\Sigma,S^*) \to C^{\infty}(\Sigma,S^*) /\G_{\Sigma}$
is the canonical projection
 and where $DB$ denotes
both the restriction of the informal  ``Lebesgue measure'' on
$C^{\infty}(\Sigma,\cG)$ onto $C^{\infty}(\Sigma,S^*)$ and the
restriction of the informal  ``Lebesgue measure'' on
$C^{\infty}(\Sigma,\ct)$ onto $C^{\infty}(\Sigma,P)$.\par However,
there are well-known topological obstructions (cf. \cite{BlTh3},
\cite{Ha3b}),  which prevent Eq. \eqref{eq_wrong} from holding in
general. Before we take a closer look at these obstructions
 in the general case
let us restrict ourselves for a while to  those (special) situations
 where Eq. \eqref{eq_wrong} does hold (this will be the case if
 $\Sigma$ is non-compact,
cf. the discussion below).
As the  operation of  $\G_{\Sigma}$ on $\cA$ is linear and as it
 leaves the subspace $\cA^{\orth}$ of $\cA$
 and the informal
 measure $DA^{\orth}$ on $\cA^{\orth}$  invariant
 we can  conclude, informally,   that
  the function $\tilde{\chi}(B):C^{\infty}(\Sigma,\cG)
\ni B \mapsto \int \chi(A^{\orth} + Bdt)  DA^{\orth} \in \bC$ is
$\G_{\Sigma}$-invariant (here $\chi$ is as in Subsec. \ref{subsec2.3}).
 Moreover,  the function
 $B \mapsto \det\bigl( \sum\nolimits_{n=0}^{\infty}
 \tfrac{(\ad(B)_{| \cG_0})^n}{(n+1)!}\bigr)$
on $C^{\infty}(\Sigma,\cG)$
 is  $\G_{\Sigma}$-invariant as well.
 In the special situations where  Eq. \eqref{eq_wrong}  holds
  we  therefore obtain, informally,
\begin{multline}  \label{eq_wrong2}
\int_{\cA} \chi(A) DA
 \sim  \int_{C^{\infty}(\Sigma,S^*)} \tilde{\chi}(B)   \det\bigl( \sum\nolimits_{n=0}^{\infty} \tfrac{(\ad(B)_{|\cG_0})^n}{(n+1)!}\bigr)      DB\\
 \overset{(*)}{=} \int_{C^{\infty}(\Sigma,P)} \tilde{\chi}(B)
\det\bigl( \sum_{n=0}^{\infty} \tfrac{(\ad(B)_{|\cG_0})^n}{(n+1)!}
\bigr)\cdot \det\bigl(-\ad(B)_{|\cG_0}\bigr)  DB\\
= \int_{C^{\infty}(\Sigma,P)}   \int
\chi(A^{\orth} + Bdt)  DA^{\orth}
\det\bigl(1_{\cG_0}-\exp(\ad(B)_{|\cG_0})\bigr)   DB
  \end{multline}
  where $1_{\cG_0}$ is the identity operator on $C^{\infty}(\Sigma,\cG_0)$.
Here step $(*)$ follows from Eqs. \eqref{eq_wrong}, \eqref{eq2.14}
and the relations $\tilde{\chi} = \tilde{\chi} \circ \pi$ and $\det( \sum\nolimits_{n=0}^{\infty} \tfrac{(\ad(\cdot)_{| \cG_0})^n}{(n+1)!})=
\det( \sum\nolimits_{n=0}^{\infty} \tfrac{(\ad(\cdot)_{| \cG_0})^n}{(n+1)!})  \circ \pi$.

\begin{remark}  \label{rm_det_interpret}
For later use (cf. the discussion at the end of Sec. \ref{subsec3.4} below)
 let us remark that the expression $\det\bigl(1_{\cG_0}-\exp(\ad(B)_{|\cG_0})\bigr)$
 on the r.h.s. of Eq. \eqref{eq_wrong2} above can be interpreted as the determinant of a suitable operator.
 More precisely, we have at an informal level
 \begin{equation} \label{eq_Rm2.1}
 \det\bigl(1_{\cG_0}-\exp(\ad(B)_{|\cG_0})\bigr) \sim \det\bigl( (\tfrac{\partial}{\partial t} + \ad(B))_{|
 C^{\infty}(\Sigma \times S^1,\cG_0)} \bigl)
\end{equation}
Eq. \eqref{eq_Rm2.1}  follows easily from Eq. \eqref{eq_det_eval} above
if one takes into account that $\tfrac{\partial}{\partial t} + \ad(B)$
agrees with $\ad(B)$ on the orthogonal complement of  $\Check{C}^{\infty}(\Sigma \times S^1,\cG_0)$ in  $C^{\infty}(\Sigma \times S^1,\cG_0)$.
\end{remark}

\bigskip

Let us now go back to the general
case where,
 because of the topological obstructions mentioned  above, Eq.  \eqref{eq_wrong}
 need not hold.
  In order to find a suitable generalization of Eq. \eqref{eq_wrong2} we now consider the
 bijection $P \times G/T \ni (B,gT) \mapsto g B  g^{-1} \in S^*$.
 Clearly, this bijection induces a  bijection
$C^{\infty}(\Sigma,P) \times C^{\infty}(\Sigma,G/T) \to
C^{\infty}(\Sigma,S^*)$, so we can identify the space $C^{\infty}(\Sigma,P) \times C^{\infty}(\Sigma,G/T)$
with the space $C^{\infty}(\Sigma,S^*)$.
After identifying these two spaces
the operation of $\G_{\Sigma}$ on $C^{\infty}(\Sigma,P) \times C^{\infty}(\Sigma,G/T) \cong
C^{\infty}(\Sigma,S^*)$ can be written in the form
$(B,\bar{g}) \cdot \Omega = (B, \Omega^{-1}  \bar{g})$
from which
 \begin{equation} \label{eq_2.15} C^{\infty}(\Sigma,S^*)/\G_{\Sigma} \cong C^{\infty}(\Sigma,P) \times ( C^{\infty}(\Sigma,G/T)/ \G_{\Sigma})
 \end{equation}
 follows.
For a proof of the following proposition see \cite{Ha3c}.

 \begin{proposition} \label{lem1}
Under the assumptions made in Subsec. \ref{subsec2.2}, i.e.
$\dim(\Sigma)=2$ and  $G$ or $\Sigma$ is simply-connected, we have
 $$C^{\infty}(\Sigma,G/T)/ \G_{\Sigma}  =  [\Sigma,G/T] $$
 \end{proposition}

 Let us now fix for the rest of this paper
  a  representative $\bar{g}_{\cl} \in
 C^{\infty}(\Sigma,G/T)$ for each homotopy class
  $\cl \in [\Sigma,G/T]$.
  For $\bar{g} = gT \in G/T$ we will denote by
  $\bar{g} B \bar{g}^{-1}$ the element  $ gBg^{-1}$ of $G$
 (which clearly does not depend on the special choice of $g$).
  Taking into account that
   $$   \det\bigl(1_{\cG_0}-\exp(\ad(B)_{| \cG_0})\bigr)
  =   \det\bigl(1_{\cG_0}-\exp(\ad((\bar{g}_{\cl} \cdot B \cdot \bar{g}_{\cl}^{-1}))_{| \cG_0})\bigr)$$
one arrives at the following generalization
 of Eq. \eqref{eq_wrong2} above
  \begin{multline} \label{eq_torusgaugefixing}
\int_{\cA} \chi(A) DA  \sim \sum_{\cl \in [\Sigma,G/T]}
 \int_{C^{\infty}(\Sigma,P)} \biggl[ \int_{\cA^{\orth}} \chi(A^{\orth} + (\bar{g}_{\cl} \cdot B \cdot \bar{g}_{\cl}^{-1}) dt)  DA^{\orth} \biggr] \\
\times  \det\bigl(1_{\cG_0}-\exp(\ad(B)_{| \cG_0})\bigr)  DB
\end{multline}
Note that because of $C^{\infty}(\Sigma,G/T)/ \G_{\Sigma} = [\Sigma,G/T]$
and the $\G_{\Sigma}$-invariance of $\tilde{\chi}(B) =  \int_{\cA^{\orth}} \chi(A^{\orth} + B  dt)  DA^{\orth}$
the expression $\int_{\cA^{\orth}} \chi(A^{\orth} + (\bar{g}_{\cl} \cdot B \cdot \bar{g}_{\cl}^{-1}) dt)  DA^{\orth}$ above does not depend on the special choice of $\bar{g}_{\cl}$.\par

If $\Sigma$ is non-compact then all continuous mappings $\Sigma
\to G/T$ are homotopic to each other. In other words, we then have
$[\Sigma,G/T] = \{[1_T]\}$ where $1_T: \Sigma \to G/T$ is the
constant map taking only the value $T \in G/T = \{gT \mid g \in
G\}$. So in this special situation Eq. \eqref{eq_torusgaugefixing}
reduces to   Eq.
\eqref{eq_wrong2}. For compact $\Sigma$, however, we will have
to work with Eq. \eqref{eq_torusgaugefixing}. Thus for compact
$\Sigma$, the functions $\bar{g}_{\cl} \cdot B \cdot
\bar{g}_{\cl}^{-1}$ will in general not take only values in $\ct$.
This  reduces the usefulness of  Eq. \eqref{eq_torusgaugefixing} considerably.
Fortunately, for many functions $\chi$
it is possible to derive an   ``Abelian version'' of
Eq. \eqref{eq_torusgaugefixing}, as we will now show.

\subsection{A useful modification  of Eq. \eqref{eq_torusgaugefixing} for compact $\Sigma$}
\label{subsec2.5}

Recall that we have fixed a point $\sigma_0 \in \Sigma$.
 Clearly, the restriction mapping
 $\G_{\Sigma} \ni \Omega \mapsto \Omega_{| \Sigma \backslash \{\sigma_0 \}} \in \G_{ \Sigma \backslash \{\sigma_0 \}}$ is injective
so we can identify $\G_{\Sigma}$ with a subgroup of $\G_{ \Sigma \backslash \{\sigma_0 \}}$.
Similarly, let us identify the spaces
$\cA^{\orth}$, $\cA^{qax}$, $C^{\infty}(\Sigma,G/T)$, and $C^{\infty}(\Sigma,\cG)$
with the obvious subspaces of
 $\cA^{\orth}_{(\Sigma \backslash \{\sigma_0 \}) \times S^1}$ resp.
 $\cA^{qax}_{(\Sigma \backslash \{\sigma_0 \}) \times S^1}$
 resp. $C^{\infty}(\Sigma \backslash \{\sigma_0 \},G/T)$ resp.
  $C^{\infty}(\Sigma \backslash \{\sigma_0 \},\cG)$.\par
As $\Sigma \backslash \{\sigma_0\}$ is noncompact
 every $\bar{g} \in C^{\infty}(\Sigma \backslash \{\sigma_0\} ,G/T)$
 is $0$-homotopic and can therefore be lifted to an element of
  $C^{\infty}(\Sigma \backslash \{\sigma_0\} ,G) = \G_{\Sigma \backslash \{\sigma_0\}}$,
  i.e. there is always a $\Omega \in \G_{\Sigma \backslash \{\sigma_0\}}$
  such that $\bar{g} = \pi_{G/T} \circ \Omega$
  where $\pi_{G/T}: G \to G/T$ is the canonical projection.
 We will now pick
 for each $\cl \in [\Sigma,G/T]$
 such a lift $\Omega_{\cl} \in \G_{\Sigma \backslash \{\sigma_0\}}$
 of  the representative $\bar{g}_{\cl} \in C^{\infty}(\Sigma, G/T) \subset
  C^{\infty}(\Sigma \backslash \{\sigma_0\}, G/T)$ of $\cl$ which we have
fixed above.\par
 Let $\chi: \cA \to \bC$ be a $\G$-invariant function.
 The space  $\cA^{qax} \subset \cA$
is clearly $\G_{\Sigma}$-invariant
so $\G_{\Sigma}$ operates on $\cA^{qax}$
and the function $\chi^{qax} := \chi_{| \cA^{qax}}$
in invariant under this operation.
Let us now make the additional assumption that  $\chi^{qax}: \cA^{qax} \to \bC$
can be extended to a
function $\overline{\chi^{qax}}: \cA^{qax}_{(\Sigma\backslash \{\sigma_0\}) \times S^1} \to \bC$
which is $\G_{ \Sigma \backslash \{\sigma_0 \}}$-invariant,
or at least $\overline{\G_{\Sigma}}$-invariant, where
$\overline{\G_{\Sigma}}$ is the subgroup
of $\G_{ \Sigma \backslash \{\sigma_0 \}}$ which is generated by
$\G_{\Sigma}$ and all  $\Omega_{\cl}$,  $\cl \in [\Sigma,G/T]$.
Then we obtain
for the integrand in the inner integral on the right-hand side of
\eqref{eq_torusgaugefixing}
\begin{multline}
\chi(A^{\orth} + (\bar{g}_{\cl} \cdot B \cdot \bar{g}_{\cl}^{-1}) dt)
= \chi^{qax}(A^{\orth} + (\bar{g}_{\cl} \cdot B \cdot \bar{g}_{\cl}^{-1}) dt)
 = \overline{\chi^{qax}}(A^{\orth} + (\Omega_{\cl} \cdot B \cdot \Omega_{\cl}^{-1}) dt)  \\
= \overline{\chi^{qax}}( (A^{\orth} \cdot \Omega_{\cl}) +  B \cdot  dt)
= \overline{\chi^{qax}}( \Omega^{-1}_{\cl} A^{\orth} \Omega_{\cl} + \Omega_{\cl}^{-1} d  \Omega_{\cl} +  B \cdot  dt)
\end{multline}
Thus, for such a function $\chi$
 we arrive at  the following useful modification of \eqref{eq_torusgaugefixing}
\begin{multline} \label{eq_Abelian_version}
\int_{\cA} \chi(A) DA  \sim \sum_{\cl \in [\Sigma,G/T]}
 \int_{C^{\infty}(\Sigma,P)} \biggl[ \int_{\cA^{\orth}} \overline{\chi^{qax}}(\Omega^{-1}_{\cl} A^{\orth} \Omega_{\cl} + \Omega_{\cl}^{-1} d  \Omega_{\cl} +  B \cdot  dt) DA^{\orth} \biggr]\\
\times   \det\bigl(1_{\cG_0}-\exp(\ad(B)_{| \cG_0})\bigr)  DB
\end{multline}

\subsection{Identification of  $ [\Sigma,G/T]$  for compact oriented surfaces $\Sigma$}
\label{subsec2.6}

Recall that we have been  assuming  that $\dim(\Sigma)=2$.
Let us now assume additionally that $\Sigma$ is oriented and compact.
Moreover, let us assume for simplicity that $G$ is simply-connected (the case
 where not $G$ is assumed to be simply-connected but $\Sigma$ is covered by Remark \ref{rm_spaet} below).
 Then there is a natural bijection from
 the set
$ [\Sigma,G/T]$ onto  $\pi_2(G/T) \cong \Ker(\exp_{| \ct}) \cong \bZ^r$ where $r = \rank(G)$,
see \cite{BlTh3} and \cite{Br_tD}, Chap. V, Sec. 7.
Instead of recalling the abstract  definition
of this  bijection  $ [\Sigma,G/T] \to \Ker(\exp_{| \ct})$
we  will  give a  more concrete description,
 which will be  more useful for our purposes.\par
 Let, for any fixed auxiliary  Riemannian metric on $\Sigma$,
$B_{\eps}(\sigma_0)$ denote the closed ball around $\sigma_0$ with radius $\eps$.
It is not difficult to see that
for each $\cl \in [\Sigma,G/T]$ the limit
\begin{equation}
\int_{\Sigma \backslash \sigma_0}  d(\Omega_{\cl}^{-1} d \Omega_{\cl}) := \lim_{\eps \to 0} \int_{\Sigma \backslash B_{\eps}(\sigma_0)}  d(\Omega_{\cl}^{-1} d \Omega_{\cl})
\end{equation}
exists and is independent of the choice of the auxiliary Riemannian metric.
Let us set
\begin{equation} n(\Omega_{\cl}) := \pi_{\ct}( \int_{\Sigma \backslash \{\sigma_0\}}  d(\Omega_{\cl}^{-1} d \Omega_{\cl})) \in \ct
\end{equation}
where  $\pi_{\ct}$ denotes the orthogonal projection $\cG \to \ct$.
Then we have
 \begin{proposition} \label{prop2.1}
\begin{enumerate}
\item $n(\Omega_{\cl})$ depends neither on the special choice of the
lift $\Omega_{\cl}$ of  $(\bar{g}_{\cl})_{| \Sigma \backslash \{\sigma_0\}}$ nor on the special
choice of the representative  $\bar{g}_{\cl} \in C^{\infty}(\Sigma, G/T)$ of $\cl$.
It only depends on $\cl$. Thus we can set $n(\cl):= n(\Omega_{\cl})$.
\item If $G$ is simply-connected then the mapping $[\Sigma,G/T] \ni \cl \mapsto n(\cl) \in \ct$
is a bijection from $[\Sigma,G/T]$ onto  $\Ker(\exp_{| \ct})$.
In particular, we then have
 \begin{equation} \label{eq_Image_of_n} \{ n(\cl) \mid \cl \in [\Sigma,G/T]\} =  \{B \in \ct \mid \exp(B) =1\}
 \end{equation}
\end{enumerate}
 \end{proposition}
 For an elementary proof of this proposition, see \cite{Ha3c}
 (cf. also Sec. 5 in \cite{BlTh3} for a very similar  result).

\begin{remark} \label{rm_spaet} \rm
The first part  of Proposition \ref{prop2.1} holds also when $G$ is
not assumed to be simply-connected. However, in this case the second
part must modified. More precisely, if $G$ is not simply-connected
then the mapping $[\Sigma,G/T] \ni \cl \mapsto n(\cl) \in \ct$ is a
bijection from $[\Sigma,G/T]$ onto the subgroup $\Gamma$ of
$\Ker(\exp_{| \ct})$ which is generated by the inverse roots, cf.
\cite{Br_tD}, Chap. V, Sec. 7.
\end{remark}

\section{Torus gauge fixing applied to  Chern-Simons models on $\Sigma \times S^1$}
\label{sec3}

\subsection{Chern-Simons models and Wilson loop observables}
\label{subsec3.1}

For the rest of this paper we will not only assume that
$\dim(\Sigma)=2$
but, additionally, that $\Sigma$ is compact and oriented.
Moreover, $G$ will be assumed to be either Abelian or
simply-connected and simple
  (in the former case we will also assume that $\Sigma \cong S^2$).\par
Without loss of generality  we can
assume that $G$ is a Lie subgroup of $U(N)$, $N \in \bN$.
The Lie algebra $\cG$ of $G$ can then be identified with the obvious Lie subalgebra of
 the Lie algebra $u(N)$ of $U(N)$.\par
For the rest of this paper  we will fix an integer  $k \in \bZ
\backslash \{0\}$ and set
 $$\lambda:= \tfrac{1}{k}.$$
  With the assumptions above  $M=\Sigma \times S^1$ is   an oriented compact
3-manifold. Thus the  Chern-Simons  action function  $S_{CS}$
corresponding to the triple $(M,G,k)$ is well-defined and given by
$$S_{CS}(A) = \tfrac{k}{4\pi} \int_M \Tr(A \wedge dA + \tfrac{2}{3} A\wedge A\wedge A), \quad
A \in \cA$$
with $\Tr:= c \cdot \Tr_{\Mat(N,\bC)}$
where $c$ is a suitable normalization constant
chosen such that the exponential $\exp(i S_{CS})$ is $\G$-invariant.
If $G$ is Abelian or if $G=SU(N)$ (which are the only two types of groups
for which we will make concrete computations below)
 this will be the case, e.g. if $c=1$,
which is the value of $c$ that we will choose for these groups.\par

From the definition of $S_{CS}$ it is obvious that $S_{CS}$ is invariant under
(orientation-preserving) diffeomorphisms.
Thus, at a heuristic level, we can expect that the
heuristic integral (the ``partition function'')
$$Z(M) := \int  \exp(i S_{CS}(A)) DA$$
is a topological invariant of the  3-manifold $M$.
Here $DA$ denotes again the informal ``Lebesgue measure'' on the space $\cA$.\par
Similarly,  we can expect that the mapping
which maps every sufficiently ``regular''
 colored link $L= ((l_1, l_2, \ldots, l_n),(\rho_1,\rho_2,\ldots,\rho_n))$ in $M$
to the heuristic integral (the ``Wilson loop observable'' associated to $L$)
\begin{equation} \label{eq_WLO}
\WLO(L)  := \frac{1}{Z(M)} \int \prod_i  \Tr_{\rho_i}\bigl(\cP \exp\bigl(\int_{l_i} A\bigr)\bigr) \exp(i S_{CS}(A)) DA
\end{equation}
is a link invariant (or, rather, an invariant of colored links).
Here  $\cP \exp\bigl(\int_{l_i} A\bigr)$ denotes the
  holonomy of $A$ around the loop $l_i$ and
 $\Tr_{\rho_i}$, $i \le n$,  the trace in the finite-dimensional representation $\rho_i$
of $G$. \par
For the rest of this paper, we will now fix
a  ``sufficiently regular'' colored link
$$L= ((l_1, l_2, \ldots, l_n),(\rho_1, \rho_2, \ldots,\rho_n))$$
  in $M$ and set $\underline{\rho}:= (\rho_1,\rho_2,\ldots,\rho_n)$.
The  ``uncolored'' link $(l_1, l_2, \ldots, l_n)$ will  also be denoted
by $L$.   In order to make precise what we mean with ``sufficiently regular'' above
 we will introduce the following definitions:\par
   Let $\pi_{\Sigma}$ (resp. $\pi_{S^1}$) denote the canonical
projection $\Sigma \times S^1 \to \Sigma$ (resp. $\Sigma \times
S^1 \to S^1$).
For each $j \le n$ we will set
$l^j_{\Sigma} := \pi_{\Sigma} \circ l_j$ and $l^j_{S^1} := \pi_{S^1} \circ l_j$.
Similarly, we will set
$c_{\Sigma}:=  \pi_{\Sigma} \circ c$ and $c_{S^1}:=  \pi_{S^1} \circ c$
for an arbitrary curve  $c$ in $\Sigma \times S^1$.
We will call $p \in \Sigma$ a ``double point'' (resp. a ``triple point'') of $L$
if the intersection of $\pi_{\Sigma}^{-1}(\{p\})$ with the union of the arcs of $l_1, l_2, \ldots l_n$
contains at least two (resp. at least three) elements.
The set of double points of $L$ will be denoted by $\DP(L)$.
We will assume in the sequel (with the exception of Subsec. \ref{subsec6.2} below
where we study ``vertical'' links)
that the link $L$ is ``admissible'' in the following sense:
\begin{enumerate}
\item[(A1)] There are only finitely many double points and no triple points of $L$
\item[(A2)] For each $p \in \DP(L)$ the corresponding tangent vectors,
i.e. the vectors $(l^i_{\Sigma})'(\bar{t})$ and $(l^{j}_{\Sigma})'(\bar{u})$ in $T_p\Sigma$
where  $\bar{t}, \bar{u} \in [0,1]$, $i, j \le n$,  are given by
$p= l^i_{\Sigma}(\bar{t}) = l^{j}_{\Sigma}(\bar{u})$,
are not parallel to each other.
\item[(A3)] For each $j \le n$
the set $I_j(t_0):= (l^j_{S^1})^{-1}(\{t_0\})$ is finite.
\item[(A4)] There is no $x \in \bigcup_j \arc(l_j)$
such that simultaneously $\pi_{S^1}(x)=t_0$ and   $\pi_{\Sigma}(x) \in \DP(L)$ holds.
\end{enumerate}

Note that from (A1) it follows that
the set $\Sigma \backslash ( \bigcup_j \arc(l^j_{\Sigma}))$ has only finitely many
connected components. We will denote these connected components
by $X_1, X_2, \ldots, X_{\mu}$, $\mu \in \bN$, in the sequel.

\subsection{The identification  $\cA^{\orth} \equiv C^{\infty}(S^1,\cA_{\Sigma})$ and the  Hilbert spaces  $\cH_{\Sigma}$,  $\cH^{\orth}$}
\label{subsec3.2}

Before we apply the results of Sec. \ref{sec2} to the Chern-Simons action function
it is useful to introduce some additional spaces.
For every real vector space $V$ let
  $\cA_{\Sigma,V}$ denote the space of smooth
  $V$-valued 1-forms on $\Sigma$. We set $\cA_{\Sigma}:= \cA_{\Sigma,\cG}$.
We will call  a  function $\alpha: S^1 \to
\cA_{\Sigma,V}$ ``smooth''  if for
every $C^{\infty}$-vector field  $X$  on $\Sigma$ the function
$ \Sigma \times S^1 \ni (\sigma,t) \mapsto \alpha(t)(X_{\sigma}) \in V$ is $C^{\infty}$
and we will set
 $C^{\infty}(S^1,\cA_{\Sigma,V}):= \{\alpha \mid \alpha:S^1 \to
\cA_{\Sigma,V} \text{ is smooth}\}$.
$\tfrac{\partial}{\partial t}$ will denote the obvious operator on
$C^{\infty}(S^1,\cA_{\Sigma,V})$.
 During the rest of this paper we will identify $\cA^{\orth}$
with $C^{\infty}(S^1,\cA_{\Sigma})$ in the obvious way.
In particular, if   $A^{\orth} \in \cA^{\orth}$ and   $t \in S^1$ then  $A^{\orth}(t)$ will denote
an element of $\cA_{\Sigma}$.\par

In the sequel  we will assume that
the $\Ad$-invariant scalar product  $(\cdot,\cdot)_{\cG}$ on $\cG$
fixed above is the one given by
\begin{equation} (A,B)_{\cG} = -\Tr(AB) \quad \text{ for all } A,B \in \cG \end{equation}
Moreover, we fix  a $(\cdot,\cdot)_{\cG}$-orthonormal basis $(T_a)_{a \le \dim(G)}$
 with the  property that $T_a \in \ct$ for all $a \le r:= \rank(G)=\dim(T)$.
($(T_a)_{a \le \dim(G)}$ will be relevant in the concrete computations in Secs \ref{sec5} and \ref{sec6}).\par

Let us also  fix  an auxiliary Riemannian metric $\mathbf g$ on $\Sigma$
for the rest of this paper.
   $\mu_{\mathbf g}$ will denote the Riemannian volume measure on $\Sigma$ associated to ${\mathbf g}$,
   $(\cdot,\cdot)_{\mathbf g,\cG}$  the fibre metric on the bundle $\Hom(T(\Sigma),\cG) \cong T^{*}(\Sigma) \otimes \cG$
  induced by ${\mathbf g}$ and  $(\cdot,\cdot)_{\cG}$,
  and $\cH_{\Sigma}$ the Hilbert space
 $\cH_{\Sigma}:= L^2$-$\Gamma(\Hom(T(\Sigma),\cG),\mu_{\mathbf g})$ of $L^2$-sections
  of the bundle $ \Hom(T(\Sigma),\cG)$ w.r.t. the measure $ \mu_{\mathbf g}$
   and the fibre metric $(\cdot,\cdot)_{\mathbf g,\cG}$.
   The scalar product $\ll\cdot,\cdot\gg_{\cH_{\Sigma}}$ of $\cH_{\Sigma}$
   is, of course, given by
  $$\ll \alpha_1, \alpha_2  \gg_{\cH_{\Sigma}} = \int_{\Sigma} (\alpha_1, \alpha_2)_{\mathbf g,\cG}
  d\mu_{\mathbf g} \quad \forall \alpha_1, \alpha_2 \in \cH_{\Sigma} $$

Finally, we set $\cH^{\orth}:= L^2_{\cH_{\Sigma}}(S^1,dt)$, i.e. $\cH^{\orth}$ is the space of $\cH_{\Sigma}$-valued
(measurable) functions on $S^1$ which are square-integrable w.r.t. $dt$.
The scalar product $\ll\cdot,\cdot\gg_{\cH^{\orth}}$ on $\cH^{\orth}$ is given by
 $$\ll A^{\orth}_1,A^{\orth}_2\gg_{\cH^{\orth}} = \int_{S^1} \ll A^{\orth}_1(t), A^{\orth}_2(t)  \gg_{\cH_{\Sigma}} dt
 \quad \text{ for all } A^{\orth}_1,A^{\orth}_2 \in \cH^{\orth}$$

By $\star$ we will denote four different operators:
firstly, the Hodge star operator $\star: \cA_{\Sigma} \to \cA_{\Sigma}$,
secondly the operator $\star: C^{\infty}(S^1,\cA_{\Sigma}) \to  C^{\infty}(S^1,\cA_{\Sigma})$
defined by $ (\star A^{\orth})(t) = \star (A^{\orth}(t))$ for all $t \in S^1$,
thirdly the operator $\cH^{\orth} \to \cH^{\orth}$
obtained by continuously extending $\star:C^{\infty}(S^1,\cA_{\Sigma}) \to  C^{\infty}(S^1,\cA_{\Sigma})$
to all of $\cH^{\orth}$ and, finally,
the Hodge operator
$\Omega^2(\Sigma,\cG) \to C^{\infty}(\Sigma,\cG)$
where $\Omega^2(\Sigma,\cG)$ denotes the space of
$\cG$-valued 2-forms on $\Sigma$.\par
The four analogous mappings obtained by replacing the surface $\Sigma$
by $\Sigma \backslash \{ \sigma_0\}$ will  also be denoted by $\star$.

\subsection{Application of formula \eqref{eq_Abelian_version}}
\label{subsec3.3}

The restriction of the Chern-Simons action function $S_{CS}$
onto the space $\cA^{qax}$ is rather simple.
More precisely,   we  have
\begin{proposition} \label{prop3.1}
Let $A^{\orth} \in \cA^{\orth}$
 and $B \in C^{\infty}(\Sigma,\cG)$. Then
  \begin{equation} \label{eq8.4}
 S_{CS}(A^{\orth} + Bdt)\\
  =  -\frac{k}{4\pi} \biggl[
   \ll A^{\orth},
 \star  \bigl( \tfrac{\partial}{\partial t} + \ad(B) \bigr)
A^{\orth} \gg_{\cH^{\orth}} - 2  \ll A^{\orth}, \star
dB\gg_{\cH^{\orth}} \biggr]
\end{equation}
\end{proposition}
\begin{proof}
It is not difficult to see that for all $A^{\orth} \in \cA^{\orth}$ and
$A^{||} \in \{A_0 dt \mid A_0 \in C^{\infty}(M,\cG)\}$ one has
$S_{CS}(A^{\orth} + A^{||}) = \tfrac{k}{4\pi} \bigl[ \int_M \Tr(A^{\orth} \wedge
dA^{\orth} ) + 2  \int_M \Tr( A^{\orth} \wedge  A^{||} \wedge
A^{\orth}) + 2  \int_M  \Tr(A^{\orth} \wedge dA^{||}) \bigr]$.
By applying this formula to  the special case
where $A^{||} = B dt$ and taking into account
  the definitions of $\star$ and
 $\ll \cdot, \cdot \gg_{\cH^{\orth}}$ the assertion follows
 (cf. Prop. 5.2 in \cite{Ha3b}).
\end{proof}

From Eq.  \eqref{eq_torusgaugefixing} we obtain
\begin{multline}
\WLO(L)  =  \frac{1}{Z(M)} \int \prod_i  \Tr_{\rho_i}\bigl(\cP \exp\bigl(\int_{l_i} A\bigr)\bigr) \exp(i S_{CS}(A)) DA \nonumber \\
 \sim  \sum_{\cl \in [\Sigma,G/T]} \int_{C^{\infty}(\Sigma,P)}   \biggl[ \int_{\cA^{\orth}} \prod_i \Tr_{\rho_i}\bigl(\cP \exp\bigl(\int_{l_i} A^{\orth}   +  (\bar{g}_{\cl}  B \bar{g}_{\cl}^{-1}) dt) \bigr)\bigr) \\
 \times \exp(i S_{CS}( A^{\orth} + (\bar{g}_{\cl}  B \bar{g}_{\cl}^{-1}) dt)) DA^{\orth} \biggr]
   \det\bigl(1_{\cG_0}-\exp(\ad(B)_{| \cG_0})\bigr)  DB
 \end{multline}
We would now like to apply formula \eqref{eq_Abelian_version} above and obtain an ``Abelian version''
of the equation above.
Before we can do this we have to extend the two $\G_{\Sigma}$-invariant functions
\begin{align} \label{eq3.3}
\cA^{qax} \ni  A^q  & \mapsto \prod_i \Tr_{\rho_i}\bigl(\cP \exp\bigl(\int_{l_i} A^q\bigr)\bigr) \in \bC\\
\cA^{qax} \ni  A^q  & \mapsto S_{CS}( A^q) \in \bC
\end{align}
to  $\overline{\G_{\Sigma}}$-invariant
 functions on $\cA^{qax}_{(\Sigma\backslash \{\sigma_0\}) \times S^1}$.\par
If $\sigma_0$ is not in the image of the loops $l^j_{\Sigma}$, which we will assume in the sequel,
 then the expression
on the right-hand side of Eq. \eqref{eq3.3}
 makes sense for arbitrary $A^q \in \cA^{qax}_{(\Sigma\backslash \{\sigma_0\}) \times S^1}$ and thus
  defines
  a $\G_{\Sigma\backslash \{\sigma_0\}}$-invariant function
  on  $\cA^{qax}_{(\Sigma\backslash \{\sigma_0\}) \times S^1}$.
The second function is just the restriction $(S_{CS})_{|\cA^{qax}}$.
  Let  $\overline{S}_{CS}:  \cA^{qax}_{|
(\Sigma \backslash \{\sigma_0 \}) \times S^1} \to \bC$ be given by
 \begin{multline*}
 \overline{S}_{CS}(A^{\orth} + B dt)\\
  :=  -  \lim_{\eps \to 0} \frac{k}{4\pi}
  \int_{S^1} \int_{\Sigma\backslash B_{\eps(\sigma_0)}} \biggl[ \bigl(A^{\orth}(t),
 \star  \bigl( \tfrac{\partial}{\partial t} + \ad(B) \bigr) A^{\orth}(t)\bigr)_{\mathbf g, \cG}
- 2  \Tr(\star dA^{\orth}(t) B) \biggr] d\mu_{\mathbf g}  dt,
\end{multline*}  for all
$A^{\orth} \in \cA^{\orth}_{(\Sigma\backslash \{\sigma_0\}) \times
S^1}$ and $B \in C^{\infty}(\Sigma\backslash \{\sigma_0\},\cG)$ for
which the limit $\eps \to 0$ exists\footnote{for example, this will
be the case if $A^{\orth} = \Omega^{-1}_{\cl} A^{\orth}_1
\Omega_{\cl} + \Omega_{\cl}^{-1} d  \Omega_{\cl} $ with $
A^{\orth}_1 \in \cA^{\orth}$ and $\cl \in [\Sigma,G/T]$},
 and by $\overline{S}_{CS}(A^{\orth} + B dt) := 0$
 otherwise.
In the special case
where $A^{\orth} \in \cA^{\orth} \subset \cA^{\orth}_{(\Sigma\backslash \{\sigma_0\}) \times S^1}$
and $B \in C^{\infty}(\Sigma,\cG) \subset C^{\infty}(\Sigma\backslash \{\sigma_0\},\cG)$
Stokes' Theorem implies that
\begin{equation} \label{eq_Stokes} \int_{S^1} \ll \star dA^{\orth}(t),  B \gg_{L^2_{\cG}(\Sigma,\mu_{\mathbf g})}  dt =  \ll A^{\orth}, \star dB\gg_{\cH^{\orth}}
\end{equation}
so $\overline{S}_{CS}$ is indeed an extension of $(S_{CS})_{|\cA^{qax}}$.
Moreover, it turns out that
$\overline{S}_{CS}$ is a  $\overline{\G_{\Sigma}}$-invariant function (for a detailed proof, see \cite{Ha3c}).
Thus we can apply  Eq. \eqref{eq_Abelian_version} and obtain
 \begin{multline}
 \WLO(L)  \sim \sum_{\cl \in [\Sigma,G/T]}
 \int_{C^{\infty}(\Sigma,P)} \biggl[ \int_{\cA^{\orth}} \prod_i  \Tr_{\rho_i}\bigl(\cP \exp\bigl(\int_{l_i} (\Omega^{-1}_{\cl} A^{\orth} \Omega_{\cl} + \Omega_{\cl}^{-1} d  \Omega_{\cl}  + Bdt) \bigr)\bigr)      \\
 \quad \quad \quad \times  \exp(i \overline{S}_{CS}(\Omega^{-1}_{\cl} A^{\orth} \Omega_{\cl} + \Omega_{\cl}^{-1} d  \Omega_{\cl} + Bdt)) DA^{\orth} \biggr] \\
 \times   \det\bigl(1_{\cG_0}-\exp(\ad(B)_{| \cG_0})\bigr)  DB
\end{multline}
 It is not difficult to see that with
 $A^{\orth}_{sing}(\cl):= \pi_{\ct}(\Omega_{\cl}^{-1} d  \Omega_{\cl})$
 we have
  \begin{multline}
\overline{S}_{CS}(\Omega^{-1}_{\cl} A^{\orth}  \Omega_{\cl}  + \Omega_{\cl}^{-1} d  \Omega_{\cl}+ Bdt) \\
=  \overline{S}_{CS}(\Omega^{-1}_{\cl} A^{\orth} \Omega_{\cl}  + \pi_{\cG_0}(\Omega_{\cl}^{-1} d  \Omega_{\cl})   + Bdt) + \tfrac{k}{2\pi} \ll  \star   dA^{\orth}_{sing}(\cl), B\gg
\end{multline}
where $\pi_{\cG_0}$ denotes the orthogonal projection $\cG \to \cG_0$
 and where we have set
 \begin{multline} \label{eq_def_llrr}
 \ll  \star   dA^{\orth}_{sing}(\cl), B\gg   := \lim_{\eps \to 0} \int_{\Sigma \backslash B_{\eps}(\sigma_0)}
 \Tr(\star dA^{\orth}_{sing}(\cl) B) d\mu_{\mathbf g}\\
   =  \lim_{\eps \to 0} \int_{\Sigma \backslash B_{\eps}(\sigma_0)}
 \Tr(dA^{\orth}_{sing}(\cl) \cdot B)
\end{multline}

 It is now tempting and, in fact, justified
 (cf. Sec. 4.2 in \cite{Ha3c})
  to make the change of variable
   $\Omega^{-1}_{\cl} A^{\orth} \Omega_{\cl} + \pi_{\cG_0}(\Omega_{\cl}^{-1} d  \Omega_{\cl}) \longrightarrow A^{\orth}$.
Note for example that,
 without loss of generality, we can assume that
each mapping $\bar{g}_{\cl} \in C^{\infty}(\Sigma \backslash \{\sigma_0\}, G/T)$ was chosen such that
$\bar{g}_{\cl} \equiv T \in G/T$
 holds on a neighborhood $U$ of the point $\sigma_0$.
Then  $(\Omega_{\cl})_{|U}$ takes only values in $T$ which implies
 $\pi_{\cG_0}(\Omega_{\cl}^{-1} d  \Omega_{\cl}) = 0$ on $U$.
So the 1-form
 $\pi_{\cG_0}(\Omega_{\cl}^{-1} d  \Omega_{\cl}) $ has no singularity
 in $\sigma_0$ and is therefore contained in $\cA^{\orth}$.
 Thus we can replace
 $\Omega^{-1}_{\cl} A^{\orth} \Omega_{\cl} + \pi_{\cG_0}(\Omega_{\cl}^{-1} d  \Omega_{\cl})$ by  $\Omega^{-1}_{\cl} A^{\orth} \Omega_{\cl} $.
Finally, it is also possible
to make the change of variable  $\Omega^{-1}_{\cl} A^{\orth} \Omega_{\cl} \longrightarrow A^{\orth}$
(taking into account that because, of the compactness of $G$, we have $\det(\Ad(\Omega_{\cl}(\sigma)))=1$
  for every $\sigma \in \Sigma$; for more details see \cite{Ha3c}).
 After this change of variable we arrive at the following equation
   \begin{multline}  \label{eq_WLO_pre}
\WLO(L)   \sim \\
 \sum_{\cl}  \int_{C^{\infty}(\Sigma,P)}  \biggl[ \int_{\cA^{\orth}} \prod_i \Tr_{\rho_i}\bigl(\cP \exp\bigl(\int_{l_i}( A^{\orth}  + A^{\orth}_{sing}(\cl) +  Bdt) \bigr)\bigr)
 \exp(i  S_{CS}( A^{\orth} + B dt)) DA^{\orth} \biggr]  \\
  \exp( i \tfrac{k}{2\pi}  \ll  \star   dA^{\orth}_{sing}(\cl), B\gg)
   \det\bigl(1_{\cG_0}-\exp(\ad(B)_{| \cG_0})\bigr)  DB
 \end{multline}

\begin{remark} \rm \label{rm3.1}
Note that the 1-forms $A^{\orth}_{sing}(\cl)$ are definitely not in
$\cA^{\orth}$ if $\cl \neq [1_T]$. Thus it is not surprising that if
one tries to make the additional change of variable $A^{\orth}  +
A^{\orth}_{sing}(\cl)  \longrightarrow A^{\orth}$ one  obtains
incorrect  expressions, cf. again Sec. 4.2 in \cite{Ha3c}.
\end{remark}

\subsection{The decomposition  $\cA^{\orth} = \hat{\cA}^{\orth} \oplus \cA_c^{\orth}$}
\label{subsec3.4}

Let us now have a closer look at the  informal measure
$\exp(i S_{CS}(A^{\orth} + Bdt)) DA^{\orth}$  in Eq. \eqref{eq_WLO_pre} above.
In view of Eq. \eqref{eq8.4} this measure is of ``Gaussian type''.
Naively, one could  try
 to identify its ``mean''  and ``covariance operator''
by  writing down the following
informal expression for $S_{CS}(A^{\orth} + B dt)$,
 pretending that  the operator  $\tfrac{\partial}{\partial t} + \ad(B)$
in Eq. \eqref{eq8.4}
is bijective:
\begin{equation} \label{eq_naive}
S_{CS}(A^{\orth} + B dt)
 = - \tfrac{k}{4\pi}   \ll A^{\orth}  - m(B) ,
 \star \bigl( \tfrac{\partial}{\partial t} + \ad(B) \bigr)
   (A^{\orth}  - m(B))  \gg_{\cH^{\orth}}
\end{equation}
where $m(B):= ( \tfrac{\partial}{\partial t} +  \ad(B))^{-1}  \cdot dB$.
However, as the use of the word ``pretend'' above
already indicates the problem with this naive ansatz
is that  the operator
$\tfrac{\partial}{\partial t} +  \ad(B): \cA^{\orth} \to \cA^{\orth}$ is neither injective nor surjective
so it is not clear what $(\tfrac{\partial}{\partial t} +  \ad(B))^{-1}$
or $m(B)= ( \tfrac{\partial}{\partial t} +  \ad(B))^{-1}  \cdot dB$ above should be.\par

In order to solve this problem let us first identify     the
kernel of $\tfrac{\partial}{\partial t} +  \ad(B)$.
It is easy to see that $ \Ker(\tfrac{\partial}{\partial t} +  \ad(B)) =  \cA_{c}^{\orth} $
where
\begin{equation}
 \cA_{c}^{\orth}  := \{ A^{\orth} \in  C^{\infty}(S^1,\cA_{\Sigma}) \mid A^{\orth}
  \text{ is constant and }\cA_{\Sigma,\ct}\text{-valued}\} \cong  \cA_{\Sigma,\ct}
\end{equation}
(here we have used the identification
$\cA_{\Sigma} \cong \cA_{\Sigma,\cG_0} \oplus \cA_{\Sigma,\ct}$).
So it is reasonable to introduce
a direct sum decomposition of $\cA^{\orth}$ of the form $\cA^{\orth} = \cC \oplus \cA^{\orth}_c$
and then restrict $\tfrac{\partial}{\partial t} +  \ad(B)$ to the space $\cC$.
This restriction is then clearly injective.
A convenient\footnote{alternatively, one can probably also work with the choice
$\cC:= \Check{\cA}^{\orth} := \{ A^{\orth} \in C^{\infty}(S^1,\cA_{\Sigma}) \mid
\int A^{\orth}(t) dt \in  \cA_{\Sigma,\cG_0} \}$.
The decomposition $\cA^{\orth} = \Check{\cA}^{\orth} \oplus \cA^{\orth}_c$
is in some sense more natural than the decomposition  $\cA^{\orth} = \hat{\cA}^{\orth} \oplus \cA^{\orth}_c$.
Moreover,  it has a satisfactory ``discrete analogue'', while
the latter decomposition doesn't. This is why in
the discretization approach developed in \cite{Ha7a,Ha7b}
we work with (the discrete analogue of) $\Check{\cA}^{\orth}$.
On the other hand, the use of the space $\Check{\cA}^{\orth}$
 in a continuum setting has certain technical disadvantages, which is
why in   \cite{Ha3b,Ha3c} and the present paper we chose to work with
$\hat{\cA}^{\orth}$.}
 choice for $\cC$, already used in \cite{Ha3b}, is  $\cC := \hat{\cA}^{\orth}$ where
\begin{equation}
\hat{\cA}^{\orth} := \{ A^{\orth} \in C^{\infty}(S^1,\cA_{\Sigma}) \mid
A^{\orth}(t_0) \in  \cA_{\Sigma,\cG_0} \}
\end{equation}
Note that the operator
$ \tfrac{\partial}{\partial t} +  \ad(B)$, when restricted onto $\hat{\cA}^{\orth}$,
is still not surjective.
We solve this problem by replacing $\hat{\cA}^{\orth}$  by the slightly bigger space\footnote{in Sec. 8 in \cite{Ha3b}  we  gave a
 detailed motivation for this ansatz
 in the special case $t_0=i_{S^1}(0)$
(note that the space  $\tilde{\cA}^{\orth}$ was denoted by
$\tilde{C}^{\infty}(S^1,\cA_{\Sigma})$ there)}
\begin{equation}
\tilde{\cA}^{\orth}:= \hat{\cA}^{\orth} \oplus  \{ A_c^{\orth} \cdot (i_{t_0}^{-1}(\cdot) - 1/2) \mid A_c^{\orth} \in   \cA_{\Sigma,\ct} \}
\end{equation}
where $i_{t_0}^{-1}$ is the inverse of the bijection
\begin{equation}
i_{t_0}: [0,1) \ni s \mapsto i_{S^1}(s) \cdot t_0 \in S^1
\end{equation}
(here $i_{S^1}$ is the mapping defined at the beginning of Sec. \ref{sec2}
  and ``$\cdot$'' denotes the standard
multiplication of $S^1 \subset \bC$).\par
We can now extend  $\tfrac{\partial}{\partial t}:\cA^{\orth} \to \cA^{\orth}$
in an obvious way to an operator $\tilde{\cA}^{\orth} \to \cA^{\orth}$
and it turns out that
 the (extended) operator  $(\tfrac{\partial}{\partial t} + \ad(B)): \tilde{\cA}^{\orth} \to  \cA^{\orth}$
is  a bijection for every $B  \in C^{\infty}(\Sigma,P)$.
 The inverse operator $ (\tfrac{\partial}{\partial t} + \ad(B))^{-1}:  \cA^{\orth} \to  \tilde{\cA}^{\orth}$ is given explicitly by
 \begin{subequations}
 \begin{equation} \label{eq_expl_inv1}
  \forall t \in S^1: \quad  \bigl( (\tfrac{\partial}{\partial t} + \ad(B))^{-1}  A^{\orth} \bigr)(t) \\
   =  \frac{1}{2} \biggl[ \int_0^{i_{t_0}^{-1}(t)}  A^{\orth}(i_{t_0}(s)) ds -
   \int_{i_{t_0}^{-1}(t)}^1  A^{\orth}(i_{t_0}(s)) ds \biggr]
   \end{equation}
 if\footnote{note that in this case $ (\tfrac{\partial}{\partial t} + \ad(B))^{-1} \cdot A^{\orth}
 =  (\tfrac{\partial}{\partial t} )^{-1} \cdot A^{\orth}$
 so it is clear that the right-hand side of Eq. \eqref{eq_expl_inv1} can not depend on $B$}
    $A^{\orth}\in C^{\infty}(S^1,\cA_{\Sigma})$ takes only values in $ \cA_{\Sigma,\ct}$
 and
  \begin{multline} \label{eq_expl_inv2}
  \forall t \in S^1: \quad \bigl( (\tfrac{\partial}{\partial t} + \ad(B))^{-1}  A^{\orth} \bigr) (t)\\
 =  \biggl( \exp(\ad(B)_{| \cG_0}) - 1_{\cG_0}\biggr)^{-1} \cdot
 \int_0^1 \exp(s \cdot \ad(B))   A^{\orth}(i_{S^1}(s)\cdot t) ds  \
   \end{multline}
   \end{subequations}
  if $A^{\orth}\in C^{\infty}(S^1,\cA_{\Sigma})$ takes only values in $ \cA_{\Sigma,\cG_0}$.
 Note that the last expression is well-defined
 because  each $B(\sigma)$, $\sigma \in \Sigma$, is an element of the (open) alcove $P$,
 from which it follows that
 $ \exp(\ad(B(\sigma))_{| \cG_0}) - 1_{\cG_0} \in \End(\cG_0)$ is invertible, cf. Remark 8.1 in
 \cite{Ha3b}.\par
  We can now define $m(B)$ rigorously by
\begin{equation} \label{eq_df_mB}
  m(B)  :=  ( \tfrac{\partial}{\partial t} +  \ad(B))^{-1} \cdot dB
  \overset{(*)}{=} (i_{t_0}^{-1}(\cdot) - 1/2) \cdot dB\in \tilde{\cA}^{\orth}
\end{equation}
(here step $(*)$ follows from  Eq. \eqref{eq_expl_inv1}).
With this definition we have
  \begin{equation}  \label{eq8.4new}
    S_{CS}(\hat{A}^{\orth} + Bdt)    =  -\frac{k}{4\pi}
   \ll \hat{A}^{\orth} - m(B),
 \star \bigl( \tfrac{\partial}{\partial t} + \ad(B) \bigr) (\hat{A}^{\orth} -
m(B))\gg_{\cH^{\orth}}
\end{equation}
for all $\hat{A}^{\orth} \in \hat{\cA}^{\orth}$ and $B \in C^{\infty}(\Sigma,P)$.
Moreover, we have
\begin{equation}  \label{eq8.4new2}
  S_{CS}(\hat{A}^{\orth} + A^{\orth}_c + Bdt)  =  S_{CS}(\hat{A}^{\orth} + Bdt)   -\frac{k}{2\pi} \ll A^{\orth}_c, \star dB \gg_{\cH_{\Sigma}}
\end{equation}
In Eq. \eqref{eq_WLO_pre}, the informal measure
``$\exp(i S_{CS}(A^{\orth} + Bdt) DA^{\orth}$''  appeared
as part of a multiple integral.
According to Eq. \eqref{eq8.4new2} we can write $\exp(i S_{CS}(A^{\orth} + Bdt) DA^{\orth}$
in the form
$(\exp(i S_{CS}(\hat{A}^{\orth} + Bdt))  D\hat{A}^{\orth}) \otimes (\exp(-i \frac{k}{2\pi} \ll A^{\orth}_c, \star dB \gg_{\cH_{\Sigma}}))DA^{\orth}_c)$
and according to Eq. \eqref{eq8.4new} the first factor is, at an informal level, a ``Gauss-type'' measure
with ``mean'' m(B),  ``covariance operator'' $C(B):  \cA^{\orth} \to \tilde{\cA}^{\orth}$ given by
 \begin{equation} \label{eq_df_CB} C(B):= -\tfrac{2\pi i}{k} (\tfrac{\partial}{\partial t} + \ad(B))^{-1} \circ \star^{-1}
   \end{equation}

Let us now plug in  the decomposition  $\cA^{\orth} = \hat{\cA}^{\orth} \oplus \cA_c^{\orth}$
into Eq. \eqref{eq_WLO_pre} above. Taking into account
 Eqs.   \eqref{eq8.4new}, \eqref{eq8.4new2} and the equality
 $  \ll \star dA^{\orth}_c,  B \gg_{L^2_{\cG}(\Sigma,\mu_{\mathbf g})}  =  \ll A^{\orth}_c, \star dB\gg_{\cH_{\Sigma}}$
we  obtain
\begin{multline}  \label{eq_WLO_0}
\WLO(L)  \sim\\
 \sum_{\cl}  \int_{\cA_c^{\orth} \times C^{\infty}(\Sigma,P)}  \biggl[ \int_{\hat{\cA}^{\orth}} \prod_i \Tr_{\rho_i}\bigl(\cP \exp\bigl(\int_{l_i}(\hat{A}^{\orth} + A^{\orth}_c  + A^{\orth}_{sing}(\cl) +  Bdt)  \bigr)\bigr)    d\hat{\mu}^{\orth}_B(\hat{A}^{\orth}) \biggr] \\
\times \bigl\{ \exp( i \tfrac{k}{2\pi}  \ll  \star   dA^{\orth}_{sing}(\cl), B\gg) \det\bigl(1_{\cG_0}-\exp(\ad(B)_{| \cG_0})\bigr)   \hat{Z}(B) \bigr\}\\
  \times \exp( i \tfrac{k}{2\pi}  \ll \star dA^{\orth}_c, B \gg_{L^2_{\ct}(\Sigma,d\mu_{\mathbf g})}) (DA_c^{\orth} \otimes    DB)
 \end{multline}
 where
\begin{align}
\hat{Z}(B) & := \int \exp(i S_{CS}( \hat{A}^{\orth} + B dt))D\hat{A}^{\orth}, \\
d\hat{\mu}^{\orth}_B(\hat{A}^{\orth}) & :=\tfrac{1}{\hat{Z}(B)} \exp(i S_{CS}( \hat{A}^{\orth} + B dt))D\hat{A}^{\orth}
\end{align}
 Note that for  $B \in C^{\infty}(\Sigma,P)$ we have\footnote{for step $(*)$ observe that
 $\tfrac{\partial}{\partial t} + \ad(B) = \tfrac{\partial}{\partial t}$
 on the orthogonal complement of
 $C^{\infty}(S^1,\cA_{\Sigma,\cG_0})$ in $\hat{\cA}^{\orth}$}, informally,
\begin{equation} \label{eq_ZB}
\hat{Z}(B) \sim \det(\tfrac{\partial}{\partial t} + \ad(B))^{- 1/2}
 \overset{(*)}{\sim} \det\bigl( L_B \bigr)^{- 1/2}
\end{equation}
where $L_B:= (\tfrac{\partial}{\partial t} + \ad(B))_{| C^{\infty}(S^1,\cA_{\Sigma,\cG_0})}$
and where $\sim$ denotes equality up to a multiplicative constant independent of $B$.
Moreover, we have\footnote{compare Remark \ref{rm_det_interpret} above
and take into account that
the operator $L_B$ on $C^{\infty}(S^1,\cA_{\Sigma,\cG_0}) = C^{\infty}(S^1, \Omega^1(\Sigma,\cG_0))$
  is the ``1-form analogue'' of the
operator $(\tfrac{\partial}{\partial t} + \ad(B))_{|
 C^{\infty}(\Sigma \times S^1,\cG_0)}$ on  $C^{\infty}(\Sigma \times S^1,\cG_0) \cong C^{\infty}(S^1, C^{\infty}(\Sigma,\cG_0))
= C^{\infty}(S^1, \Omega^0(\Sigma,\cG_0))$
 appearing in  Remark \ref{rm_det_interpret}} (again informally)
\begin{equation} \label{eq_ad'}
\det(L_B) = \det\bigl(1'_{\cG_0}-\exp(\ad(B)'_{| \cG_0})\bigr)
\end{equation}
where $1'_{\cG_0}$ is the identity operator on $\cA_{\Sigma,\cG_0}$
and $\ad(B)'_{| \cG_0 }:\cA_{\Sigma,\cG_0} \to \cA_{\Sigma,\cG_0}$
is defined ``pointwise'',
i.e. $(\ad(B)'_{| \cG_0} A_c)(\sigma) = \ad(B(\sigma))_{| \cG_0} \cdot A_c(\sigma)$
for all $A_c \in \cA_{\Sigma,\cG_0}$  and  $\sigma \in \Sigma$.
(We use the notation $1'_{\cG_0}$ and $\ad(B)'_{| \cG_0}$ in order to distinguish
these operators from the operators $1_{\cG_0}$
and  $\ad(B)_{| \cG_0}$ on $C^{\infty}(\Sigma,\cG_0)$).

\subsection{Evaluation of $\det\bigl(1_{\cG_0}-\exp(\ad(B)_{| \cG_0})\bigr)   \hat{Z}(B)$}
\label{subsec3.5}

Let us now  make  rigorous sense of the heuristic expression
  \begin{equation}  \label{eq_heuristic_expr}
\det\bigl(1_{\cG_0}-\exp(\ad(B)_{| \cG_0})\bigr)   \hat{Z}(B)  \sim
\frac{\det\bigl(1_{\cG_0}-\exp(\ad(B)_{| \cG_0})\bigr)}{\det\bigl(1'_{\cG_0}-\exp(\ad(B)'_{| \cG_0})\bigr)^{1/2}}
\end{equation}
for $B \in C^{\infty}(\Sigma,P)$.
The detailed analysis in Sec. 6 of \cite{BlTh1}
suggests that in the simplest case,
i.e. the case of constant\footnote{which is the only case
of relevance in \cite{BlTh1}, cf. our Subsec. \ref{subsec6.2} below}
 $B \equiv b$, $b \in P$,
the  expression on the r.h.s. of \eqref{eq_heuristic_expr}
should be replaced by the more complicated (and rigorous) expression
\begin{equation} \label{eq_renDet_const}
\bigl(\det\bigl(1_{\cG_0}-\exp(\ad(b)_{| \cG_0})\bigr))^{\chi(\Sigma)/2}    \times
  \exp\bigl( i \tfrac{c_G}{2\pi}  \ll \star dA^{\orth}_c + \star dA^{\orth}_{sing}(\cl),
   b \gg_{L^2_{\ct}(\Sigma,d\mu_{\mathbf g})}\bigr)
\end{equation}
where $c_G$ is the dual Coxeter number\footnote{This gives rise to the so-called ``charge shift'' $k \to k + c_G$.
Let us mention that the prediction that such a charge shift will appear here is
contested by some authors, cf. Remark B.2 in \cite{Ha7b}.
If one does not believe that this charge shift will appear
one will have to drop the second factor on the r.h.s. of \eqref{eq_renDet_const}} of $G$.
For example, for $G=SU(N)$ we have $c_G = N$.\par
In Subsec. \ref{subsec6.3} below, not only constant functions $B$ will appear  but  more general
``step functions'', i.e. functions $B$
 which are constant on each of the
connected components $X_1, X_2, \ldots, X_{\mu}$ of the set
$ \Sigma \backslash ( \bigcup_j \arc(l^j_{\Sigma}))$, cf. Subsec. \ref{subsec3.1}.

\begin{remark} \rm \label{rm3.3}
Of course, these ``step functions''  are not
well-defined elements of $C^{\infty}(\Sigma,P)$.
Thus it is actually necessary to use
an additional regularization procedure in Subsec. \ref{subsec6.3}
by which the step functions are replaced
by certain  smooth approximations (later one has to perform
a limit procedure).
As the implementation of this additional regularization procedure
 is on one hand straightforward and, on the other hand,  would give rise to some rather clumsy
notation which would distract the reader from the main line of argument of this paper
we have decided not to include this additional regularization procedure here
but to postpone it to a subsequent paper.
\end{remark}
The expression \eqref{eq_renDet_const} and the results that we will obtain in Subsec. \ref{subsec6.3} below
strongly suggest
that for  such  ``step functions'' $B $
the  expression  \eqref{eq_heuristic_expr} should be replaced by
\begin{equation} \label{eq_renDet_step}  \prod_{t=1}^{\mu} \bigl(\det\bigl(1_{\cG_0}-\exp(\ad(b_t)_{| \cG_0})\bigr)\bigr)^{\chi(X_t)/2}
  \times  \exp\bigl( i \tfrac{c_G}{2\pi}  \ll \star dA^{\orth}_c + \star dA^{\orth}_{sing}(\cl),
  B \gg_{L^2_{\ct}(\Sigma,d\mu_{\mathbf g})}\bigr)
\end{equation}
where $b_t \in P$, $t \le \mu$, are given by
$B_{| X_t} \equiv b_t$.\par
If we want to work with Eq. \eqref{eq_WLO_0}
 we have to make sense of  \eqref{eq_heuristic_expr}
  for all   $B \in  C^{\infty}(\Sigma,P)$
  even if later only special $B$ will play a role.
  In view of \eqref{eq_renDet_step}
  we suggest that for general   $B \in  C^{\infty}(\Sigma,P)$
 the expression
\eqref{eq_heuristic_expr}  should be replaced
by the (metric dependent) expression (cf. Remark \ref{rm3.2})
\begin{equation} \label{eq_det_neu_0} \det\nolimits_{reg}\bigl(1_{\cG_0}-\exp(\ad(B)_{| \cG_0})\bigr)
  \times   \exp\bigl( i \tfrac{c_G}{2\pi}  \ll \star dA^{\orth}_c + \star dA^{\orth}_{sing}(\cl),
  B \gg_{L^2_{\ct}(\Sigma,d\mu_{\mathbf g})}\bigr)
\end{equation}
where
\begin{multline} \label{eq_det_neu}
\det\nolimits_{reg}\bigl(1_{\cG_0}-\exp(\ad(B)_{| \cG_0})\bigr):= \\
\prod_{t=1}^{\mu} \exp\biggl(\tfrac{1}{vol(X_t)} \int_{X_t} \ln\bigl(\det\bigl(1_{\cG_0}-\exp(\ad(B(\sigma))_{| \cG_0})\bigr)\bigr)
  d\mu_{\mathbf g}(\sigma) \biggr)^{\chi(X_t)/2}
\end{multline}
With this Ansatz we finally arrive at  the following
heuristic formula for the WLOs which will be fundamental for the rest of this paper.
\begin{multline}   \label{eq_WLO_end}
\WLO(L)   \sim  \\
\sum_{\cl \in [\Sigma,G/T]}  \int_{\cA_c^{\orth} \times \cB}  \biggl[ \int_{\hat{\cA}^{\orth}} \prod_i \Tr_{\rho_i}\bigl(\cP \exp\bigl(\int_{l_i}(\hat{A}^{\orth} + A^{\orth}_c  + A^{\orth}_{sing}(\cl) +  Bdt)  \bigr)\bigr)    d\hat{\mu}^{\orth}_B(\hat{A}^{\orth}) \biggr] \\
\times \bigl\{ \exp( i \tfrac{k+ c_G}{2\pi}  \ll  \star   dA^{\orth}_{sing}(\cl), B\gg)  \det\nolimits_{reg}\bigl(1_{\cG_0}-\exp(\ad(B)_{| \cG_0})\bigr)  \bigr\}\\
 \times \exp\bigl( i \tfrac{k+ c_G}{2\pi}  \ll \star dA^{\orth}_c,
 B \gg_{L^2_{\ct}(\Sigma,d\mu_{\mathbf g})}\bigr) (DA_c^{\orth} \otimes    DB)
 \end{multline}
 where
\begin{equation}
\cB:=C^{\infty}(\Sigma,P)
\end{equation}
  Eq. \eqref{eq_WLO_end} can be considered to be the generalization of formula (7.1)
   in \cite{BlTh1} to arbitrary links (cf. also Sec. 7.6 in \cite{BlTh1}).

\begin{remark} \rm  \label{rm3.2}
It would be desirable
to find a more thorough justification  (which is independent
of the considerations in Subsec. \ref{subsec6.3}
below) for replacing   expression \eqref{eq_heuristic_expr}
by \eqref{eq_det_neu_0}.
In particular, such a justification
will have to explain/answer why  -- for making sense of the expression \eqref{eq_heuristic_expr}
-- one has to use a regularization scheme
 that depends   on the   link $L$ even though  the expression
  \eqref{eq_heuristic_expr}  does not, cf. Remark 4.2 in \cite{Ha3c}
  for some more details.
\end{remark}

\subsection{The explicit Computation of the WLOs: overview}
\label{subsec3.6}

We will divide the evaluation of the right-hand side of
Eq. \eqref{eq_WLO_end} into the following three steps:

\begin{itemize}

\item {\bf  Step 1:} Make  sense of the integral functional $\int \cdots d\hat{\mu}^{\orth}_B$

\item {\bf  Step 2:} Make sense of the ``inner'' integral  $\int_{\hat{\cA}^{\orth}} \prod_i \Tr_{\rho_i}( \cP \exp(\int_{l_i} \hat{A}^{\orth} + A^{\orth}_c + A^{\orth}_{sing}(\cl) + Bdt) ) d\hat{\mu}^{\orth}_B(\hat{A}^{\orth})$ in Eq. \eqref{eq_WLO_end}
 and compute its value.

\item {\bf  Step 3:} Make sense of the total expression on the right-hand side of Eq. \eqref{eq_WLO_end}
and compute its value.
\end{itemize}

\section{The Computation of the WLOs: Step 1}
\label{sec4}

In  Sec. 8 in \cite{Ha3b} we gave a rigorous implementation $\Phi^{\orth}_{B}$
of  the integral functional $\int \cdots d\hat{\mu}^{\orth}_B$.
Here we briefly recall the construction of $\Phi^{\orth}_{B}$.
Eqs.  \eqref{eq8.4new},  \eqref{eq8.4new2} \eqref{eq_df_mB}, and  \eqref{eq_df_CB} suggest that
   the heuristic ``measure'' $\hat{\mu}^{\orth}_B$ on  $\hat{\cA}^{\orth}$
is of ``Gaussian type'' with ``mean'' $m(B)$ and
``covariance operator'' $C(B)$.
From the results in  Sec. 8 in \cite{Ha3b}
it follows\footnote{Note that the Hilbert space $\cH$ appearing in Subsec. 8.2. in \cite{Ha3b}
  is naturally isomorphic to our  Hilbert space $\cH^{\orth}$
  and we can therefore identify the two  Hilbert spaces with each other}
 that the operator $C(B): \cA^{\orth} \to \tilde{\cA}^{\orth} \subset \cH^{\orth}$
is a bounded and  symmetric (densely defined)
operator on  $\cH^{\orth} = L^2_{\cH_{\Sigma}}(S^1,dt)$.
 This allows us to use the  standard approach of white noise analysis
 and to realize the integral functional
 $\int \cdots d\hat{\mu}^{\orth}_B$ rigorously
 as a  generalized distribution $\Phi^{\orth}_{B}$ on the topological dual $\cN^*$ of
 a suitably chosen  nuclear subspace $\cN$ of $\cH^{\orth}$.
We will not go into details here. Let us mention here   only the following points:
\begin{enumerate}
\item It turned out in \cite{Ha3b}
that the nuclear space $\cN$
which was chosen there using a standard procedure
 coincides with the space $\cA^{\orth}$.
  Thus the operator  $C(B)$ can be considered to be an operator
$\cN \to \cH^{\orth}$.

\item The statement that $\Phi^{\orth}_{B}$ is a generalized distribution $\cN^*$
 means that $\Phi^{\orth}_{B}$ is a continuous linear functional
 $(\cN) \to \bC$
 where the topological space  $(\cN)$ (``the space of test functions'')
 is defined in a suitable way.
 We will not give a full definition of  $(\cN)$ here as this is rather technical.
 For our purposes it is enough  to know that
 each test function $\psi \in (\cN)$
 is a continuous mapping $\cN^* \to \bC$
  and that
$(\cN)$ contains the trigonometric exponentials
$\exp(i(\cdot,j)): \cN^* \to \bC$, $j \in \cN$,
and the polynomial functions
$\prod_{i=1}^n (\cdot,j_i): \cN^* \to \bC$,  $j_1, j_2, \ldots, j_n \in \cN$.
Here  $(\cdot,\cdot): \cN^* \times \cN \to \bR$  denotes the canonical pairing.

\item The generalized distribution  $\Phi^{\orth}_{B}$
was defined in \cite{Ha3b} as the unique continuous linear functional $(\cN) \to \bC$
with the property that
 \begin{equation} \label{eq_defPhiB}
 \Phi^{\orth}_{B}(\exp(i(\cdot,j)))   = \exp(i \ll j,m(B) \gg_{\cH^{\orth}} ) \exp(- \tfrac{1}{2}  \ll j, C(B)  j\gg_{\cH^{\orth}} )
 \end{equation}
holds for all $j \in \cN$.
Note that $\Phi^{\orth}_{B}(\exp(i(\cdot,j)))$
is the analogue of the Fourier transformation of the Gauss-type ``measure'' $\hat{\mu}^{\orth}_B$
and, at a heuristic level, one expects that this Fourier transform
is indeed given by the right-hand side of Eq. \eqref{eq_defPhiB}.

\item  The ``moments'' of  $\Phi^{\orth}_{B}$,
 i.e. the expressions
 $\Phi^{\orth}_{B}(\prod_{i=1}^n (\cdot,j_i))$
 with fixed $j_1, j_2, \ldots, j_n \in \cN$ can be computed easily, using  similar arguments as
 in the proof of Proposition 3 in \cite{Ha2}.
 In particular, the first and second moments are given by
\begin{equation} \label{eq_erstes_moment} \Phi^{\orth}_{B}((\cdot,j_1)) = \ll j_1,m(B) \gg_{\cH^{\orth}}
\end{equation}
and
\begin{equation} \label{eq_zweites_moment} \Phi^{\orth}_{B}((\cdot,j_1) \cdot (\cdot,j_2))
 = \ll j_1, C(B) \ j_2 \gg_{\cH^{\orth}} +   \ll j_1,m(B) \gg_{\cH^{\orth}} \cdot  \ll j_2,m(B) \gg_{\cH^{\orth}}
\end{equation}
for all $j_1, j_2 \in \cN$.\par
The higher moments are given by
 expressions that are totally analogous to the expressions
that appear in the classical Wick theorem for
the moments of a Gaussian probability measure on a  Euclidean space.

\item Clearly,  the linear functional $\Phi^{\orth}_{B}: (\cN) \to \bC$
induces a linear function
$(\cN) \otimes_{\bC} \Mat(N,\bC) \to\Mat(N,\bC)$
in an obvious way, which will also be denoted by $\Phi^{\orth}_{B}$.
 \end{enumerate}

\section{The Computation of the WLOs: Step 2 }
\label{sec5}

In order to make sense of   $\int_{\hat{\cA}^{\orth}} \prod_i \Tr_{\rho_i}( \cP \exp(\int_{l_i} \hat{A}^{\orth} + A^{\orth}_c + A^{\orth}_{sing}(\cl) + Bdt) ) d\hat{\mu}^{\orth}_B(\hat{A}^{\orth})$
we proceed in the following way:
\begin{itemize}
\item{We regularize $\prod_i \Tr_{\rho_i}( \cP \exp(\int_{l_i} \hat{A}^{\orth} + A^{\orth}_c + A^{\orth}_{sing}(\cl) + Bdt) )$
 by using ``smeared loops'' $l_i^{\eps}$. Later we let $\eps \to 0$.}

\item{Then we introduce ``deformations''  $\Phi^{\orth}_{B,\phi_s}$
of  $\Phi^{\orth}_{B}$
w.r.t. a suitable family $(\phi_s)_{s>0}$ of diffeomorphisms of $\Sigma \times S^1$ such that $\phi_s \to \id_{\Sigma \times S^1}$ uniformly as $s \to 0$   (``Framing'')}

 \item Finally we prove that the limit\footnote{for Abelian $G$
 we have $A^{\orth}_{sing}(\cl)=0$ for $\cl \in [\Sigma,G/T] = \{[1_T]\}$
 so in this case we will use the notation
 $\WLO(L,\phi_s;A^{\orth}_c, B)$ instead of $\WLO(L,\phi_s;A^{\orth}_c, A^{\orth}_{sing}(\cl), B)$}
 \begin{multline}  \label{eq_WLO_AB} \WLO(L,\phi_s;A^{\orth}_c, A^{\orth}_{sing}(\cl), B) := \\
   \lim_{\eps \to 0}  \Phi^{\orth}_{B,\phi_s}\biggl( \prod_i \Tr_{\rho_i}( \cP \exp(\int_{l^{\eps}_i} (\cdot) + A^{\orth}_c + A^{\orth}_{sing}(\cl) + Bdt) \biggr)
\end{multline}
 exists  and we compute this limit explicitly for small $s>0$.
\end{itemize}

\subsection{Abelian $G$ and general  $L$}
\label{subsec5.1}

Let us start with considering the case where $G$ is Abelian, i.e. a torus,
and where  $\Sigma = S^2$.
For Abelian $G$ we have  $G = T$, $P=\ct$, $\cG_0 = \{0\}$, $c_G=0$, and $[\Sigma,G/T] = \{[1_T]\}$.
Thus we can drop the expression $\det\nolimits_{reg}\bigl(1_{\cG_0}-\exp(\ad(B)_{| \cG_0})\bigr)$
and we can choose  $\Omega_{\cl}=1$, for $\cl = [1_T]$, from which  $A^{\orth}_{sing}(\cl)=0$
follows.
Accordingly, Eq. \eqref{eq_WLO_end} simplifies and we obtain
\begin{multline} \label{eq_original_AbWLO}
\WLO(L)  \sim   \int_{\cA_c^{\orth} \times C^{\infty}(\Sigma,\ct)}  \biggl[ \int_{\hat{\cA}^{\orth}} \prod_i \Tr_{\rho_i}\bigl(\cP \exp\bigl(\int_{l_i}(\hat{A}^{\orth} + A^{\orth}_c   +  Bdt)  \bigr)\bigr)    d\hat{\mu}^{\orth}_B(\hat{A}^{\orth}) \biggr] \\
\times
  \exp( i \tfrac{k}{2\pi}   \ll \star dA^{\orth}_c, B \gg_{L^2_{\ct}(\Sigma,d\mu_{\mathbf g})}) (DA_c^{\orth} \otimes    DB)
 \end{multline}
 For simplicity, we will only consider  the special case  where $G=U(1)$
and where every $\rho_i$ is equal to the fundamental representation $\rho_{U(1)}$
of $U(1)$.
In this case we can choose the basis $(T_a)_{a \le \dim(G)}$
to consist of the single element $T_1 = i \in  u(1)$.
Clearly, we have
$$\Tr_{\rho_{U(1)}}( \cP \exp(\int_{l} \hat{A}^{\orth} + A^{\orth}_c  + Bdt) ) =
\exp(\int_{l} \hat{A}^{\orth})  \exp(\int_{l} A^{\orth}_c)
  \exp(\int_{l} Bdt)  $$
  for every loop $l$.\par
 Let us now replace in Eq. \eqref{eq_original_AbWLO}
  the integral functional $\int \cdots d\hat{\mu}^{\orth}_B$
by the functional $\Phi^{\orth}_{B}$ which we have introduce in Sec. \ref{sec4}.
As we pointed out in Sec. \ref{sec4},  $\Phi^{\orth}_{B}$
is a generalized distribution on the topological dual $\cN^*$ of  $\cN = \cA^{\orth}$.
A general element $\hat{A}^{\orth} \in \cN^*$
will not be a smooth function,
 so $\int_{l} \hat{A}^{\orth} =  \int_0^1  \hat{A}^{\orth}(l'(s)) ds$
 does not make sense in general.
 In \cite{Ha3b} we solved this problem by replacing
 $\hat{A}^{\orth}(l'(s))$, $s \in [0,1]$, by
 $T_1 (\hat{A}^{\orth}, f_1^{l^{\eps}}(s))$, $\eps >0$,
 for  suitable elements $f_1^{l^{\eps}}(s)$ of $\cN = \cA^{\orth}$
 which were defined using parallel transport w.r.t. the Levi-Civita connection
 of $(\Sigma,\mathbf g)$ (here $(\cdot,\cdot)$ denotes again the canonical pairing
 $\cN^* \times \cN \to \bR$).
 However, this Ansatz requires the use of some  rather clumsy notation which distracts from the
 main points of the computation.
 For this reason we will proceed in a different way in the present paper.
 Here we will just concentrate on the special situation when the following condition is fulfilled:
 \begin{enumerate}
 \item[(S)] There is an open subset $U$ of $\Sigma$
 which is diffeomorphic to $\bR^2$
 and which ``contains'' all the $l^j_{\Sigma}$, i.e. which fulfills
 $\arc(l^j_{\Sigma}) \subset U$, $j \le n$.
  \end{enumerate}
 In this case $U$ inherits an  group structure from  $\bR^2$
  and we can then use this group structure $+:U \times U\to U$
   rather than parallel   transport w.r.t. the Levi-Civita connection
    for the definition of the functions  $f_1^{l^{\eps}}(s)$, $s \in [0,1]$, $\eps >0$.
 Moreover, by ``identifying'' $U$ with  $\bR^2$  we can simplify
 our notation.\par
 Let us fix a Dirac family $(\delta^{\eps}_{S^1})_{\eps >0}$ on $S^1$
  in the point $1 = i_{S^1}(0) \in S^1$
  and a  Dirac family $(\delta^{\eps}_{\Sigma})_{\eps >0}$  on $U$
  in the point $(0,0) \in U \cong \bR^2$.
  Then we obtain a Dirac family
     $ (\delta^{\eps})_{\eps>0}$ on $U \times S^1$
     (and thus also on $\Sigma \times S^1$)
     in the point $((0,0),1)$
     given by $\delta^{\eps}(\sigma,t)=\delta^{\eps}_{\Sigma}(\sigma) \delta^{\eps}_{S^1}(t)$.
    We now define  $    f_1^{l^{\eps}}(s) \in \cA^{\orth}_{U \times S^1}$
    by
    $    f_1^{l^{\eps}}(s) =  T_1     l'_{\Sigma}(s) \delta^{\eps}(\cdot - l(s)) $
    where we have used the
   identification $\cA^{\orth}_{U \times S^1}  \cong C^{\infty}(U \times S^1,\bR^2 \otimes \cG)$
    (induced by the identification $U \cong \bR^2$)
   and    where ``$-$'' denotes the subtraction associated to the product
  group structure  $+:(U \times S^1) \times (U \times S^1) \to U \times S^1$.
  As  $f_1^{l^{\eps}}(s)$ has compact support and
  as the subspace of $\cA^{\orth}_{U \times S^1}$ which consists
  of all elements with compact support can  be embedded naturally
  into the space $\cA^{\orth}$ we can consider  $f_1^{l^{\eps}}(s)$
  as an element  of $\cA^{\orth}$.
 Instead of using the notation $f_1^{l^{\eps}}(s)$
   we will use the more suggestive notation $T_1     l'_{\Sigma}(s) \delta^{\eps}(\cdot - l(s))$
   in the sequel and
 we set for every $ \hat{A}^{\orth} \in \cN^*$
\begin{align} \label{eq_def_loopsmearing}
\int_{l^{\eps}_i} \hat{A}^{\orth} & :=  T_1 (\hat{A}^{\orth} , T_1   \int_0^1  (l^i_{\Sigma})'(s) \delta^{\eps}(\cdot - l_i(s)) ds), \nonumber \\
 \cP \exp\bigl(\int_{l^{\eps}_i}(\hat{A}^{\orth} + A^{\orth}_c   +  Bdt)  \bigr) &:=
\exp( \int_{l^{\eps}_i} \hat{A}^{\orth} )   \exp(\int_{l_i}
A^{\orth}_c) \exp(\int_{l_i} Bdt)
\end{align}
and
 \begin{equation} \label{eq_WLOLAB_def}  \WLO(L;A^{\orth}_c, B) :=    \lim_{\eps \to 0}  \Phi^{\orth}_{B}\biggl( \prod_i \Tr_{\rho_i}( \cP \exp(\int_{l^{\eps}_i} (\cdot) + A^{\orth}_c   + Bdt) ) \biggr)
\end{equation}
provided that the limit on the right-hand side exists.\par

 \begin{remark} \rm  \label{rm5.1}
 Informally, we have  $\int_{l} \hat{A}^{\orth} =  \int_0^1  \hat{A}^{\orth}(l'(s)) ds
      =  T_1 ( \hat{A}^{\orth}, T_1   \int_0^1  l'_{\Sigma}(s) \delta(\cdot - l(s)) ds)$
    where $\delta$ denotes the informal ``Dirac function'' on $U \times S^1$ in the point $((0,0),1)$.
   So ``loop smearing'' just amounts to replacing the ill-defined
    expression $\delta(\cdot - l(s))$ by the test function $ \delta^{\eps}(\cdot - l(s))$
  \end{remark}
If we insert  Eq. \eqref{eq_def_loopsmearing} into Eq. \eqref{eq_WLOLAB_def}
 above we  obtain
\begin{multline} \label{eq_6.7}
 \WLO(L;A^{\orth}_c, B)    =   \prod_j   \exp(\int_{l_j} A^{\orth}_c)
  \exp(\int_{l_j} Bdt)  \\
  \times \lim_{\eps \to 0}
  \Phi^{\orth}_{B}\biggl(  \exp\bigl( T_1 \bigl(  \cdot ,\sum_i T_1   \int_0^1  (l^i_{\Sigma})'(s) \delta^{\eps}(\cdot - l_i(s)) ds\bigr) \bigr)\biggr)
 \end{multline}

From $T_1 = i$ and Eq. \eqref{eq_defPhiB} we obtain
\begin{multline} \label{eq_6.8}
 \Phi^{\orth}_{B}\biggl(  \exp\bigl( T_1 \bigl(  \cdot ,\sum_i T_1   \int_0^1  (l^i_{\Sigma})'(s) \delta^{\eps}(\cdot - l_i(s)) ds\bigr) \bigr)\biggr)\\
   =    \prod_{j,k} \exp( - \frac{1}{2} \ll  T_1   \int_0^1  (l^j_{\Sigma})'(s) \delta^{\eps}(\cdot - l_j(s)) ds,
           C(B) \cdot \bigl(T_1   \int_0^1  (l^k_{\Sigma})'(u) \delta^{\eps}(\cdot - l_k(u)) du\bigr) \gg_{\cH^{\orth}})\\
         \times \prod_i \exp( T_1 \ll m(B),     T_1   \int_0^1  (l^i_{\Sigma})'(t) \delta^{\eps}(\cdot - l_i(t)) dt\gg_{\cH^{\orth}})
          \end{multline}
Clearly, we have
\begin{multline} \label{eq_6.9} \lim_{\eps \to 0} T_1 \ll m(B),     T_1   \int_0^1  (l^i_{\Sigma})'(t) \delta^{\eps}(\cdot - l_i(t)) dt\gg_{\cH^{\orth}}\\
 =   \int_0^1    (i_{t_0}^{-1} (l^i_{S^1}(t))- 1/2) dB((l^i_{\Sigma})'(t)) dt
 =  \int_0^1   l^i_{\bR}(t) \tfrac{d}{dt} B(l^i_{\Sigma}(t)) dt
 \end{multline}
 where we have set   $l^i_{\bR}:= i_{t_0}^{-1} \circ l^i_{S^1} - 1/2$.
 Thus we obtain from Eqs. \eqref{eq_WLOLAB_def}--\eqref{eq_6.9}
 \begin{multline} \label{eq_stil_verbesserung}
  \WLO(L;A^{\orth}_c, B)  =  \bigl( \prod_{j,k} \exp( - \frac{1}{2}  T(l_j,l_k)) \bigr)   \bigl(  \prod_j  \exp(\int_{l_j} A^{\orth}_c) \bigr)\\
  \times     \bigl\{ \prod_j   \exp(\int l^j_{\bR}(t) \tfrac{d}{dt} B(l^j_{\Sigma}(t)) dt)  \exp(\int_{l_j} Bdt) \bigr\}
 \end{multline}
 where we have set
 \begin{multline} T(l_j,l_k):=\\
  \lim_{\eps \to 0} \ll  T_1   \int_0^1  (l^j_{\Sigma})'(s) \delta^{\eps}(\cdot - l_j(s)) ds,
           C(B) \cdot \bigl(T_1   \int_0^1  (l^k_{\Sigma})'(u) \delta^{\eps}(\cdot - l_k(u)) du\bigr) \gg_{\cH^{\orth}})
  \end{multline}
  provided that the limit $T(l_j,l_k)$
 exists for each pair $(l_j,l_k)$.
  Taking into account that  $(l^j_{S^1})'(t)  = (l^j_{\bR})'(t)$
 we see that
 \begin{multline} \label{eq_6.10}  \int l^j_{\bR}(t) \tfrac{d}{dt} B(l^j_{\Sigma}(t)) dt +  \int_{l_j} Bdt
 =  \int_0^1  \bigl\{    l^j_{\bR}(u) \tfrac{d}{du} B(l^j_{\Sigma}(u)) + B(l^j_{\Sigma}(u)) \cdot (l^j_{\bR})'(u) \bigr\} du \\
= \sum_{i=0}^{ n_j+1} \int_{s^j_i}^{s^j_{i+1}}   \frac{d}{du} \bigl[ l^j_{\bR}(u) \cdot B(l^j_{\Sigma}(u))
 \bigr] du  =  \sum_{i=1}^{n_j} \sgn(l^j_{S^1};s^j_{i})  \cdot B(l^j_{\Sigma}(s^j_{i}))
 \end{multline}
 where
 $(s^j_i)_{0 \le i \le n_j+1}$ denotes the strictly increasing
 sequence of $[0,1]$ given by $s^j_0:=0$, $s^j_{n_j+1}=1$,  and $ \{s^j_i \mid 1\le i \le n_j\} = I_j(t_0):=
  (l^j_{S^1})^{-1}(\{t_0\})$ with  $n_j:= \# I_j(t_0)$
 and where\footnote{in the special case where $0 \in I_j(t_0)$
 the definition of $\sgn(l^j_{S^1};s^j_{i})$ has to be modified in the obvious way}
 we have set  $\sgn(l^j_{S^1};s^j_{i}) :=  \lim_{s \uparrow s^j_{i}} l^j_{\bR}(s)-  \lim_{s \downarrow s^j_{i}} l^j_{\bR}(s) \in \{-1,0,1\}$.
 Thus  it follows that the last factor in Eq. \eqref{eq_stil_verbesserung} equals
\begin{equation}
\prod_j  \exp\bigl(\sum_{u \in I_j(t_0)} \sgn(l^j_{S^1};u)  \cdot B(l^j_{\Sigma}(u))\bigr)
\end{equation}
 Let us now evaluate the expression
 $T(l_j,l_k)$ for fixed $j, k \le n$.
We will first concentrate on the case where $j \neq k$.
As $C(B)  = -\tfrac{2\pi i}{k} (\tfrac{\partial}{\partial t} + \ad(B))^{-1} \circ \star^{-1}
= -\tfrac{2\pi i}{k} (\tfrac{\partial}{\partial t})^{-1} \circ (- \star)
= \tfrac{2\pi i}{k} \star \circ (\tfrac{\partial}{\partial t})^{-1}$
we obtain from Eq. \eqref{eq_expl_inv1}, setting $l:= l_j$, $\tilde{l}:=l_k$,
\begin{multline} \label{eq_formel_fuer_T} T(l,\tilde{l}) = \tfrac{2\pi i}{k} \lim_{\eps \to 0}     \int_0^1  \int_0^1  \biggl[  \int_{S^1}    \delta^{\eps}_{S^1}(t - l_{S^1}(s))
    \bigl( \tfrac{\partial}{\partial t}^{-1}  \delta^{\eps}_{S^1}(t - \tilde{l}_{S^1}(u)) \bigr) \ dt\\
   \times \ll  T_1   l'_{\Sigma}(s) \delta^{\eps}_{\Sigma}(\cdot - l_{\Sigma}(s)), \star \
     (T_1    \tilde{l}'_{\Sigma}(u) \delta^{\eps}_{\Sigma}(\cdot - \tilde{l}_{\Sigma}(u))) \gg_{\cH_{\Sigma}}
     \biggr] ds \ du
\end{multline}
where $\tfrac{\partial}{\partial t}^{-1}$ denotes the operator
given by
$ \bigl( \tfrac{\partial}{\partial t}^{-1}  f \bigr)(t)
   =  \frac{1}{2} \bigl[ \int_0^{i_{t_0}^{-1}(t)}  f(i_{S^1}(s) \cdot t_0) ds - \int_{i_{t_0}^{-1}(t)}^1  f(i_{S^1}(s) \cdot t_0) ds \bigr]$
for all $f \in C^{\infty}(S^1,\bR)$ and $t \in S^1$.
Clearly, for fixed $s,u \in [0,1]$ and  sufficiently small $\eps$ we have
\begin{multline}
 \int_{S^1}    \delta^{\eps}_{S^1}(t - l_{S^1}(s)) \
    \bigl( \tfrac{\partial}{\partial t}^{-1}   \delta^{\eps}_{S^1}(t - \tilde{l}_{S^1}(u))\bigr) \ dt
     = - \frac{1}{2} \bigl[ 1_{l_{S^1}(s) < \tilde{l}_{S^1}(u)} - 1_{l_{S^1}(s) > \tilde{l}_{S^1}(u)} \bigr]
\end{multline}
 where $>$ denotes the order relation on $S^1$
which is obtained by transport of the standard order relation on $[0,1)$
with the mapping $i_{t_0}$ (and which therefore depends on the choice of $t_0\in S^1$).\par
For simplicity, let us assume that $\mathbf g$
was chosen such
that, when restricted onto a suitable open neighborhood $V$ of $\bigcup_j \arc(l^j_{\Sigma})$,
it coincides with the restriction of the standard Riemannian metric on $U \cong \bR^2$ onto $V$.
Then  we have
\begin{multline}
\ll  T_1   l'_{\Sigma}(s) \delta^{\eps}_{\Sigma}(\cdot - l_{\Sigma}(s)), \star \
     (T_1    \tilde{l}'_{\Sigma}(u) \delta^{\eps}_{\Sigma}(\cdot - \tilde{l}_{\Sigma}(u))) \gg_{\cH_{\Sigma}} \\
     = -\Tr(T_1 T_1) \bigl\langle l'_{\Sigma}(s), \star \ \tilde{l}'_{\Sigma}(u) \bigr\rangle_{\bR^2} \int_{\Sigma}  \delta^{\eps}_{\Sigma}(\sigma - l_{\Sigma}(s)) \delta^{\eps}_{\Sigma}(\sigma - \tilde{l}_{\Sigma}(u))
     d\mu_{\mathbf g}(\sigma)\\
    = \bigl(l'_{\Sigma}(s)_1 \tilde{l}'_{\Sigma}(u)_2 - l'_{\Sigma}(s)_2 \tilde{l}'_{\Sigma}(u)_1\bigr) \int_{\Sigma}  \delta^{\eps}_{\Sigma}(\sigma - l_{\Sigma}(s)) \delta^{\eps}_{\Sigma}(\sigma - \tilde{l}_{\Sigma}(u))
     d\mu_{\mathbf g}(\sigma)
\end{multline}
Here the last step follows because
 the Hodge operator  $\star:\bR^2 \to \bR^2$ appearing above
is just given by  $\star \ (x_1,x_2) = (x_2, -x_1)$.
One can show (cf. \cite{Ha3b,Ha1}) that
for every smooth function $f:[0,1] \times [0,1] \to \bR$ one has
\begin{multline} \label{eq_count_crossings}
 \lim_{\eps \to 0} \int_0^1   \int_0^1 \biggl[ f(s,u)
\int_{\Sigma}  \delta^{\eps}_{\Sigma}(\sigma - l_{\Sigma}(s)) \delta^{\eps}_{\Sigma}(\sigma - \tilde{l}_{\Sigma}(u))
     d\mu_{\mathbf g}(\sigma) \biggr] ds \ du\\
     = \sum_{\bar{s}, \bar{u} \text{ with } l_{\Sigma}(\bar{s}) = \tilde{l}_{\Sigma}(\bar{u})}
     f(\bar{s},\bar{u}) \frac{1}{| l'_{\Sigma}(\bar{s})_1 \tilde{l}'_{\Sigma}(\bar{u})_2 - l'_{\Sigma}(\bar{s})_2 \tilde{l}'_{\Sigma}(\bar{u})_1|}
\end{multline}

Combining Eqs. \eqref{eq_formel_fuer_T}--\eqref{eq_count_crossings}  we obtain
\begin{equation}
\exp( - \tfrac{1}{2}T(l,\tilde{l})) =  \exp(\tfrac{\pi i}{k} \LK(l,\tilde{l}))
\end{equation}
where we have set
\begin{equation} \LK(l,\tilde{l}) :=  \frac{1}{2} \sum_{\bar{s}, \bar{u} \text{ with } l_{\Sigma}(\bar{s}) = \tilde{l}_{\Sigma}(\bar{u})}\eps(\bar{s}, \bar{u})
\end{equation}
with
$$\eps(\bar{s}, \bar{u}) :=  \bigl[ 1_{l_{S^1}(\bar{s}) < \tilde{l}_{S^1}(\bar{u})} - 1_{l_{S^1}(\bar{s}) > \tilde{l}_{S^1}(\bar{u})} \bigr] \sgn(  l'_{\Sigma}(\bar{s})_1 \tilde{l}'_{\Sigma}(\bar{u})_2 - l'_{\Sigma}(\bar{s})_2 \tilde{l}'_{\Sigma}(\bar{u})_1) \in \{-1,1\}  $$
Clearly,
$\LK(l,\tilde{l})$  depends on the choice of the point $t_0 \in S^1$.
Anyhow, it is closely  related to the linking number $\Link(l,\tilde{l})$
of $l$ and $\tilde{l}$ (which does, of course, not depend on $t_0$).
 The precise relationship will be given in Proposition \ref{lem2} below.\par

Until now
we have only studied the expression $T(l_j,l_k)$
in the case $j \neq k$.
The reason why we have excluded the case  $j=k$
so far is that  in a naive treatment of the case  $j=k$  the so-called ``self-linking problem''
 would appear.
 One can avoid the ``self-linking problem''
 by introducing  an additional regularization procedure
which is called ``framing''.
By a ``framing'' of the link  $L=(l_1,l_2,\ldots,l_n)$
we will understand in the sequel (cf. Remark \ref{rm5.2})
a family $(\phi_s)_{s>0}$ of diffeomorphisms of $M$
such that  $\phi_s \to  \id_M$ uniformly on $M$ (or, at least uniformly on  $\bigcup_i \arc(l_i)$)
as $s \to 0$.
We will call a framing $(\phi_s)_{s>0}$ ``admissible''
iff it has the following properties:
\begin{enumerate}
\item[(F1)] Each $\phi_s$ preserves  the orientation of $M$
  and also the volume of $M$ (if $M$ is equipped with the Riemannian metric $\mathbf g_M$
   induced by $\mathbf g$)
   \item[(F2)] Each $\phi_s$ is ``compatible with the torus gauge'' in the
sense that $\phi^*_s( \cA^{\orth}) = \cA^{\orth}$
\item[(F3)] Each two-component link  $(l_j, \phi_s \circ l_j)$, $j \le n$,
 is admissible for all sufficiently small $s>0$.
\end{enumerate}

From condition (F2) it follows  that each $\phi_s$ induces a  diffeomorphism $\bar{\phi}_s:\Sigma \to \Sigma$
and a linear isomorphism $(\phi_s)_*:\cA^{\orth} \to \cA^{\orth}$
in a natural way (cf.  Sec. 9.3 in  \cite{Ha3b}.
Note that $(\phi_s)_*$ does not coincide with $(\phi_s^{-1})^*$).
\begin{remark} \rm \label{rm5.2}
Normally, by a
 ``framing'' of a link $L=(l_1,l_2,\ldots,l_n)$
one understands a family $(X_1,X_2,\ldots,X_n)$ where each $X_i$
is a smooth normal vector field on $\arc(l_i)$, i.e.
$X_i$ is a mapping $\arc(l_i) \to TM$
such that $X_i(l_i(s)) \in T_{l_i(s)}M$, $s \in [0,1]$,
 is normal (w.r.t. to $\mathbf g_M$)
 to the tangent vector
$l'_i(s) \in T_{l_i(s)}M$.
One can always find a global vector field $X$ on $M$
such that $X_{| \arc(l_i)} = X_i$.
As $M$ is compact, $X$ induces a global flow $(\phi_s)_{s \in \bR}$ on $M$.
Clearly, $\phi_s \to  \id_M$ as $s \to 0$
so $X$ induces
a ``framing'' in the above sense.
\end{remark}

With the help of the framing $(\phi_s)_{s>0}$
we can now solve the self-linking problem.
The simplest\footnote{The standard way of dealing with  the self-linking problem
consists in replacing
some of the loops $l_i$ that appear in the singular terms by their
 ``deformations''  $\phi_s \circ l_i$ where $s$ is chosen small enough.
 If one proceeds in this way one  has to deal with each singular term separately.
 Moreover, the replacement of $l_i$ by $\phi_s \circ l_i$
 has to be made ``by hand'' in the middle of the computations
 rather the before beginning the computations.
 Clearly, this is not very elegant.}
 way to do this
is the following:
We introduce for each $\phi_s$ a  ``deformed'' versions $\Phi^{\orth}_{B,\phi_s}$ of  $\Phi^{\orth}_{B}$.
$\Phi^{\orth}_{B,\phi_s}$ is the unique continuous linear functional $(\cN) \to \bC$ such that
 \begin{equation}
 \Phi^{\orth}_{B,\phi_s}(\exp(i(\cdot,j)))
   = \exp(i \ll j,m(B) \gg_{\cH^{\orth}} ) \exp(- \tfrac{1}{2}  \ll (\phi_s)_*(j), C(B)  j\gg_{\cH^{\orth}} )
 \end{equation}
for every $j \in \cN = \cA^{\orth}$
where $(\phi_s)_*:\cA^{\orth} \to  \cA^{\orth}$ is the  linear isomorphism
mentioned above.
We then obtain a ``framed'' version of $\WLO(L;A^{\orth}_c, B)$  by setting
 \begin{equation} \label{eq_defWLOAB}  \WLO(L,\phi_s;A^{\orth}_c, B) :=     \lim_{\eps \to 0}  \Phi^{\orth}_{B,\phi_s}( \prod_i \Tr( \cP \exp(\int_{l^{\eps}_i} (\cdot) + A^{\orth}_c   + Bdt) )
\end{equation}
Carrying out similar computations as above (for details, see \cite{Ha3b,Ha1})
we then obtain
\begin{multline}  \label{eq_deformedPhi}
 \lim_{\eps \to 0}
 \Phi^{\orth}_{B,\phi_s}( \prod_i \exp(  T_1 (  \cdot , T_1   \int_0^1  (l^i_{\Sigma})'(s) \delta^{\eps}(\cdot - l_i(s)) ds
  ))) \\
  =  \biggl( \prod_{j} \exp\bigl(\int_0^1   l^j_{\bR}(t) \tfrac{d}{dt} B(l^j_{\Sigma}(t)) dt\bigr)\biggr)   \biggl( \prod_{j,k}  \exp(\pi i \lambda \LK(l_j, \phi_s \circ l_k))\biggr)
  \end{multline}
  if $s$ is sufficiently small.
  From Eqs. \eqref{eq_def_loopsmearing}, \eqref{eq_defWLOAB}, \eqref{eq_deformedPhi}, and  \eqref{eq_6.10} above
 we finally obtain (taking into account that for $j \neq k$ one has
$\LK(l_j, \phi_s \circ l_k) = \LK(l_j, l_k)$
if $s$ is sufficiently small)
 \begin{multline} \label{eq_maintheorem}
  \WLO(L,\phi_s;A^{\orth}_c,B)
= \biggl( \prod_j   \exp( \lambda \pi i  \LK(l_j, \phi_s \circ l_j)) \biggr) \biggl(  \prod_{j\neq k} \exp( \lambda \pi i \LK(l_j,l_k)) \biggr) \\
  \times \biggl( \prod_j  \exp(\int_{l^j_{\Sigma}} A^{\orth}_c)\biggr)   \exp\bigl(\sum_{m \in  \cM(t_0)} \eps_m   B(\sigma_m)\bigr)
  \end{multline}
  if $s>0$ is sufficiently small.
  Here we have set
\begin{equation}\cM(t_0) := \bigcup_{j=1}^n \cM_j(t_0), \quad \text{with }
 \cM_j(t_0)  := \{ (j,u) \mid  u \in I_j(t_0)\}
\quad \text{ for } j \le n
\end{equation}
and
\begin{equation} \sigma_m  := l_{\Sigma}^{j_m}(u_m), \quad
\eps_m  := \sgn(l_{S^1}^{j_m};u_m) \quad \text{ for } m \in \cM(t_0)
\end{equation}
where $j_m$ and $u_m$ are given by
$m=(j_m,u_m)$.\par
Of course, Eq. \eqref{eq_maintheorem}
can also derived in the general case, i.e. in the case when assumption (S) above, which we
have made in order to simplify the notation, is not fulfilled.

  \subsection{Non-Abelian $G$ and   $L$ has  no double points}
\label{subsec5.2}

Let us now consider the case where $G$
is a Non-Abelian (simple and simply-connected compact) Lie group.
In the present subsection we will only consider the special situation where
$\DP(L) =  \emptyset$, i.e. where the link
$L= ((l_1, l_2, \ldots, l_n),(\rho_1,\rho_2,\ldots,\rho_n))$
has no double points.
For simplicity we will give a detailed computation
only for   the group $G=SU(N)$ and we will assume that
 the ``colors'' $(\rho_1,\rho_2,\ldots,\rho_n)$ of the link $L$
all coincide with the fundamental representation $\rho_{SU(N)}$
of $G=SU(N)$ (see Remark \ref{rm_Nachtrag}  below for the case of general $G$ and general link colors).\par

Let us  fix an admissible framing $(\phi_s)_{s>0}$ of $L$
with the following two extra properties:
\begin{enumerate}
\item[(H1)] For all $j \le n$ and all sufficiently small $s>0$
 the set of ``twist framing double points'' of $(l_j, \phi_s \circ l_j)$ (in the sense
 of Remark \ref{rm5.3} (1) below)
is empty.
 \item[(H2)]  For  every $\sigma \in \arc(l^j_{\Sigma})$
 which is not an $l_j$-self-crossing double point\footnote{in which case $\bar{\phi}_s(\sigma) \in \arc(l^j_{\Sigma})$ would follow}   of $(l_j, \phi_s \circ l_j)$ (in the sense
 of Remark \ref{rm5.3} (1) below)
  the points
$\bar{\phi}_s(\sigma)$ and $\bar{\phi}^{-1}_s(\sigma)$
lie in different connected components
of $\Sigma \backslash \arc(l^j_{\Sigma})$
provided that $s>0$ is sufficiently small.
\end{enumerate}
Such a framing will be called ``horizontal''.
\begin{remark} \rm  \label{rm5.3}
\begin{enumerate}
\item In order to explain what the two notions ``twist framing double points'' and ``self-crossing double point''
which we have used above mean we first note that if $(l,\tilde{l})$ is an admissible  link in  $\Sigma \times S^1$
and   $p = \pi_{\Sigma}(x) = \pi_{\Sigma}(y)$
where $x \in \arc(l)$, $y \in \arc(\tilde{l})$ then if  $\tilde{l}$ is ``close'' to $l$  normally  also $y$ will be ``close'' to $x$.
But there is one exception: If $p$ is ``close'' to a double point of $l$, $y$ need not be ``close'' to $x$.
In the first case we
 call
 $p$ a ``twist double point'' of $(l,\tilde{l})$ and in the second case
a ``$l$-self-crossing double point'' (this distinction can be made precise in a very similar
 way as in Def. 16 in \cite{Ha2}).
\item As a  motivation for the use of the term  ``horizontal'' we remark that
if an admissible framing
$(\phi_s)_{s>0}$ is induced by a tuple of vector fields
$(X_1,X_2,\ldots,X_n)$ like in Remark \ref{rm5.2} above
then  for $(\phi_s)_{s>0}$ to be horizontal it is sufficient that
each vector field $X_j$ is ``horizontal'' in the sense
that $dt(X_j)=0$.\par
We would like to emphasize
that  here we do \underline{not}
 follow the terminology of \cite{Ha2}
where the $\bR^3$-analogue of this type of framing was not
called ``horizontal'' but
``strictly vertical''.
\end{enumerate}
\end{remark}

Using the Piccard-Lindeloef series expansion
we obtain
\begin{multline}
\cP \exp(\int_{l_j} \hat{A}^{\orth} + A^{\orth}_c + A^{\orth}_{sing}(\cl) + Bdt) \\
=  \sum_{m=0}^{\infty}
 \int_{\triangle_m}  \bigl[ D^{l_j}_{u_1}(\hat{A}^{\orth}+ A^{\orth}_c + A^{\orth}_{sing}(\cl) + Bdt ) \cdots D^{l_j}_{u_m}(\hat{A}^{\orth}+ A^{\orth}_c + A^{\orth}_{sing}(\cl) + Bdt )\bigr] \ du
\end{multline}
where we have set $\triangle_m :=  \{ u \in [0,1]^m \mid u_1 \ge u_2 \ge \cdots \ge
u_m \}$ and
\begin{align*}
D^{l_j}_{u}(\hat{A}^{\orth}+ A^{\orth}_c + A^{\orth}_{sing}(\cl) + Bdt ) & :=
( \hat{A}^{\orth} +   A^{\orth}_c + A^{\orth}_{sing}(\cl) + Bdt)(l_j'(u))
 \end{align*}
This holds if  $\hat{A}^{\orth}$, $A^{\orth}_c$,  and $B$
are smooth.
In order to be able to work also with general $\hat{A}^{\orth} \in \cN^*$
we  now use again loop smearing.
As in Subsec. \ref{subsec5.1}
we will assume again
for simplicity that condition (S) above is fulfilled
so that we can use the notation $\delta^{\eps}(\cdot - l_j(u))$
and make the identification $\cA^{\orth}_{U \times S^1}  \cong C^{\infty}(U \times S^1,\bR^2 \otimes \cG)$.
Recall  that the orthogonal-basis
$(T_a)_{a \le \dim(G)}$ of $\cG$
was chosen such that
$T_a \in \ct$ for all $a \le r= \rank(G)$.\par
 Let us now replace all terms of the form  $ \hat{A}^{\orth} (l'(u))$
 by
 $\sum\nolimits_a T_a ( \hat{A}^{\orth},  T_a   (l_{\Sigma})'(u) \delta^{\eps}(\cdot - l(u)))$
 (here $(\cdot,\cdot)$ denotes again the canonical pairing
 $\cN^* \times \cN \to \bR$).
 In particular,  we replace $D^{l_j}_{u}(\hat{A}^{\orth}+ A^{\orth}_c + A^{\orth}_{sing}(\cl) + Bdt )$
 by
 \begin{multline*}
D^{l^{\eps}_j}_{u}(\hat{A}^{\orth}+ A^{\orth}_c + A^{\orth}_{sing}(\cl) + Bdt ) := \\
 \sum\nolimits_a T_a ( \hat{A}^{\orth},  T_a    (l^j_{\Sigma})'(u) \delta^{\eps}(\cdot - l_j(u))) + (A^{\orth}_c + A^{\orth}_{sing}(\cl) + Bdt)(l'_j(u))
 \end{multline*}
 where $\eps >0$ and we replace
$ \cP \exp(\int_{l_j} \hat{A}^{\orth} + A^{\orth}_c + A^{\orth}_{sing}(\cl) + Bdt)$
by
\begin{multline}
\cP \exp(\int_{l^{\eps}_j} \hat{A}^{\orth} + A^{\orth}_c + A^{\orth}_{sing}(\cl) + Bdt) \\
:=  \sum_{m=0}^{\infty} \int_{\triangle_m}  \bigl[ D^{l^{\eps}_j}_{u_1}(\hat{A}^{\orth}+ A^{\orth}_c + A^{\orth}_{sing}(\cl) + Bdt ) \cdots D^{l^{\eps}_j}_{u_m}(\hat{A}^{\orth}+ A^{\orth}_c + A^{\orth}_{sing}(\cl) + Bdt )\bigr] \ du
\end{multline}
 for $\hat{A}^{\orth} \in \cN^*$,  $A^{\orth}_c \in \cA^{\orth}_c$, $\cl \in [\Sigma,G/T]$, $B \in \cB$.
We then have
\begin{multline}   \label{eq_PicLin}
\prod_{j=1}^n \Tr_{\rho_j}(\cP \exp(\int_{l^{\eps}_j} \hat{A}^{\orth} + A^{\orth}_c + A^{\orth}_{sing}(\cl) + Bdt)) \\
= \prod_{j=1}^n  \Tr_{\rho_j}\biggl[ \sum_{m_j=0}^{\infty} \int_{\triangle_{m_j}} \prod_{i_j =1}^{m_j}  \biggl[ D^{l^{\eps}_j}_{u_{i_j}}(\hat{A}^{\orth}+ A^{\orth}_c + A^{\orth}_{sing}(\cl) + Bdt )\biggr]  \ du \biggr]
\end{multline}

We will now apply the functional  $\Phi^{\orth}_{B,\phi_s}$ on both sides of the previous equation.
From the assumption that  $\DP(L) = \emptyset$
and that the framing $(\phi_s)_{s >0}$ is horizontal it  follows
for all sufficiently small $s>0$ that the following statement is true:
For all sufficiently small  $\eps>0$ the functions $\psi_1, \ldots, \psi_n$ on $\cN^*$ given by
$$ \psi_j := \Tr_{\rho_j}(\cP \exp(\int_{l^{\eps}_j} (\cdot) + A^{\orth}_c + A^{\orth}_{sing}(\cl) + Bdt))$$
are ``independent'' w.r.t. the $\Phi^{\orth}_{B,\phi_s}$
in the sense that
\begin{equation} \label{eq_independence} \Phi^{\orth}_{B,\phi_s}\bigl(\prod_j \psi_j\bigr)  =\prod_j  \Phi^{\orth}_{B,\phi_s}\bigl( \psi_j \bigr)
\end{equation}
holds  (cf. Appendix A.).
Thus we can interchange $\Phi^{\orth}_{B,\phi_s}$ with $\prod_{j=1}^n$ in Eq. \eqref{eq_PicLin}.
We have assumed above that each representation $\rho_j$ equals
the fundamental representation $\rho_{SU(N)}$ of $G=SU(N)$.
Thus, each $\Tr_{\rho_j}$ can be replaced by $\Tr(\cdot):=\Tr_{\Mat(N,\bC)}(\cdot)$.
Clearly, $\Phi^{\orth}_{B,\phi_s}$ ``commutes''\footnote{More precisely, we have
$\Phi^{\orth}_{B,\phi_s} \circ \Tr(\cdot) = \Tr(\cdot) \circ \Phi^{\orth}_{B,\phi_s}$
where
$\Phi^{\orth}_{B,\phi_s}:(\cN) \otimes \Mat(N,\bC) \to \Mat(N,\bC)$
and $\Tr:\Mat(N,\bC) \to \bC$ on the r.h.s.
and $\Phi^{\orth}_{B,\phi_s}:(\cN)  \to \bC$ and
$\Tr: (\cN) \otimes \Mat(N,\bC) \to   (\cN)$
on the l.h.s.} with $\Tr(\cdot)$
and so we can  interchange $\Phi^{\orth}_{B,\phi_s}$
and $\Tr(\cdot)$.
 By interchanging
 $\Phi^{\orth}_{B,\phi_s}$ also with   $\sum_{m_j}$, $\int_{\triangle_{m_j}} du$, and $\prod_{i=1}^{m_j}$,
 which  can be justified rigorously,
 we obtain
\begin{multline}
  \Phi^{\orth}_{B,\phi_s}\biggl( \prod_j \Tr(\cP \exp(\int_{l^{\eps}_j} (\cdot) + A^{\orth}_c + A^{\orth}_{sing}(\cl) + Bdt)) \biggr) \\
= \prod_j  \Tr\biggr[ \sum_{m_j} \int_{\triangle_{m_j}}  \prod_{i=1}^{m_j}    \Phi^{\orth}_{B,\phi_s}\biggl( D^{l^{\eps}_j}_{u_i}(\cdot + A^{\orth}_c + A^{\orth}_{sing}(\cl) + Bdt ) \biggr)  \ du \biggr]
\end{multline}
Thus,  by  applying $\lim_{\eps \to 0}$ on both  sides of the previous equation
and interchanging  the $\lim_{\eps \to 0}$ -limit with $\sum_{m_j}$ and $\int_{\triangle_{m_j}} \cdots du$
 (this can be justified in a similar way as the analogous steps
 in the proof of Theorem 4 in \cite{Ha2})
 and using Eq. \eqref{eq_WLO_AB}
  we obtain  for sufficiently small $s>0$
\begin{align} \label{eq_WLOAB}
  & \WLO(L,\phi_s;A^{\orth}_c, A^{\orth}_{sing}(\cl), B) \nonumber \\
  &   =  \lim_{\eps \to 0} \Phi^{\orth}_{B,\phi_s}\biggl( \prod_j \Tr(\cP \exp(\int_{l^{\eps}_j} (\cdot) + A^{\orth}_c + A^{\orth}_{sing}(\cl) + Bdt)) \biggr)  \nonumber \nonumber \\
& = \prod_j  \Tr\biggr[ \sum_{m_j} \int_{\triangle_{m_j}}  \prod_{i=1}^{m_j}   \lim_{\eps \to 0}  \Phi^{\orth}_{B,\phi_s}\biggl( D^{l^{\eps}_j}_{u_i}(\cdot + A^{\orth}_c + A^{\orth}_{sing}(\cl) + Bdt ) \biggr)  \ du \biggr] \nonumber \\
 & \overset{(*)}{=} \prod_j \Tr\biggl[\exp\biggl(\int_0^1 du  \biggl[  \lim_{\eps \to 0}  \Phi^{\orth}_{B,\phi_s}\biggl( D^{l^{\eps}_j}_{u}(\cdot + A^{\orth}_c + A^{\orth}_{sing}(\cl) + Bdt)  \biggr) \biggr] \biggr) \biggr]
\end{align}
In step $(*)$ we have taken into account that
 $\lim_{\eps \to 0}  \Phi^{\orth}_{B,\phi_s}\biggl( D^{l^{\eps}_j}_{u_i}(\cdot + A^{\orth}_c + A^{\orth}_{sing}(\cl) + Bdt )\biggr)  \in \ct$ (cf. Eq. \eqref{eq_Dl_value} below)
 so all the factors in the $\prod_{i=1}^{m_j} \cdots$ product in the last
 but one line  in Eq. \eqref{eq_WLOAB} commute with each other and  the
Piccard-Lindeloef series is reduces to the exponential expression
in the last line of Eq. \eqref{eq_WLOAB}.\par
Let us  set $l^j_{\bR}:= i_{t_0}^{-1} \circ l^j_{S^1} - 1/2$, $j \le n$.
 Then  we have for fixed $s>0$ and $u \in [0,1]$
\begin{multline} \label{eq_Dl_value}   \lim_{\eps \to 0}  \Phi^{\orth}_{B,\phi_s}\biggl( D^{l^{\eps}_j}_{u}(\cdot + A^{\orth}_c + A^{\orth}_{sing}(\cl) + Bdt) \biggr)\\
 \overset{(*)}{=}
  \lim_{\eps \to 0}   \sum\nolimits_{a=1}^{r} T_a \ll m(B),  T_a    (l^j_{\Sigma})'(u) \delta^{\eps}(\cdot - l_j(u)) \gg_{\cH^{\orth}}
 +  (A^{\orth}_c + A^{\orth}_{sing}(\cl) + Bdt)(l'_j(u))\\
 \overset{(+)}{=}  \bigl\{    l^j_{\bR}(u) \tfrac{d}{du} B(l^j_{\Sigma}(u)) + B(l^j_{\Sigma}(u)) \cdot (l^j_{\bR})'(u) \bigr\} + (A^{\orth}_c + A^{\orth}_{sing}(\cl))(l'_{\Sigma}(u))
\end{multline}
Here step $(*)$ follows from Eq. \eqref{eq_erstes_moment}
and step $(+)$ follows in a similar way as Eq. \eqref{eq_6.9} above.
From Eqs. \eqref{eq_WLOAB} and  \eqref{eq_Dl_value} we now
obtain (taking into account Eq. \eqref{eq_6.10} above)
 \begin{multline} \label{eq_WLOABexpl} \WLO(L,\phi_s;A^{\orth}_c, A^{\orth}_{sing}(\cl),B) \\
  = \prod_{j=1}^n \Tr\bigg[ \exp(\int_{l^j_{\Sigma}}   A^{\orth}_c)    \exp(\int_{l^j_{\Sigma}}   A^{\orth}_{sing}(\cl))  \exp(\ \sum_{m \in  \cM_j(t_0)} \eps_m  B(\sigma_m))  \biggr]
 \end{multline}
where $ \cM_j(t_0)$ and  $\eps_m$, $\sigma_m$  for $m \in   \cM_j(t_0)$
  are defined as at the end of Subsec. \ref{subsec5.1}.

\begin{remark} \label{rm_Nachtrag}  \rm
Most of the steps in the computation above
can be generalized immediately
 to the case where $G$ is an arbitrary Non-Abelian (simple and simply-connected compact) Lie group
  and each $\rho_i$, $i \le n$, an arbitrary finite-dimensional representation of $G$.
The only exception is the
step where we interchange
$\Phi^{\orth}_{B,\phi_s}$ and $\Tr$.
$\Tr$ enters the computation because
it is the obvious extension
of  $\Tr_{\rho_{SU(N)}}: G \to \bC$
to a linear mapping  $\Mat(N,\bC) \to \bC$.
By contrast,
$\Tr_{\rho_i}:G \to \bC$
will in general not admit an extension to a linear mapping $\Mat(N,\bC) \to \bC$.
Luckily, this complication can be circumvented
by  taking into account that
for every finite-dimensional representation $\rho:G \to \GL(V)$
(and, in particular, for $\rho = \rho_1, \rho_2, \ldots, \rho_n$)
one has
 \begin{equation} \label{eq_derived_rep} \Tr_{\rho}(\cP \exp(\int_l A)) =  \Tr_{\End(V)}(\rho(\cP \exp(\int_l A)))
 = \Tr_{\End(V)}(\cP \exp(\int_l \rho_*(A)))
 \end{equation}
where $\rho_*:\cG \to \gl(V)$ is the derived representation of $\rho$.
Clearly, the mappings
$\Phi^{\orth}_{B,\phi_s}$ and  $\Tr_{\End(V)}$
can be ``interchanged''\footnote{more precisely, we can make use of the equality
$\Phi^{\orth}_{B,\phi_s} \circ \Tr_{\End(V)} = \Tr_{\End(V)} \circ \Phi^{\orth}_{B,\phi_s}$
where
$\Phi^{\orth}_{B,\phi_s}:(\cN) \otimes \End(V) \to \End(V)$
and $\Tr:\End(V) \to \bC$ on the r.h.s.
and $\Phi^{\orth}_{B,\phi_s}:(\cN)  \to \bC$ and
$\Tr: (\cN) \otimes\End(V) \to   (\cN)$
on the l.h.s. are the obvious mappings}.
The generalization of
 Eq. \eqref{eq_WLOABexpl}
at which one then arrives is the obvious one, i.e.
 in the general case
one simply has
\begin{multline} \label{eq_WLOABexpl_gen} \WLO(L,\phi_s;A^{\orth}_c, A^{\orth}_{sing}(\cl),B) \\
  = \prod_{j=1}^n \Tr_{\rho_j}\bigg[ \exp(\int_{l^j_{\Sigma}}   A^{\orth}_c)    \exp(\int_{l^j_{\Sigma}}   A^{\orth}_{sing}(\cl))  \exp(\ \sum_{m \in  \cM_j(t_0)} \eps_m  B(\sigma_m))  \biggr]
 \end{multline}

\end{remark}

\subsection{Non-Abelian $G$ and general $L$}
\label{subsec5.3}

As in Subsec. \ref{subsec5.2}
let  $G$ be a Non-Abelian (simple and simply-connected compact) Lie group.
We will now consider the case where $L$ is a general link in $\Sigma$.
(In order to simplify the notation we will again restrict
ourselves to the  case where $G=SU(N)$ and where
 the ``colors'' $(\rho_1,\rho_2,\ldots,\rho_n)$ of the link $L$
all coincide with the fundamental representation $\rho_{SU(N)}$
of $G=SU(N)$).
We will  briefly sketch a strategy for evaluating  \eqref{eq_WLO_AB}
 which  is  similar to the strategy  used in Sect. 6 in \cite{Ha2}.
We first  ``cut'' the loops of $L$ into finitely many subcurves
in such a way  that the following
relations are fulfilled
 for every $c \in \cC(L)$ where $\cC(L)$ denotes the set of curves which are obtained by cutting the loops in $L$
 (it is not difficult to see that this is always possible if $L$ is
admissible)
\begin{itemize}
\item{$\DP(c) = \emptyset$ and  $\pi_{\Sigma}(x) \notin \DP(L)$  if $x \in \Sigma \times S^1$ is an endpoint of $c$.}
\item{There is at most one $c'  \in \cC(L)$, $c \neq c'$, such that
$\DP_o(c,c') \neq \emptyset$ and if there is such a $c'$ then $\# \DP(c,c') =1$.}
\end{itemize}
where $\DP_o(c,c'):= \DP(c,c') \backslash \{\pi_{\Sigma}(x) \mid
 x \in \Sigma \times S^1 \text{ is  an endpoint of } c \text{ or } c'\}$.
A ``1-cluster'' of $L$ is a set of the form $\{c\}$, $c \in \cC(L)$, such that $\DP_o(c,c') = \emptyset$ for all $c' \in \cC(L)$ with $c' \neq c$. A ``2-cluster''
 of $L$ is a set of the form $\{c,c'\}$, $c,c' \in \cC(L)$, $c \neq c'$, such that
 $\DP_o(c,c') \neq \emptyset$.\par
The set of 1-clusters (resp. 2-clusters) of $L$ will be denoted by $Cl_1(L)$ (resp. $Cl_2(L)$).
From the properties of $\cC(L)$ above it immediately follows that
the set $Cl(L)$ defined by $Cl(L):= Cl_1(L) \cup Cl_2(L)$  is a partition of $\cC(L)$. If $cl = \{c_1,c_2\} \in Cl_2(L)$ we write $c_1 < c_2$ iff the pair
$(\hat{c}_2,\hat{c}_1)$  is positively oriented where $\hat{c}_i$, $i \in \{1,2\}$, denotes the
tangent vector of $\pi_{\Sigma} \circ c_i$ in the unique double point $p$ of $(c_1,c_2)$.\par
Let   $\eps>0$, $A^{\orth}_c \in \cA^{\orth}_c$, $\cl \in [\Sigma, G/T]$,
 $B \in \cB$ be fixed
and let $cl \in  Cl(L)$.
We set
\begin{equation} P^{cl^{\eps}}(\cdot + A^{\orth}_c + A^{\orth}_{sing}(\cl) + Bdt)
:= \otimes_{i=1}^{\# cl}
 \cP \exp(\int_{c_i^{\eps}} (\cdot + A^{\orth}_c  + A^{\orth}_{sing}(\cl) + Bdt))
\end{equation}
where we have set $\# cl :=1$
(resp. $\# cl :=2$) if $cl  \in  Cl_1(L)$ (resp. $cl  \in  Cl_2(L)$)
and where $c_1$ is given (resp. $c_1, c_2$ are given) by
$cl=\{c_1\}$ (resp.   $cl = \{c_1,c_2\}$ where $c_1 < c_2$).\par
 It is not difficult to see that there is a linear form $\beta_L$ on  $\otimes_{cl \in Cl(L)} \bigl(\otimes^{\# cl} \Mat(N,\bC)\bigr)$
 such that for all $\eps>0$ we have
$\prod_i \Tr( \cP \exp(\int_{l^{\eps}_i} \cdot + A^{\orth}_c  + A^{\orth}_{sing}(\cl)  + Bdt)
 = \beta_L \circ \bigl( \otimes_{cl \in Cl(L)} P^{cl^{\eps}}( \cdot + A^{\orth}_c  + A^{\orth}_{sing}(\cl) + Bdt) \bigr)$.
 If $s>0$ is chosen  small enough we have
\begin{multline} \label{eqinle5}
  \Phi^{\orth}_{B,\phi_s} \biggl( \bigl( \otimes_{cl \in Cl(L)} P^{cl^{\eps}}(\cdot + A^{\orth}_c  + A^{\orth}_{sing}(\cl)+ Bdt) \bigr) \biggr)  \\
 =  \otimes_{cl \in Cl(L)}   \Phi^{\orth}_{B,\phi_s}\biggl(  P^{cl^{\eps}}(\cdot + A^{\orth}_c + A^{\orth}_{sing}(\cl) + Bdt)  \biggr)
\end{multline}
 for all sufficiently small $\eps>0$.
This follows in a similar way as Eq. \eqref{eq_independence} above (cf. also
Eq. 6.3 in  \cite{Ha2}).
Eq. \eqref{eqinle5} implies
\begin{multline*}
    \Phi^{\orth}_{B,\phi_s} \biggl( \prod_i \Tr( \cP \exp(\int_{l^{\eps}_i} \cdot + A^{\orth}_c  + A^{\orth}_{sing}(\cl) + Bdt) \biggr) \\
  = \beta_L\biggl( \otimes_{cl \in Cl(L)}  \Phi^{\orth}_{B,\phi_s}\biggl(  P^{cl^{\eps}}(\cdot + A^{\orth}_c + A^{\orth}_{sing}(\cl) + Bdt) \biggl)  \biggl)
\end{multline*}
 for all sufficiently small $s>0$ and $\eps>0$.
One can  show that the limits
\begin{equation} \label{eq_def_Rcl}
 R^{cl}(\phi_s; A^{\orth}_c, A^{\orth}_{sing}(\cl), B) :=  \lim_{\eps \to 0}   \Phi^{\orth}_{B,\phi_s}\biggl(  P^{cl^{\eps}}(\cdot + A^{\orth}_c + A^{\orth}_{sing}(\cl) + Bdt) \biggl)
\end{equation}
exist.  Consequently, we obtain
\begin{align} \label{eq_pre_statesum} \WLF(L, \phi_s; A^{\orth}_c, A^{\orth}_{sing}(\cl), B)
 & = \beta_L( \otimes_{cl \in Cl(L)}R^{cl}(\phi_s; A^{\orth}_c, A^{\orth}_{sing}(\cl), B))
\end{align}
The values of $R^{cl}(\phi_s; A^{\orth}_c, A^{\orth}_{sing}(\cl), B)$
can be computed explicitly using similar techniques as in \cite{Ha2}.
In the special case when the framing $(\phi_s)_{s >0}$
is horizontal and $\# cl =1$
the values of $R^{cl}(\phi_s; A^{\orth}_c, A^{\orth}_{sing}(\cl), B)$
 can be computed  in a very similar way as
those of the expression $\WLO(L,\phi_s;A^{\orth}_c, A^{\orth}_{sing}(\cl), B)$ appearing
 in Eq. \eqref{eq_WLOAB}.
By contrast, the computation of $R^{cl}(\phi_s; A^{\orth}_c,
A^{\orth}_{sing}(\cl), B)$ for $\# cl =2$ is rather tedious. We will
postpone these computations to a future paper.

\section{The Computation of the WLOs: Step 3}
\label{sec6}

We will now evaluate the whole expression
on the right-hand side
of Eq. \eqref{eq_WLO_end} in a couple of special cases
and then make some remarks concerning the general case.\par

\subsection{Special case 1: $G=U(1)$ and general $L$}
\label{subsec6.1}

Let us  go back to the situation  of Subsec. \ref{subsec5.1} above,
i.e. $G=T = U(1)$ and $\Sigma=S^2$.  We want to  evaluate the
expression
\begin{multline} \label{eq_WLOAbel}
 \WLO(L,\phi_s):=   \int_{\cA_c^{\orth} \times \cB}  \WLO(L,\phi_s;A^{\orth}_c, Bdt)\\
 \times \exp( i \tfrac{k}{2\pi}   \ll \star dA^{\orth}_c, B \gg_{L^2_{\ct}(\Sigma,d\mu_{\mathbf g})}) (DA_c^{\orth} \otimes  DB)
\end{multline}
where $\cB=C^{\infty}(\Sigma,\ct)$
and where $\WLO(L,\phi_s;A^{\orth}_c, B)$ is as in Eq. \eqref{eq_defWLOAB}.
In order to achieve this
 we  use the identification $\cA_c^{\orth} \cong  \cA_{\Sigma,\ct} (= \cA_{\Sigma}) $,
  plug in  the right-hand side of Eq. \eqref{eq_maintheorem} into Eq. \eqref{eq_WLOAbel}
 and make use of the Hodge decomposition
\begin{equation} \label{eq_Hodge}
 \cA_{\Sigma,\ct}  = \cA_{ex} \oplus \cA_{harm} \oplus \cA^*_{ex}
\end{equation}
 where $\cA_{ex}  := \{d f \mid f \in C^{\infty}(\Sigma,\ct)  \}$,
  $\cA^*_{ex}  := \{ \star d f \mid f \in C^{\infty}(\Sigma,\ct)\}$, and
  $\cA_{harm}  := \{ A \in \cA_{\Sigma,\ct} \mid dA = d(\star A) = 0\}$.
 As we have assumed here that $\Sigma = S^2$ holds we have $H^1_{\bR}(\Sigma) = 0$ which implies $\cA_{harm} = \{0\}$,
i.e. Eq. \eqref{eq_Hodge} reduces to $\cA_{\Sigma,\ct} = \cA_{ex}
\oplus \cA^*_{ex} $. Accordingly, we can replace the $\int \cdots
DA_c^{\orth}$ integration in Eq. \eqref{eq_WLOAbel} by the
integration $ \int \int \cdots DA_{ex} DA^*_{ex}$ where $DA_{ex}$,
$DA^*_{ex}$ denote the ``Lebesgue measures'' on $\cA_{ex}$ and
$\cA^*_{ex}$. Clearly, we have
 $\int_{l^j_{\Sigma}} A_{ex} = 0$ and
$\ll \star  dA_{ex}, B \gg_{L^2_{\ct}(\Sigma,d\mu_{\mathbf g})} = 0$
for every $A_{ex} \in \cA_{ex}$. This means that the integrand in the modification
of Eq. \eqref{eq_WLOAbel} just described,
does not depend on the variable $A_{ex}$.
Thus the $\int \cdots DA_{ex}$-integration produces just a constant
and we obtain
 \begin{multline}  \label{eq_WLO_intermediate}
 \WLO(L,\phi_s)   \sim
  \prod_j   \exp( \lambda \pi i \LK(l_j,\phi_s \circ l_j))  \prod_{j\neq k} \exp( \lambda \pi i \LK(l_j,l_k)) \\
\times  \int_{\cB} \int_{\cA^*_{ex}} \biggl[ \prod_j \exp(\int_{l^j_{\Sigma}} A^*_{ex}) \biggr]
 \biggl[  \prod_{m \in \cM(t_0)} \exp(\eps_m B(\sigma_m))\biggr] \\
   \times \exp( i \tfrac{k}{2\pi}  \ll \star dA^*_{ex},B\gg_{L^2_{\ct}(\Sigma,d\mu_{\mathbf g})})  DA^*_{ex} DB
\end{multline}

Let us assume for a while that $l^j_{\Sigma}$ is a Jordan loop in $\Sigma = S^2$.
Then there are exactly two connected components
 $K_+$ and $K_-$
of $\Sigma \backslash \arc(l^j_{\Sigma})$.
Here  $K_+$ (resp. $K_-$) denotes
 the connected component
 of $\Sigma \backslash \arc(l^j_{\Sigma})$
  with the property that
 the orientation on  $\partial K_+ =  \partial K_- = \arc(l^j_{\Sigma})$
 which is induced by that on $K_+$ (resp. $K_-$)
 coincides with (resp. is opposite to) the orientation on  $\arc(l^j_{\Sigma})$
 which is obtained from the standard orientation of $S^1$  by transport with
     $ l^j_{\Sigma}: S^1 \to    \arc(l^j_{\Sigma})$.
Stokes' Theorem implies
\begin{multline*}
\int_{l^j_{\Sigma}} A^*_{ex} = \frac{1}{2}( \int_{\partial K_+} A^*_{ex}  + \int_{\partial K_-} A^*_{ex})\\
=  \frac{1}{2}( \int_{K_+} d A^*_{ex} -  \int_{K_-} d A^*_{ex}) =
T_1 \ll  \star d A^*_{ex},   T_1  \ind(l^j_{\Sigma};\cdot)
\gg_{L^2_{\ct}(\Sigma,d\mu_{\mathbf g})}
\end{multline*}
where we have  set $\ind(l^j_{\Sigma};\cdot):= \tfrac{1}{2} (1_{K_+} - 1_{K_-})$.
This formula  can be generalized  to the situation
where   $l^j_{\Sigma}$ is not necessarily
a Jordan loop but any smooth loop in $\Sigma = S^2$ with the property
 that  $\Sigma \backslash \arc(l^j_{\Sigma})$ has only finitely many connected components.
In this case we can ``decompose''   $l^j_{\Sigma}$ into finitely many
 Jordan loops $l^j_{\Sigma,1}, \ldots, l^j_{\Sigma,m}$.
More precisely, we can find a finite sequence of  (piecewise smooth)
Jordan loops $l^j_{\Sigma,1}, \ldots, l^j_{\Sigma,m}$
such that $\arc(l^j_{\Sigma}) = \bigcup_{i=1}^m \arc(l^j_{\Sigma,i})$
and $\arc(l^j_{\Sigma,i}) \cap \arc(l^j_{\Sigma,i'}) \subset \DP(l_j)$ if $i \neq i'$.
Then we have $\int_{l^j_{\Sigma}} A^*_{ex} = \sum_{i =1}^m \int_{l^j_{\Sigma,i}} A^*_{ex}$.
So, setting
\begin{equation} \label{eq_def_ind_or}
 \ind(l^j_{\Sigma};\cdot):= \sum_{i =1}^m \ind(l^j_{\Sigma,i};\cdot)
\end{equation}
we obtain again
\begin{equation} \label{eq_defind}
 \int_{l^j_{\Sigma}} A^*_{ex} =
T_1 \ll  \star d A^*_{ex},   T_1  \ind(l^j_{\Sigma};\cdot) \gg_{L^2_{\ct}(\Sigma,d\mu_{\mathbf g})}
\end{equation}
\begin{remark} \rm  \label{rm6.1}
One can show that
for all $\sigma \in \Sigma \backslash \arc(l^j_{\Sigma})$ with $\sigma \neq \sigma_0$
we have
\begin{equation}\label{eq_char_ind}
\ind(l^j_{\Sigma};\sigma) - \ind(l^j_{\Sigma};\sigma_0) = \ind(l^j_{\Sigma \backslash \{\sigma_0\}};\sigma)
\end{equation}
where $\ind(l^j_{\Sigma \backslash \{\sigma_0\}};\sigma)$ denotes the
index of the point $\sigma$
with respect to the loop
$l^j_{\Sigma \backslash \{\sigma_0\}}:
[0,1] \ni t  \mapsto l^j_{\Sigma}(t) \in \Sigma \backslash \{\sigma_0\} = S^2 \backslash \{\sigma_0\} \cong \bR^2$
(or the ``winding number'' of $l^j_{\Sigma \backslash \{\sigma_0\}}$ around $\sigma$).
Eq. \eqref{eq_char_ind} characterizes $\ind(l^j_{\Sigma};\cdot)$ on
 $\Sigma \backslash \arc(l^j_{\Sigma})$ completely
up to an additive constant.
 Clearly, this additive constant does not affect the validity of
 Eq. \eqref{eq_defind}.
 This means that if we had  defined $\ind(l^j_{\Sigma};\cdot)$  by
 \begin{equation} \label{eq_def_ind_new}
 \ind(l^j_{\Sigma};\sigma) :=
 \begin{cases} \ind(l^j_{\Sigma \backslash \{\sigma_0\}};\sigma) & \text{ if  $\sigma \neq \sigma_0 $
 and $\sigma \notin  \arc(l^j_{\Sigma})$}\\
 0 & \text{ if  $\sigma = \sigma_0$ or  $\sigma \in  \arc(l^j_{\Sigma})$}  \\
 \end{cases}
 \end{equation}
 then Eq. \eqref{eq_defind} would still hold.
 This alternative definition of  $\ind(l^j_{\Sigma};\cdot)$   (resp. a suitable generalization of it)
 will be useful in Subsec. \ref{subsec6.3} below.
\end{remark}
We will now  evaluate the right-hand side of Eq. \eqref{eq_WLO_intermediate}
at a heuristic level.
In \cite{Ha6} we will sketch how a  rigorous treatment can be obtained. Recall that $T_1 = i$. Thus we have
\begin{multline}
 \biggl[ \prod_j \exp(\int_{l_j} A^*_{ex}) \biggr]
  \exp( i \tfrac{k}{2\pi}  \ll \star dA^*_{ex},B\gg_{L^2_{\ct}(\Sigma,d\mu_{\mathbf g})}) \\
 =  \exp( i \tfrac{k}{2\pi}  \ll \star dA^*_{ex},B+ \tfrac{2\pi}{k} \sum_j  T_1  \ind(l^j_{\Sigma};\cdot)
   \gg_{L^2_{\ct}(\Sigma,d\mu_{\mathbf g})})
 \end{multline}
 Note that
   $ \ll \star dA^*_{ex},B+ \tfrac{2\pi}{k} \sum_j  T_1  \ind(l^j_{\Sigma};\cdot)\gg_{L^2_{\ct}(\Sigma,d\mu_{\mathbf g})}$
 vanishes for all $A^*_{ex} \in \cA^*_{ex}$
 if and only if  $B+ \tfrac{2\pi}{k} \sum_j  T_1  \ind(l^j_{\Sigma};\cdot)$
 is a constant function,
 i.e. iff there is a $b \in t$ with $B= b- \tfrac{2\pi}{k} \sum_j  T_1  \ind(l^j_{\Sigma};\cdot)$.
  So we obtain, informally,
 \begin{multline} \label{eq6.8}
  \int_{\cB} \int_{\cA^*_{ex}} \biggl[ \prod_j \exp(\int_{l^j_{\Sigma}} A^*_{ex}) \biggr]
 \biggl[  \prod_{m \in \cM(t_0)} \exp(\eps_m B(\sigma_m))\biggr] \\
   \exp( i \tfrac{k}{2\pi}  \ll \star dA^*_{ex},B\gg_{L^2_{\ct}(\Sigma,d\mu_{\mathbf g})})  DA^*_{ex} DB \\
   =  \int_{\cB} \biggl[   \biggl[ \int_{\cA^*_{ex}} \exp( i \tfrac{k}{2\pi}  \ll \star dA^*_{ex},B+ \tfrac{2\pi}{k} \sum_j  T_1  \ind(l^j_{\Sigma};\cdot)
   \gg_{L^2_{\ct}(\Sigma,d\mu_{\mathbf g})})  DA^*_{ex} \biggr] \\
   \times   \prod_{m \in \cM(t_0)} \exp(\eps_m B(\sigma_m)) \biggr] DB \\
= \int_{\ct} db \biggl[ \int  \delta(B-(b-  \tfrac{2\pi}{k} \sum_j  T_1  \ind(l^j_{\Sigma};\cdot)))     \prod_{m \in \cM(t_0)} \exp(\eps_m B(\sigma_m)) DB \biggr]\\
=   \biggl( \prod_{m \in \cM(t_0)} \exp( - \eps_m   2\pi\lambda \sum_j  T_1  \ind(l^j_{\Sigma};\sigma_m)) \biggr) \biggl(  \int_{\ct} db  \prod_{m \in \cM(t_0)} \exp(\eps_m b)\biggr)
\end{multline}
 Informally, we have
\begin{equation} \label{eq_windnumb_arg}
\int_{\ct}  \prod_m \exp(\eps_m b)\  db = \int_{\ct}  \exp(\sum_m \eps_m b) \  db
=  \delta(\sum_m \eps_m) = \delta(\sum_j \wind(l^j_{S^1}))
\end{equation}
because $\sum_m \eps_m = \sum_j \wind(l^j_{S^1})$
where $\wind(l^j_{S^1})$ is the winding number of $l^j_{S^1}$.\par
For evaluating the other factor in Eq. \eqref{eq6.8}
we now  use the 2-dimensional analogue of the framing procedure of
Sec. \ref{sec5}.
We replace
 the expression $\ind(l^j_{\Sigma};\sigma_m)$ by $ \tfrac{1}{2} \bigl[ \ind(l^j_{\Sigma};\bar{\phi}_s(\sigma_m)) + \ind(l^j_{\Sigma};\bar{\phi}^{-1}_s(\sigma_m)) \bigr]$ where  $\bar{\phi}_s:\Sigma \to \Sigma$ is as in
the paragraph preceding Remark \ref{rm5.2} above. Taking into
account that $T_1 =i$ we then obtain
\begin{multline} \label{eq_link_form_1}
 \WLO(L,\phi_s)  \sim\\
  \biggl(\prod_j   \exp( \lambda \pi i \LK(l_j,\phi_s \circ l_j)) \biggr) \biggl(  \prod_{j\neq k} \exp( \lambda \pi i \LK(l_j,l_k))\biggr)  \delta(\sum_j \wind(l^j_{S^1})) \\
\times \prod_j  \prod_{m \in \cM(t_0)} \exp(  -\pi i \lambda
 \eps_m \bigl[ \ind(l^j_{\Sigma};\bar{\phi}_s(\sigma_m)) + \ind(l^j_{\Sigma};\bar{\phi}^{-1}_s(\sigma_m)) \bigr])
\end{multline}

We will now make use of the following proposition, which is not difficult to prove.
\begin{proposition} \label{lem2}
If $l$ and $\tilde{l}$ are loops in $\Sigma \times S^1$ which are $0$-homologous
and which have the additional property that  $(l,\tilde{l})$ is admissible in the sense
of Subsec. \ref{subsec3.1}
 then the linking number $\Link(l,\tilde{l})$ of the pair
$(l,\tilde{l})$ is well-defined and we have
\begin{equation} \label{eq_lemma}
\Link(l,\tilde{l}) = \LK(l,\tilde{l}) - \sum_{u \in I} \eps_u \ind(\tilde{l}_{\Sigma}; \sigma_u) -
\sum_{u \in \tilde{I}} \tilde{\eps}_u \ind(l_{\Sigma}; \tilde{\sigma}_u)
\end{equation}
where we have set
$ \sigma_{u}  := l_{\Sigma}(u)$, $\eps_{u} := \sgn(l_{S^1};u)$ for $u \in I:=  l_{S^1}^{-1}(\{t_0\})$
and  $ \tilde{\sigma}_{u}  := \tilde{l}_{\Sigma}(u)$, $\tilde{\eps}_{u} := \sgn(\tilde{l}_{S^1};u)$
for $ u \in \tilde{I}:= \tilde{l}_{S^1}^{-1}(\{t_0\})$
\end{proposition}
From this proposition it follows that
for sufficiently small $s>0$ we have
 \begin{equation}  \label{eq_link_form_4}
  \Link(l_j,\phi_s \circ l_j) =  \LK(l_j,\phi_s \circ l_j) -
 \sum_{m \in  \cM_j(t_0)}
  \eps_m \bigl[ \ind(l^j_{\Sigma};\bar{\phi}_s(\sigma_m)) +  \ind(l^j_{\Sigma};\bar{\phi}^{-1}_s(\sigma_m))  \bigr]
 \end{equation}
and that
 \begin{equation}  \label{eq_link_form_3}
 \sum_{j\neq k}  \Link(l_j,l_k) =  \sum_{j\neq k} \LK(l_j,l_k) - \sum_j \sum_{m \in \cM(t_0) \backslash  \cM_j(t_0)} 2 \eps_m \ind(l^j_{\Sigma};\sigma_m)
\end{equation}
(here we have used that
for every $m \in \cM(t_0) \backslash  \cM_j(t_0)$ and every sufficiently small $s>0$
one has $\ind(l^j_{\Sigma};\bar{\phi}_s(\sigma_m)) + \ind(l^j_{\Sigma};\bar{\phi}^{-1}_s(\sigma_m))
= 2 \ind(l^j_{\Sigma};\sigma_m)$).
 So, if every $l_j$ is $0$-homologous (in which case $\sum_{j=1}^n  \wind(l^j_{S^1}) = \sum_{j=1}^n  0 =0$ holds)
  we finally obtain from Eqs. \eqref{eq_link_form_1}, \eqref{eq_link_form_4}, and \eqref{eq_link_form_3}
\begin{equation}\label{eq_main'}
 \WLO(L,\phi_s)  \sim \biggl( \prod_j   \exp( \lambda \pi i \Link(l_j,\phi_s \circ l_j)) \biggr)
 \biggl( \prod_{j\neq k} \exp( \lambda \pi i \Link(l_j,l_k)) \biggr)
\end{equation}
for sufficiently small $s>0$.
This is exactly the expression that was obtained by other methods, see, e.g., \cite{AS,LS}.

\begin{remark} \rm \label{rm6.2}
 Eq. \eqref{eq_lemma} only holds when we use the original definition
of $\ind(l^j_{\Sigma};\cdot)$ given in Eq. \eqref{eq_def_ind_or}.
If we had defined $\ind(l^j_{\Sigma};\cdot)$  by  Eq. \eqref{eq_def_ind_new} instead
we would have obtained a correction factor of the form
$\exp(C \cdot \sum_m \eps_m)$ in Eq. \eqref{eq_main'}
where $C$ is a suitable constant. Of course,
if every loop is $0$-homologous we have  $\wind(l^j_{S^1})=0$, $j \le n$, and thus
also $\sum_m \eps_m = \sum_{j=1}^n  \wind(l^j_{S^1}) = 0$.
So the correction factor is  trivial and we obtain again Eq. \eqref{eq_main'}

\end{remark}

 \subsection{Special case 2: $G=SU(2)$ and $L$ consists of vertical loops with arbitrary colors}
 \label{subsec6.2}

 Let us now consider the case where $G=SU(N)$  and where $\Sigma$ is an arbitrary
 compact oriented surface. (Later we will restrict ourselves to the special case $N=2$).
 We will  assume in the present subsection
 that in  the colored link $L= ((l_1,l_2,\ldots,l_n), (\rho_1,\rho_2,\ldots,\rho_n))$,
   which we have fixed in Subsec. \ref{subsec3.1}
   each $l_i$, $i \le n$, is a vertical loop (cf. Subsec. \ref{subsec2.1})
 above the point $\sigma_i$, $i\le n$. The colors $\rho_i$ can be arbitrary.
 In this  situation we have
 \begin{multline*}
 \prod_i \Tr_{\rho_i}( \cP \exp(\int_{l_i} \hat{A}^{\orth} +
 A^{\orth}_c + A^{\orth}_{sing}(\cl) + Bdt) )
  = \prod_i \Tr_{\rho_i}(  \exp(\int_{l_i}  Bdt)
   = \prod_i \Tr_{\rho_i}(\exp(B(\sigma_i)))
  \end{multline*}
 so we can conclude, informally,
 that  the integral  $\int_{\hat{\cA}^{\orth}} \prod_i \Tr_{\rho_i}(  \cP \exp(\int_{l_i} \hat{A}^{\orth} + A^{\orth}_c + A^{\orth}_{sing}(\cl) + Bdt) ) d\hat{\mu}^{\orth}_B(\hat{A}^{\orth})$
appearing in Eq.  \eqref{eq_WLO_end} coincides with $\prod_i \Tr_{\rho_i}(\exp(B(\sigma_i)))$.
 For $G=SU(N)$, for which $c_G = N$,
 we thus obtain\footnote{note that in contrast to the situation
 in Subsec. \ref{subsec6.1} and  Subsec. \ref{subsec6.2}
   no framing is necessary in the present subsection.
   For this reason we will denote the WLOs just by $\WLO(L)$ instead of $\WLO(L;\phi_s)$} from Eq. \eqref{eq_WLO_end}
\begin{multline} \label{eq_reform_BlTh}
 \WLO(L) \sim
  \sum_{\cl} \int_{\cB}   \bigl(  \prod_j \Tr_{\rho_j}\bigl[
\exp( B(\sigma_j)) \bigr]  \bigr)
 \biggl(  \int_{\cA^{\orth}_c}  \exp( i \tfrac{k+N}{2\pi}  \ll \star dA^{\orth}_c , B \gg_{L^2_{\ct}(\Sigma,d\mu_{\mathbf g})})  DA^{\orth}_c   \biggr) \\
  \exp( i \tfrac{k+ N}{2\pi}  \ll  \star   dA^{\orth}_{sing}(\cl), B\gg)
  \det\nolimits_{reg}\bigl(1_{\cG_0}-\exp(\ad(B)_{| \cG_0})\bigr) DB
 \end{multline}
  Eq. \eqref{eq_reform_BlTh} can be considered as a reformulation of
    Eq. (7.24) in \cite{BlTh1}.
 The evaluation of Eq. \eqref{eq_reform_BlTh} which we will give now
 differs only slightly from the analogous treatment   given in Secs. 7.1--7.6 in \cite{BlTh1}.\par
   Let us use again  the  Hodge decomposition
\eqref{eq_Hodge} of $\cA^{\orth}_c \cong \cA_{\Sigma,\ct}$. In the
present
 subsection we do not assume  that $\Sigma \cong S^2$ holds
 so  the space $\cA_{harm} \cong H^1_{\bR}(\Sigma) \otimes \ct$ need not vanish.
 After replacing the
$\int \cdots DA_c^{\orth}$-integration in Eq. \eqref{eq_reform_BlTh}  by
$ \int \int \int \cdots DA_{ex} DA_{harm} DA^*_{ex}$,
where $DA_{ex}$, $DA_{harm}$, $DA^*_{ex}$ denote the ``Lebesgue measures''
on the obvious spaces, we obtain
 \begin{multline*}
    \int_{\cA^{\orth}_c}  \exp( i \tfrac{k+N}{2\pi}
     \ll \star dA^{\orth}_c , B \gg_{L^2_{\ct}(\Sigma,d\mu_{\mathbf g})})  DA^{\orth}_c
  \sim  \int_{\cA^*_{ex}}  \exp( i \tfrac{k+N}{2\pi}  \ll \star d A^*_{ex} , B \gg_{L^2_{\ct}(\Sigma,d\mu_{\mathbf g})})  DA^*_{ex}
\end{multline*}
because the $\int \cdots DA_{ex}$- and $\int \cdots DA_{harm}$-integrations  are trivial.
 Taking into account  that
  $ \ll \star d A^*_{ex} , B \gg_{L^2_{\ct}(\Sigma,d\mu_{\mathbf g})}$
  vanishes for every $A^*_{ex}$ if and only if $B \in C^{\infty}(\Sigma,P)$ is constant
 we obtain
\begin{align*} \WLO(L) &   \sim \sum_{\cl} \int_{\ct}  db \int_{\cB}   DB   \biggr[ \delta(B - b)\bigl(  \prod_j \Tr_{\rho_j}\bigl[ \exp( B(\sigma_j)) \bigr]  \bigr)\\
 & \quad   \times\exp\bigl( i \tfrac{k+N}{2\pi}  \ll  \star   dA^{\orth}_{sing}(\cl), B\gg\bigr)  \det\nolimits_{reg}\bigl(1_{\cG_0}-\exp(\ad(B)_{| \cG_0})\bigr)  \biggr]\\
 &  =  \sum_{\cl} \int_{P} db   \bigl( \prod_j \Tr_{\rho_j}\bigl[
\exp(b) \bigr] \bigr)
    \exp( i \tfrac{k+N}{2\pi}  \ll  \star   dA^{\orth}_{sing}(\cl), b \gg) \\
     & \quad \quad \times \det\bigl(1_{\cG_0}-\exp(\ad(b)_{|\cG_0})\bigr)^{\chi(\Sigma)/2}
 \end{align*}
where $db$ is the  Lebesgue measure on $\ct$.
Here we have used that
$$\det\nolimits_{reg}\bigl(1_{\cG_0}-\exp(\ad(B)_{| \cG_0})\bigr) = \det\bigl(1_{\cG_0}-\exp(\ad(b)_{|\cG_0})\bigr)^{\chi(\Sigma)/2}$$
if $B$ equals the constant function $b$, cf. Eq. \eqref{eq_det_neu}.\par
 From Eq. \eqref{eq_def_llrr} and the definition of   $n(\cl)$ and
  $A^{\orth}_{sing}(\cl)$
  it follows immediately that
 \begin{equation} \label{eq_Asing_Ausw}
  \tfrac{k+N}{2\pi}  \ll  \star   dA^{\orth}_{sing}(\cl), b\gg =    \tfrac{k+N}{2\pi} n(\cl) \cdot b
  \end{equation}
  where ``$\cdot$'' denotes the scalar product on $\ct$ induced by $(\cdot,\cdot)_{\cG}$.
  From $\exp(\ad(b)_{|\cG_0}) =  \Ad(\exp(b))_{|\cG_0}$
   we obtain
  \begin{equation} \label{eq_finiteintegral} \WLO(L)  \sim \sum_{\cl} \int_{P} db  \  f(\exp(b))     \exp( i \tfrac{k+N}{2\pi} \ n(\cl) \cdot b)
   \end{equation}
   where we have set $ f(t) :=  \bigl( \prod_j \Tr_{\rho_j}(t) \bigr)  \det\bigl(1_{\cG_0}-\Ad(t)_{|\cG_0}\bigr)^{\chi(\Sigma)/2} $, $t \in T$.\par
  For simplicity, let us restrict ourselves
   to the the special case $N=2$, i.e. $G=SU(2)$.
   We can then choose $T$ to be the maximal torus
   $\{\bigl( \begin{matrix}   e^{i \theta} & 0 \\ 0 &   e^{-i \theta}
    \end{matrix} \bigr) \mid \theta \in [0,2\pi]\}$
       and $P \subset \ct=  \bR \tau =  \{  \theta \cdot \tau  \mid \theta \in \bR\} $  to be the (open) alcove
          \begin{equation} \label{eq6.22}
    P:= \{  \theta \cdot \tau  \mid \theta \in (0,\pi) \} \quad \text{ where } \tau  :=  \bigl( \begin{matrix}   i  & 0 \\ 0 &  -i
    \end{matrix} \bigr)
    \end{equation}
   Taking into account  Eq. \eqref{eq_Image_of_n} and
    \begin{align}
& \label{eq_6.22neu} \{ n(\cl) \mid \cl \in [\Sigma,G/T]\} = \Ker(\exp_{| \ct}) = 2 \pi  \bZ \cdot \tau\\
&\label{eq_6.23neu}  \tau \cdot \tau  = - \Tr(\tau \tau) =2\\
 \label{eq_sinx2}
& \det\bigl(1_{\cG_0}-\Ad(\exp(x\cdot \tau)))_{|\cG_0}\bigr) = \sin(x)^2\\
& \Tr_{\rho_j}(\exp(x\cdot \tau))= \frac{\sin(d_j x)}{\sin(x)}
\end{align}
where $d_j$ is the dimension of the representation $\rho_j$
    we obtain, informally,
   \begin{align} \label{eq_comp_BlTh}
   \WLO(L) & \sim  \sum_{m = - \infty}^{\infty} \int_{(0,\pi)}   e^{i m (k+2) 2  x} f(e^{x \tau})  \ dx  \nonumber\\
   & =   \int  1_{(0,\pi)}(x)   \bigl(  \sum_{m = - \infty}^{\infty} e^{i m (k+2) 2  x} \bigr)
    f(e^{x \tau})  \ dx  \nonumber \\
    & \overset{(*)}{=} \int  1_{(0,\pi)}(x) \delta_{\frac{\pi}{k+2}\bZ}(x) f(e^{x \tau}) \ dx  =  \sum_{l = 1}^{k+1} f(e^{\frac{ \pi}{k+2} l \tau})  \nonumber \\
   & =  \sum_{l = 1}^{k+1 }    \prod_j \frac{\sin(\frac{l d_j  \pi}{k+2} )}{\sin(\frac{ l \pi}{k+2})}
  \sin^{2-2g}(\tfrac{ l \pi}{k+2})
   \end{align}
   where $\delta_{\frac{\pi}{k+2}\bZ}$ is the periodic delta-function associated to the lattice
    $\frac{\pi}{k+2}\bZ$ in $\bR$
      and where  $g$ denotes the genus of $\Sigma$.
     In step $(*)$ we have pretended that we can apply the Poisson summation formula
          $\int \phi(x) \bigl(  \sum_{m = - \infty}^{\infty} e^{i m (k+2) 2  x} \bigr)  dx
      = \int \phi(x)  \delta_{\frac{\pi}{k+2}\bZ}(x)  dx$,
       which holds, e.g., if $\phi$ is a smooth function of rapid decrease.
      Clearly, the  function $1_{(0,\pi)}(x) f(e^{x \tau})$ is not smooth,
      it even has a singularity at the points $x=0$ and $x=\pi$ if $\chi(\Sigma)<0$.
      In a rigorous treatment of the above derivation
      where, among other things, ``loop smearing'' is used in a suitable way this
      complication can probably be avoided.

 \subsection{Special case 3: $G=SU(2)$ and $L$ has standard colors and no double points}
\label{subsec6.3}

Let us consider again the case where $G=SU(N)$. (Later we will restrict ourselves to the special case $N=2$).
We will now assume that the colored link $L=((l_1,l_2,\ldots,l_n),(\rho_1,\rho_2,\ldots,\rho_n))$
 which we have fixed in Subsec.
  \ref{subsec3.1} is admissible and
   has no double points and that each $\rho_j$ is equal to the fundamental representation
   $\rho_{SU(N)}$ of $SU(N)$.\par
As $G=SU(N)$ is simply-connected $\Sigma$ can be an arbitrary
(oriented compact) surface. Note, however, that the case   $\Sigma
\not\cong S^2$ is slightly more complicated than the case  $\Sigma
\cong S^2$. Firstly, in the  Hodge decomposition \eqref{eq_Hodge} of
$\cA^{\orth}_c \cong \cA_{\Sigma,\ct}$ the space $\cA_{harm}$ is
not trivial if $\Sigma \not\cong S^2$. Secondly, in the case $\Sigma
\not\cong S^2$ the definition of
 the functions $\ind(l^j_{\Sigma};\cdot)$
for  general loops $l^j_{\Sigma}$ in $\Sigma$ is also more complicated
 than in the case $\Sigma \cong S^2$.\par
In order to circumvent these complications in the present paper
   we will make the additional assumption  that for the link $L$ considered each
 $l^j_{\Sigma}$ is 0-homotopic. From this and  $\DP(L) = \emptyset$
  it then follows that
  $\Sigma \backslash \arc(l^j_{\Sigma})$ will have exactly two connected components
and  we can then define the functions $\ind(l^j_{\Sigma};\cdot)$ for arbitrary $\Sigma$
in a similar way as in Subsec. \ref{subsec6.1} above for the case  $\Sigma = S^2$.
As in the case $\Sigma = S^2$ there is a certain freedom in defining
$\ind(l^j_{\Sigma};\cdot)$. It turns out that it has several advantages to define
$\ind(l^j_{\Sigma};\cdot)$ in analogy to Remark \ref{rm6.1}, i.e.
to fix the additive constant mentioned in Remark \ref{rm6.1}
by demanding  that
\begin{equation} \label{eq_ind_conv} \ind(l^j_{\Sigma};\sigma_0)= 0
\end{equation}
holds.  As in Subsec. \ref{subsec5.2} let $(\phi_s)_{s>0}$ be a horizontal framing of $L$.
In view of Eq. \eqref{eq_WLO_end} and Eq. \eqref{eq_WLO_AB}
let us now set
\begin{multline*}
\WLO(L;\phi_s ) := \\
\sum_{\cl \in [\Sigma,G/T]}  \int_{\cA_c^{\orth} \times \cB} \WLO(L,\phi_s;A^{\orth}_c, A^{\orth}_{sing}(\cl), B)
 \exp( i \tfrac{k+ c_G}{2\pi}  \ll  \star   dA^{\orth}_{sing}(\cl), B\gg)  \\ \times \det\nolimits_{reg}\bigl(1_{\cG_0}-\exp(\ad(B)_{| \cG_0})\bigr)
 \exp( i \tfrac{k+ c_G}{2\pi}  \ll \star dA^{\orth}_c, B \gg_{L^2_{\ct}(\Sigma,d\mu_{\mathbf g})} (DA_c^{\orth} \otimes    DB)
\end{multline*}
From Eq. \eqref{eq_WLOABexpl} we obtain (for small $s>0$)
\begin{multline} \label{eq_P_Version} \WLO(L;\phi_s )  \sim\\
  \sum_{\cl \in [\Sigma,G/T]} \int_{\cB} \biggl[ \int_{\cA^{\orth}_c}
 \prod_j \Tr\bigl[\exp(\ \sum_{m \in  \cM_j(t_0)} \eps_m  B(\sigma_m))    \exp(\int_{l^j_{\Sigma}}   A^{\orth}_c)   \exp(\int_{l^j_{\Sigma}}   A^{\orth}_{sing}(\cl))   \bigr]  \\
\times  \exp( i \tfrac{k+N}{2\pi}   \ll \star dA^{\orth}_c , B \gg_{L^2_{\ct}(\Sigma,d\mu_{\mathbf g})}) ) DA^{\orth}_c \biggr]\\
\times  \exp( i \tfrac{k+N}{2\pi}  \ll \star dA^{\orth}_{sing}(\cl) , B \gg) ) \det\nolimits_{reg}\bigl(1_{\cG_0}-\exp(\ad(B)_{| \cG_0})\bigr) DB
 \end{multline}
  Let us again use  the Hodge decomposition
 $\cA^{\orth}_c \cong \cA_{\Sigma,\ct} =  \cA_{ex} \oplus \cA_{harm} \oplus \cA^*_{ex} $
 and replace the
$\int \cdots DA_c^{\orth}$-integration  by
$ \int \int \int \cdots DA_{ex} DA_{harm} DA^*_{ex}$.
Clearly,
  $\int_{l^j_{\Sigma}}   A_{ex} = 0$ and from the assumption that
 each $l_{\Sigma}^j$ is 0-homotopic  it follows that also $\int_{l^j_{\Sigma}}   A_{harm} = 0$ for all  $A_{harm} \in \cA_{harm}$.
  Thus the $\int \cdots DA_{ex}$- and $\int \cdots DA_{harm}$-integrations are
 trivial and we  obtain
\begin{multline} \WLO(L;\phi_s )  \sim\\
 \sum_{\cl \in [\Sigma,G/T]} \int_{\cB} \biggl\{ \int_{\cA^*_{ex}}   \prod_j \Tr\bigl[
\exp( \sum_{m \in  \cM_j(t_0)} \eps_{m}  B(\sigma_{m}))  \exp(\int_{l^j_{\Sigma}}   A^*_{ex})   \exp(\int_{l^j_{\Sigma}}   A^{\orth}_{sing}(\cl))   \bigr]  \\
 \times  \exp( i \tfrac{k+N}{2\pi}  \ll \star d A^*_{ex} , B \gg_{L^2_{\ct}(\Sigma,d\mu_{\mathbf g})})
DA^*_{ex} \biggr\} \\
\times  \exp( i \tfrac{k+N}{2\pi}   \ll \star dA^{\orth}_{sing}(\cl) , B \gg)\det\nolimits_{reg}\bigl(1_{\cG_0}-\exp(\ad(B)_{| \cG_0})\bigr) DB
 \end{multline}
 From a straightforward generalization of Eq. \eqref{eq_defind}  we obtain
 (recall that $T_a \in \ct$ for $a \le r$)
\begin{equation}   \int_{l^j_{\Sigma}}   A^*_{ex} =  \sum_{a =1}^r T_a \ll \star d  A^*_{ex}, T_a  \ind(l^j_{\Sigma};\cdot)\gg_{L^2_{\ct}(\Sigma,d\mu_{\mathbf g})}
\end{equation}
 Taking into account  that\footnote{note that, as $\rho_{SU(N)}$ is the fundamental representation of
 $SU(N)$,   all the weights that appear in the
 character associated to  $\rho_{SU(N)}$ have multiplicity one} for
 $b \in \ct$ we have
\begin{equation}
\Tr(\exp(b)) = \Tr_{\rho_{SU(N)}}(\exp(b)) = \sum_{\alpha \in W_{\rho_{SU(N)}}} \exp(\alpha(b))
\end{equation}
where $W_{\rho_{SU(N)}}$ is the set of infinitesimal weights $\alpha:\ct \to i\bR$
of $\rho_{SU(N)}$
and setting
\begin{equation} \label{eq_def_A}  A:=  W_{\rho_{SU(N)}} \times \cdots \times   W_{\rho_{SU(N)}}
\end{equation}
(for an element $(\alpha_1, \ldots,  \alpha_n) \in A$
we will  often use the shorthand
$\underline{\alpha}$)
we thus have
\begin{multline}
 \prod_j \Tr\bigl[
\exp\bigl( \sum_{m \in  \cM_j(t_0)} \eps_m  B(\sigma_m) \bigr)  \exp\bigl(\int_{l^j_{\Sigma} } A^*_{ex} \bigr) \exp\bigl(\int_{l^j_{\Sigma}}   A^{\orth}_{sing}(\cl)\bigr)   \bigr]  \\
= \sum_{\underline{\alpha} \in A}  \prod_{j=1}^n
 \exp\bigl( \alpha_j(  \sum_{m \in  \cM_j(t_0)} \eps_m  B(\sigma_m) +  \int_{l^j_{\Sigma}}   A^*_{ex} +   \int_{l^j_{\Sigma}} A^{\orth}_{sing}(\cl)) \bigr)\\
 = \sum_{\underline{\alpha} \in A}
 \exp\bigl(\sum_j \sum_{m \in  \cM_j(t_0)} \eps_m \alpha_j(   B(\sigma_m))  + \int_{l^j_{\Sigma}} \alpha_j(  A^{\orth}_{sing}(\cl)) \bigr)\\
 \times  \exp\bigl(- i \ll \star d  A^*_{ex},  \Ind(L,\underline{\alpha}) \gg_{L^2_{\ct}(\Sigma,d\mu_{\mathbf
 g})}\bigr)
\end{multline}
where we have set
\begin{equation}
\Ind(L,\underline{\alpha}):= - \sum_{j'} \sum_{a \le r} \frac{1}{i} \alpha_{j'}(T_a) \ T_a  \ind(l^{j'}_{\Sigma};\cdot)
\end{equation}

Note that the function $\frac{1}{i} \alpha_j$ takes values in $\bR$.
 So $\Ind(L,\underline{\alpha})$ is a well-defined element of
 $L^2_{\ct}(\Sigma,d\mu_{\mathbf g})$.\par
We now regularize the expressions  $B(\sigma_{m})$ using  ``framing''
 as in Subsec. \ref{subsec6.1}.
  This amounts to
 replacing  $B(\sigma_{m})$ by
 $\tfrac{1}{2} \bigl[ B(\bar{\phi}_s(\sigma_m)) + B(\bar{\phi}_s^{-1}(\sigma_m))\bigr]$.
Then we obtain (for sufficiently small $s>0$)
\begin{multline} \WLO(L;\phi_s )\\
 \sim \sum_{\cl \in [\Sigma,G/T]} \int_{\cB} \sum_{\underline{\alpha} \in A}
 \biggl\{ \int_{\cA^*_{ex}}      \exp( i   \ll \star dA^*_{ex},\tfrac{k+N}{2\pi} B - \Ind(L,\underline{\alpha}) )\gg_{L^2_{\ct}(\Sigma,d\mu_{\mathbf g})})
DA^*_{ex} \biggr\} \\
\times \exp\biggl(  \sum_{j}  \sum_{m \in  \cM_j(t_0)} \eps_m   \tfrac{1}{2} \alpha_j\bigl( B(\bar{\phi}_s(\sigma_m)) + B(\bar{\phi}_s^{-1}(\sigma_m)) \biggr)\det\nolimits_{reg}\bigl(1_{\cG_0}-\exp(\ad(B)_{| \cG_0})\bigr) \\
\times \exp\bigl(\sum_j  \int_{l^j_{\Sigma}} \alpha_j( A^{\orth}_{sing}(\cl)) \bigr)
 \exp( i \tfrac{k+N}{2\pi}   \ll \star dA^{\orth}_{sing}(\cl) , B \gg) DB
 \end{multline}

Similarly as in Subsec. \ref{subsec6.2} we can argue, informally\footnote{we expect that
it is possible to avoid this heuristic argument
and to give a fully rigorous treatment instead, cf. point (4) in Subsec. \ref{subsec7.2}}, that
$\int_{\cA^*_{ex}}
   \exp( i   \ll\star dA^*_{ex},\tfrac{k+N}{2\pi} B - \Ind(L,\underline{\alpha}))  \gg_{L^2_{\ct}(\Sigma,d\mu_{\mathbf g})})
DA^*_{ex}$ vanishes unless
$ B - \tfrac{2\pi} {k+N} \Ind(L,\underline{\alpha})$
is a constant function taking\footnote{that $B - \tfrac{2\pi} {k+N} \Ind(L,\underline{\alpha})$
must take values in $P$ follows from $\Ind(L,\underline{\alpha})(\sigma_0)=0$, cf.  Eq. \eqref{eq_ind_conv}}  values in $P$.
In other words: the aforementioned integral vanishes unless
there is a $b \in P$ such that
$B = b + \tfrac{2\pi}{k+N} \Ind(L,\underline{\alpha})$
holds.
Accordingly, let us replace the  $\int \cdots DB$-integration
 by the integration
$$\int_{P} db \biggl[ \int_{\cB}  \cdots \delta(B - b - \tfrac{2\pi}{k+N} \Ind(L,\underline{\alpha})) DB \biggr]$$

Let us  set
$\eps_j :=  \sum_{m \in  \cM_j(t_0)}    \eps_{m} = \wind(l_{S^1}^j)$
and choose for each $j$ a fixed element of
$\{ \sigma_{m} \mid m \in  \cM_j(t_0)\}$ which we will denote $\sigma_j$
(if $ \cM_j(t_0)$ is empty we choose   an arbitrary point
of $\arc(l^j_{\Sigma})$ for $\sigma_j$).
Moreover, let $P^{\Sigma}$ denote the set of all mappings $\Sigma \to P$
and $1_{P^{\Sigma}}$ the corresponding indicator function, i.e. for a function $B:\Sigma \to \ct$
we have
$$ 1_{P^{\Sigma}}(B)  =
\begin{cases} 1 & \text{ if } \Image(B) \subset P\\
0 & \text{ otherwise }
\end{cases}
$$
We obtain (for small $s>0$)
   \begin{align} \label{eq_general_for_WLO_0}
  & \WLO(L;\phi_s) \nonumber \\
 & \sim   \sum_{\underline{\alpha} \in A}  \sum_{\cl \in [\Sigma,G/T]}
 \int_{P} db
\biggl(  1_{P^{\Sigma}}(B)      \exp\bigl(  \sum_{j}    \eps_{j} \frac{1}{2} \bigl[ \alpha_j\bigl( B(\bar{\phi}_s(\sigma_{j})))
  +   \alpha_j\bigl(B(\bar{\phi}^{-1}_s(\sigma_{j})))   \bigr]\bigr)  \nonumber \\
& \quad \times  \biggl[ \exp\bigl(\sum_j  \int_{l^j_{\Sigma}}\alpha_j( A^{\orth}_{sing}(\cl)) \bigr)
 \exp( i \tfrac{k+N}{2\pi}   \ll  \star   dA^{\orth}_{sing}(\cl),\tfrac{2\pi}{k+N} \Ind(L,\underline{\alpha})\gg)
 \biggr]   \nonumber \\
 & \quad  \times \det\nolimits_{reg}\bigl(1_{\cG_0}-\exp(\ad(B)_{| \cG_0})\bigr)
   \exp( i \tfrac{k+N}{2\pi}   \ll  \star   dA^{\orth}_{sing}(\cl),b\gg)
   \biggr)_{| B=b + \tfrac{2\pi}{k+N} \Ind(L,\underline{\alpha})} \nonumber \\
 & \overset{(*)}{=}  \sum_{\underline{\alpha} \in A}  \sum_{\cl \in [\Sigma,G/T]}
 \int_{P} db \biggl(
 1_{P^{\Sigma}}(B)      \exp\bigl(  \sum_{j}    \eps_{j} \frac{1}{2} \bigl[ \alpha_j\bigl( B(\bar{\phi}_s(\sigma_{j})))
  +   \alpha_j\bigl(B(\bar{\phi}^{-1}_s(\sigma_{j})))   \bigr]\bigr)  \nonumber \\
& \quad  \times \biggl[ 1 \biggr]      \times \det\nolimits_{reg}\bigl(1_{\cG_0}-\exp(\ad(B)_{| \cG_0})\bigr)
   \exp( i \tfrac{k+N}{2\pi} n(\cl) \cdot b)
 \biggr)_{| B=b + \tfrac{2\pi}{k+N} \Ind(L,\underline{\alpha})} \nonumber \\
& \overset{(**)}{=}  \sum_{\underline{\alpha} \in A} \sum_{\cl}  \int_P db  \bigl( \exp( i \tfrac{k+N}{2\pi} n(\cl) \cdot b) \bigr)
   \biggl(  1_{P^{\Sigma}}(B)  \exp\bigl(  \sum_{j}    \eps_{j} \frac{1}{2} \bigl[ \alpha_j\bigl( B(\bar{\phi}_s(\sigma_{j})) +
 B(\bar{\phi}^{-1}_s(\sigma_{j})))    \bigr] \bigr) \nonumber \\
& \quad  \times  \prod_{t=1}^{\mu} \det\bigl(1_{\cG_0}-\Ad(\exp(B(\sigma_{X_t})))_{|\cG_0}\bigr) ^{\chi(X_t)/2}
\biggr)_{| B=b + \tfrac{2\pi}{k+N} \Ind(L,\underline{\alpha})}
\end{align}
Here step $(*)$ follows from Eq. \eqref{eq_Asing_Ausw}
and the relation
\begin{equation} \label{eq_Asing_cancel}
\sum_{a=1}^{r} T_a  \ll  \star   dA^{\orth}_{sing}(\cl), T_a \ind(l_{\Sigma}^j;\cdot)  \gg  \\
=  \int_{l_{\Sigma}^j} A^{\orth}_{sing}(\cl)
\end{equation}
which is not difficult to show\footnote{if we had not defined
$\ind(l_{\Sigma}^j;\cdot) $ such that
Eq. \eqref{eq_ind_conv} holds then we would have  to replace
 Eq. \eqref{eq_Asing_cancel} by the equation
$\sum_{a=1}^r T_a  \ll  \star   dA^{\orth}_{sing}(\cl), T_a \ind(l_{\Sigma}^j;\cdot)  \gg
= n(\cl) \cdot \ind(l_{\Sigma}^j;\sigma_0)  + \int_{l_{\Sigma}^j} A^{\orth}_{sing}(\cl)$
}. Step $(**)$ follows
  from
 \begin{equation} \label{eq_simplification1}
\det\nolimits_{reg}\bigl(1_{\cG_0}-\exp(\ad(B)_{| \cG_0})\bigr)
= \prod_{t=1}^{\mu} \det\bigl(1_{\cG_0}-\Ad(\exp(B(\sigma_{X_t})))_{|\cG_0}\bigr) ^{\chi(X_t)/2}
\end{equation}
(here we have fixed $\sigma_{X_t}  \in X_t$ for each $t \le \mu$).
Eq. \eqref{eq_simplification1} follows
from Eq.  \eqref{eq_det_neu}
if we take  into account that each function $B= b+   \tfrac{2\pi}{k+N}     \Ind(L,\underline{\alpha})$
is a ``step function'' in the sense of
Subsec. \ref{subsec3.5}.\par

For simplicity, let us now restrict ourselves to the special case
where $N=2$, i.e. $G=SU(2)$.
Then,
after informally interchanging  the  $\sum_{\cl \in [\Sigma,G/T]}$-summation with the $\int_P$-integration
and taking into account Eq.  \eqref{eq6.22}
and the relation
$1_{P^{\Sigma}}(B) = 1_P(B(\sigma_0))  \cdot 1_{P^{\Sigma}}(B)  = 1_P(x \tau)  \cdot 1_{P^{\Sigma}}(B) $
 for $B=x \tau + \tfrac{2\pi}{k+2} \Ind(L,\underline{\alpha})$,
 which holds because $\Ind(L,\underline{\alpha})(\sigma_0)=0$, we obtain from Eq. \eqref{eq_general_for_WLO_0}
 \begin{align}
  & \WLO(L;\phi_s )   \nonumber \\
 &  \sim
   \sum_{\underline{\alpha} \in A}  \int_{0}^{\pi} dx
    \bigl( \sum_{\cl}  \exp( i \tfrac{k+2}{2\pi} n(\cl) \cdot x \tau) \bigr)  1_P(x \tau)  \nonumber \\
 &  \quad \times  \biggl(  1_{P^{\Sigma}}(B)  \exp\bigl(  \sum_{j}    \eps_{j} \frac{1}{2} \bigl[ \alpha_j\bigl( B(\bar{\phi}_s(\sigma_{j})) +
 B(\bar{\phi}^{-1}_s(\sigma_{j})))    \bigr] \bigr) \nonumber \\
& \quad  \times  \prod_{t=1}^{\mu} \det\bigl(1_{\cG_0}-\Ad(\exp(B(\sigma_{X_t})))_{|\cG_0}\bigr) ^{\chi(X_t)/2}
\biggr)_{| B=x  \tau + \tfrac{2\pi}{k+2} \Ind(L,\underline{\alpha})} \nonumber \\
 & \overset{(+)}{=}   \sum_{\underline{\alpha} \in A}  \sum_{l=1}^{k+1}
   \biggl(  1_{P^{\Sigma}}(B)
   \exp\bigl(  \sum_{j}    \eps_{j} \frac{1}{2} \bigl[ \alpha_j\bigl(B(\bar{\phi}_s(\sigma_{j}))
    +  B(\bar{\phi}^{-1}_s(\sigma_{j}))\bigr)    \bigr] \bigr) \nonumber \\
& \quad \times    \prod_{t=1}^{\mu} \det\bigl(1_{\cG_0}-\Ad(\exp(B(\sigma_{X_t})))_{|\cG_0}\bigr)^{\chi(X_t)/2}
\biggr)_{| B=  \tfrac{ \pi}{k+2} (  l  \tau +  2  \Ind(L,\underline{\alpha}))}
\end{align}
In step $(+)$ we have used
that $\{ \exp( i \tfrac{k+2}{2\pi} (n(\cl) \cdot x\tau) \bigr) \mid \cl \in [\Sigma,G/T]\}
 = \{ \exp( i 2m(k+2) x \bigr) \mid m \in \bZ\}$
(cf. Eqs. \eqref{eq_6.22neu} and \eqref{eq_6.23neu})
and as in Subsec. \ref{subsec6.2} above we have again pretended that
we can apply the  Poisson summation formula.
As in Subsec. \ref{subsec6.2} we expect that in a rigorous treatment
where, among other things, ``loop smearing'' and ``point smearing'' are used in a suitable way
this argument can be made rigorous, cf. \cite{Ha6}.\par

 For every $l \in \{1,2,\ldots,k+1\}$ and $\underline{\alpha} \in A$
 let us now set
 \begin{equation}\label{def_xi} \xi_{l,\underline{\alpha}}:= l - \sum_{j'}  \frac{1}{i}  \alpha_{j'}(\tau) \ind(l^{j'}_{\Sigma};\cdot)
\end{equation}
Clearly, we can choose $T_1 = \tfrac{1}{\sqrt{2}} \tau$.
Taking into account that $\xi_{l,\underline{\alpha}}$ takes
values in $\bZ$ and that
 $   \xi_{l,\underline{\alpha}} \cdot \tau  = l \tau   +  2\Ind(L,\underline{\alpha})$
 we obtain (with the help of Eq. \eqref{eq_sinx2})
\begin{align} & \WLO(L;\phi_s )    \sim   \sum_{\underline{\alpha} \in A}  \sum_{l=1}^{k+1}
   \biggl( 1_{\Image(\xi_{l,\underline{\alpha}}) \subset \{ 1,2,\dots, k+1\} }
       \prod_{t=1}^{\mu}  \sin(\tfrac{\pi}{k+2}  \xi_{l,\underline{\alpha}}(\sigma_{X_t}))^{\chi(X_t)}      \nonumber \\
 & \times \exp\bigl( \tfrac{\pi}{k+2}  \tfrac{1}{2} \sum_{j}   \alpha_j(\tau) \eps_{j} (\xi_{l,\underline{\alpha}}(\bar{\phi}_s(\sigma_{j}))     +   \xi_{l,\underline{\alpha}}(\bar{\phi}^{-1}_s(\sigma_{j}))) \bigr)  \biggr)
\end{align}

Taking into account that
 for sufficiently small $s>0$ we have
$ \ind(l^{j'}_{\Sigma};\bar{\phi}_s(\sigma_{j}))- \ind(l^{j'}_{\Sigma};\bar{\phi}_s^{-1}(\sigma_{j}))
 = 0$ if $j\neq j'$,
 and $ \ind(l^{j'}_{\Sigma};\bar{\phi}_s(\sigma_{j}))- \ind(l^{j'}_{\Sigma};\bar{\phi}_s^{-1}(\sigma_{j}))\in \{-1,1\}$  (cf. condition (H2) in Subsec. \ref{subsec5.2})
and thus $(\ind(l^{j'}_{\Sigma};\bar{\phi}_s(\sigma_{j}))- \ind(l^{j'}_{\Sigma};\bar{\phi}_s^{-1}(\sigma_{j})))^2=1$  if $j=j'$
we obtain from Eq. \eqref{def_xi} (for arbitrary $l$)
\begin{equation} \label{eq_nach_alpha_aufgeloest}
\alpha_j(\tau)
= -i \bigl(\ind(l^{j}_{\Sigma};\bar{\phi}_s(\sigma_{j})) - \ind(l^{j}_{\Sigma};\bar{\phi}_s^{-1}(\sigma_{j}))\bigr)   \bigl( \xi_{l,\underline{\alpha}}(\bar{\phi}_s(\sigma_{j}))
  - \xi_{l,\underline{\alpha}}(\bar{\phi}_s^{-1}(\sigma_{j}))
\end{equation}
 Thus we obtain
\begin{align} \label{eq_last_ind_WLO}
  & \WLO(L;\phi_s )    \sim   \sum_{(l,\underline{\alpha}) \in \Pairs_{adm}}    \prod_{t=1}^{\mu}  \sin(\tfrac{\pi}{k+2}  \xi_{l,\underline{\alpha}}(\sigma_{X_t}))^{\chi(X_t)}  \nonumber\\
 & \times \exp\bigl( -\tfrac{\pi i}{k+2} \tfrac{1}{2}  \sum_{j}   \eps_{j}  \bigl(\ind(l^{j}_{\Sigma};\bar{\phi}_s(\sigma_{j})) - \ind(l^{j}_{\Sigma};\bar{\phi}_s^{-1}(\sigma_{j}))\bigr) (\xi_{l,\underline{\alpha}}(\bar{\phi}_s(\sigma_{j}))^2     -   \xi_{l,\underline{\alpha}}(\bar{\phi}^{-1}_s(\sigma_{j})))^2 \bigr)
\end{align}
where we have set
\begin{equation}
\Pairs_{adm}:= \{ (l, \underline{\alpha}) \in   \{1,\ldots,k+1\} \times  A \mid
 \Image(\xi_{l,\underline{\alpha}}) \subset \{ 1,2,\dots, k+1\} \}
\end{equation}

\bigskip

 We will now show that the right-hand side of Eq. \eqref{eq_last_ind_WLO}
  reduces to expression \eqref{Turaevs_state_sum} in  Appendix B (up to a multiplicative constant
  depending only on the charge $k$).
First we observe that  each $(l, \underline{\alpha}) \in \Pairs_{adm}$
determines an area coloring $\eta_{l,\underline{\alpha}}$ of $sh(L)$ with colors
in $I_{k+2}$  (cf. Appendix B) given by
\begin{equation} \label{eq_def_eta}
\eta_{l,\underline{\alpha}}(X_t) = \tfrac{1}{2}(\xi_{l,\underline{\alpha}}(\sigma_{X_t})-1)
\end{equation}
with $\sigma_{X_t}$ as above.
It is well-known  in the ``physical interpretation'' of the framework  in Appendix B (cf., e.g., \cite{PoRe})
that the color $1/2 \in I_{k+2}$ corresponds to the fundamental representation $\rho_{SU(2)}$
of $SU(2)$.
As we have only considered
links where all the loops $l_1, l_2 , \ldots, l_n$ carry the standard representation $\rho_{SU(2)}$
one should expect that the constant ``coloring''
 $col_{1/2}:\{l_1, l_2 , \ldots, l_n\} \to \{0,1/2, \ldots, k/2\}$
 taking only the value $1/2$
 will play a role in the sequel.
 The next proposition (in which we use the notation of Appendix B)
 shows that this is indeed the case.
\begin{proposition} \label{lem3}
For each  $(l, \underline{\alpha}) \in \Pairs_{adm}$
the area coloring $\eta_{l,\underline{\alpha}}$ is admissible w.r.t. $col_{1/2}$
and the mapping
$\Xi:   \Pairs_{adm} \ni (l, \underline{\alpha})  \mapsto \eta_{l,\underline{\alpha}}
\in \ad(sh(L);col_{1/2})$
is a  bijection.
\end{proposition}
\begin{proof} $\Xi$ is injective:
 Let us assume without loss of generality that
 $\sigma_0 \in X_{\mu}$.
 Then we have (cf. Eq. \eqref{eq_ind_conv})
 $$l= \xi_{l,\underline{\alpha}}(\sigma_0) = 2\eta_{l,\underline{\alpha}}(X_{\mu})+1$$
 so $l$ is uniquely determined by $\eta_{l,\underline{\alpha}}$.
 Moreover, from Eqs. \eqref{eq_nach_alpha_aufgeloest} and \eqref{eq_def_eta}
it follows that also $\underline{\alpha}$ is uniquely determined by $\eta_{l,\underline{\alpha}}$,
so  $\Xi$ is injective.\par
$\Xi(\Pairs_{adm}) \subset \ad(sh(L);col_{1/2})$: Let $(l, \underline{\alpha}) \in \Pairs_{adm}$ and
let $e \in E(L)$.
 As we only consider the special case $\DP(L)=\emptyset$ where
 $ E(L)  = \{l^1_{\Sigma}, l^2_{\Sigma} , \ldots, l^n_{\Sigma}\}$
 we have $e= l^j_{\Sigma}$ for some fixed $j \le n$.
We have to prove that the triple $(\bar{i},\bar{j},\bar{k}) \in I_{k+2}^3$ given by
$$\bar{i}=1/2, \quad \bar{j}=\eta(X_1(e)), \quad  \bar{k}=\eta(X_2(e))$$
fulfills the relations  \eqref{eq_relations_ijk_1}--\eqref{eq_relations_ijk_4}
in Appendix B with $\bar{r}=k+2$.
Here $X_1(e)$ and  $X_2(e)$ are defined as in Appendix B.
In order to see this first note that for sufficiently small $s>0$ we have
\begin{align} \label{eq_align_in_lemma2}
&  \bar{j} - \bar{k} =  \eta(X_1(e)) - \eta(X_2(e)) \nonumber \\
& = \tfrac{1}{2}\bigl(\xi_{l,\underline{\alpha}}(\bar{\phi}_s(\sigma_{j})) -  \xi_{l,\underline{\alpha}}(\bar{\phi}_s^{-1}(\sigma_{j})) \bigr)\nonumber \\
& =    \tfrac{1}{2} \tfrac{\alpha_j(\tau)}{i} \bigl(\ind(l^{j}_{\Sigma};\bar{\phi}_s(\sigma_{j})) - \ind(l^{j}_{\Sigma};\bar{\phi}_s^{-1}(\sigma_{j}))\bigr)
\end{align}
But $\tfrac{\alpha_j(\tau)}{i} \in \{-1,1\}$ and $\bigl(\ind(l^{j}_{\Sigma};\bar{\phi}_s(\sigma_{j})) - \ind(l^{j}_{\Sigma};\bar{\phi}_s^{-1}(\sigma_{j}))\bigr) \in \{-1,1\}$
so we obtain
\begin{equation} \label{eq_step1/2}  | \bar{j} - \bar{k}| =\tfrac{1}{2}
\end{equation}
As  $\bar{i} = \tfrac{1}{2}$ this implies relations \eqref{eq_relations_ijk_1}
and \eqref{eq_relations_ijk_4}.
Moreover, Eq. \eqref{eq_step1/2} implies that
at least one of the two numbers  $\bar{j}, \bar{k} \in I_{k+2} = \{0, 1/2,1, \ldots, k/2\}$
must lies even in $\{1/2,1, \ldots, (k-1)/2  \}$.
Relations  \eqref{eq_relations_ijk_2}
and \eqref{eq_relations_ijk_3} now follow easily.\par
$\Xi(\Pairs_{adm}) \supset \ad(sh(L);col_{1/2})$:
Let $\eta \in  \ad(sh(L);col_{1/2})$. Let us assume without loss of generality
that $\sigma_0 \in X_{\mu}$.
Let $l:= 2\eta(X_{\mu})+1$ and
let $\alpha_j: \ct \to \bC$ be given by
\begin{equation} \label{eq_def_alphaj}
\alpha_j(\tau) = -i \sgn(X^{+}_j; l^{j}_{\Sigma}))  2
\bigl( \eta(X^{+}_j) - \eta(X^{-}_j))
\end{equation}
where we have set
$X^{+}_j:=X_1(l^j_{\Sigma})$, $X^{-}_j:=X_2(l^j_{\Sigma})$ for   $l^j_{\Sigma} \in E(L) = \{l^1_{\Sigma}, l^2_{\Sigma} , \ldots, l^n_{\Sigma}\}$
and where $\sgn(X^{+}_j; l^{j}_{\Sigma})$
is defined as  in Remark  \ref{rmB.1}.
From \eqref{eq_relations_ijk_1}--\eqref{eq_relations_ijk_4} it follows  that
$l \in \{1,2\ldots, k+1\}$
and $ \underline{\alpha}:=(\alpha_1,\ldots,\alpha_n) \in A$ so $(l, \underline{\alpha}) \in  \Pairs_{adm}$.
Finally,  from Eqs.  \eqref{eq_nach_alpha_aufgeloest},  \eqref{eq_def_alphaj}, \eqref{eq_def_eta} and
\begin{equation} \label{eq_sgn_ind}
\sgn(X^{\pm}_j; l^{j}_{\Sigma})) = \pm  \bigl(\ind(l^{j}_{\Sigma};\bar{\phi}_s(\sigma_{j})) - \ind(l^{j}_{\Sigma};\bar{\phi}_s^{-1}(\sigma_{j}))\bigr)
\end{equation}
(which holds if $s$ was chosen sufficiently small)
we see that $\eta=\eta_{l,\underline{\alpha}}$ holds.
\end{proof}
In the sequel we will set $ \ad(sh(L)) :=  \ad(sh(L);col_{1/2})$.
Let $X^{\pm}_j$, $j \le n$, be defined as in the last part of the proof of Proposition \ref{lem3}.
Taking into account  Eqs. \eqref{eq_def_eta}, \eqref{eq_sgn_ind}
and Proposition \ref{lem3} we now obtain from Eq. \eqref{eq_last_ind_WLO} (provided that $s$ was chosen sufficiently small)
\begin{align}
  & \WLO(L;\phi_s )    \\
 & \sim \sum_{\eta \in ad(sh(L))}    \bigl( \prod_{t=1}^{\mu}  \sin(\tfrac{\pi}{k+2}       (2 \eta(X_t)+1))^{\chi(X_t)} \bigr)  \nonumber \\
 &  \quad \quad \times   \exp\bigl( -\tfrac{\pi i}{k+2} \tfrac{1}{2}  \sum_{j}    \eps_{j}  \sgn(X^+_j; l^{j}_{\Sigma})
  \cdot 4 ((\eta( X^{+}_j)+ 1/2)^2 -  (\eta( X^{-}_j)+ 1/2)^2)  \bigr) \nonumber\\
  & = \sum_{\eta \in ad(sh(L))}    \bigl( \prod_{t=1}^{\mu}  \sin(\tfrac{\pi}{k+2}       (2 \eta(X_t)+1))^{\chi(X_t)} \bigr)   \nonumber\\
 &  \quad \quad   \quad \quad  \times \biggl\{ \bigl(\prod_j \exp\bigl( -\tfrac{\pi i}{k+2} 2  \eps_{j}  \sgn(X^+_j; l^{j}_{\Sigma})
 (\eta( X^{+}_j)^2 + \eta( X^{+}_j)\bigr)   \nonumber\\
 &  \quad \quad   \quad \quad  \times \bigl(\prod_j \exp\bigl( -\tfrac{\pi i}{k+2} 2  \eps_{j}  \sgn(X^-_j; l^{j}_{\Sigma})
 (\eta( X^{-}_j)^2 + \eta( X^{-}_j)\bigr)  \biggr\} \nonumber\\
 & = \sum_{\eta \in ad(sh(L))}   \prod_{t=1}^{\mu} \biggl( \sin(\tfrac{\pi}{k+2}       (2 \eta(X_t)+1))^{\chi(X_t)} \biggr)  \nonumber\\
 & \quad  \quad \quad \times \prod_{t=1}^{\mu} \exp\biggl( -\tfrac{\pi i}{k+2} 2 \biggl(\sum\nolimits_{j \text{ with } \arc(l^j_{\Sigma}) \subset \partial X_t}  \eps_{j}   \sgn(X_t;l^j_{\Sigma})\biggr) \eta( X_t) (\eta( X_t) + 1) \biggr) \nonumber\\
& = \sum_{\eta \in ad(sh(L))}   \prod_{t=1}^{\mu}  \biggl( \sin(\tfrac{\pi}{k+2}  (2 \eta(X_t)+1))^{\chi(X_t)} \times \exp\bigl( -\tfrac{\pi i}{k+2} 2 x_t   (\eta( X_t) \cdot (\eta( X_t) + 1)) \biggr) \nonumber\\
& = \! \! \! \sum_{\eta \in ad(sh(L))}  \prod_{t=1}^{\mu} (v_{\eta(X_t)})^{\chi(X_t)}  \sin(\tfrac{\pi}{k+2})^{\chi(X_t)} (-1)^{\chi(X_t) 2\eta(X_t)}   \exp(2 x_t u_{\eta(X_t)})   (-1)^{x_t 2\eta(X_t)}
\end{align}
where $u_i$, $v_j$, $x_t$ are given as in Eqs. \eqref{eqB.2}, \eqref{eqB.3}, and \eqref{eq_formel_xt} in Appendix B.\par
As each $l^j_{\Sigma}$ is -- by assumption -- a  Jordan loop which is 0-homotopic
it follows that
$$\chi(X_t) =  \# \{j \le n \mid  \arc(l^j_{\Sigma}) \subset \partial X_t\}   \quad \mod 2$$
 for each $t \le \mu$.
 So in the special case where all $\eps_j$ are odd we have
 $$\chi(X_t) = x_t  \quad \mod 2$$
 for each $t \le \mu$.
 If at least one $\eps_j$ is even then the last equation does not hold in general
 but using a simple induction over the number of indices $j$ for which $\eps_j$ is
 even it follows  that one always has
$$\sum_t \chi(X_t) 2\eta(X_t) = \sum_t x_t 2\eta(X_t) \quad \mod 2$$
Moreover, we have
$$  \prod_{t} \sin(\tfrac{\pi}{k+2})^{\chi(X_t)} = \sin(\tfrac{\pi}{k+2})^{\chi(\Sigma)}=\sin(\tfrac{\pi}{k+2})^{2-2g}$$
For sufficiently small $s>0$ we therefore obtain
\begin{equation} \label{eq_WLO_final}
\WLO(L,\phi_s) \sim \sin(\tfrac{\pi}{k+2})^{2-2g} \sum_{\eta \in ad(sh(L))} \prod_{t=1}^{\mu} (v_{\eta(X_t)})^{\chi(X_t)} \exp(2 x_t u_{\eta(X_t)})
\end{equation}
Apart from the constant factor $\sin(\tfrac{\pi}{k+2})^{2-2g}$, which depends only on the charge $k$
but not on the link $L$,
the right-hand side of Eq. \eqref{eq_WLO_final}
 coincides exactly with the right-hand side of Eq.
 \eqref{Turaevs_state_sum} in Appendix B.
 In particular, $\WLO(L,\phi_s)$ does not
   depend on the special choice of the points $t_0$ and $\sigma_0$
   at the beginning of Sec. \ref{sec2}.

\medskip

\section{Outlook and Conclusions}
\label{sec7}

\subsection{Generalizing the computations of Subsec.~\ref{subsec6.3} to links with double points}
\label{subsec7.1}

In order to complete the computation of the WLOs for $G=SU(2)$
and general links (with standard colors)
one has to carry out the following steps:\par
Firstly, one has to prove that the limits
\eqref{eq_def_Rcl} exist and one has to calculate their values.
Secondly, one has to evaluate expression \eqref{eq_pre_statesum} explicitly,
for example by rewriting it
  in terms of ``state sums'' similar to the ones that appear
 in Eq. (6.2) in \cite{Ha2}.
  Finally, one has to perform the $\int \cdots DA^{\orth}_c$ and $\int \cdots DB$ integrations,
  which can probably be done  in a  similar way as in Subsec. \ref{subsec6.3}.
 We consider it to be  likely that after completing these steps one will  finally
  arrive at an expression  for the WLOs
  that is given by the right-hand side of \eqref{Turaevs_general_state_sum}.\par

One word of caution is appropriate here, though:
       it is not not totally impossible that
     something similar will happen as in the axial gauge approach to Chern-Simons models on $\bR^3$,
  cf.  \cite{FK, ASen, Ha1, Ha2}.
  In \cite{Ha2} it turned out that, in the Non-Abelian case,
  the expressions for the WLOs obtained for links with double points
   depended on the precise way in which the loop smearing regularization
  procedure was implemented.
  This ``loop smearing dependence'' destroys topological invariance
  and it is thus not surprising that the final expressions
  obtained for the WLOs in \cite{Ha2} do not  fully coincide
  with the knot polynomial expressions that were expected in the standard literature.
  (For special non-integer values of the charge $k$ topological invariance could be recovered in \cite{Ha2}
   within a restricted class of loop smearing procedures, called ``axis dependent loop smearing''.
   However, the values of $k$ for which this happens are exactly those that make the relevant knot polynomials
      trivial).\par
    If one interprets the loop smearing dependence
    as a reflection of the fact that
    axial gauge is in a certain sense a rather ``singular'' gauge
    then it would seem natural to worry
    that a similar  loop smearing dependence problem (LSD problem)
    might appear in the
    torus gauge setting of the present paper
    when evaluating the WLOs of links with double points.
   (After all ``torus gauge fixing'' and ``axial gauge fixing''
    share the aspect of being ``singular'' gauges).\par
   On the other hand it is clear that
   the ``singularity'' of axial gauge fixing alone cannot be ``the cause''
   for the LSD problem.
   Clearly, axial gauge fixing is equally singular if the structure group $G$ of the model
   is Abelian but as we saw in \cite{Ha2}
   some of the additional algebraic relations that hold
   in the Abelian case prevent the LSD problem from appearing.
Instead we prefer to interpret the LSD problem
as a reflection of the idea that something  is ``wrong''
with Chern-Simons models on non-compact manifolds.
For example, the non-compactness of the manifold $\bR^3$
has the unpleasant effect that
the expression $S_{CS}(A)$ is not defined for every $A \in \cA$.
Following \cite{FK} we therefore assumed in \cite{Ha2}
 in several computations that the 1-form  $A$
had compact support (or, alternatively, was of  rapid decrease).
If one could make this assumption  consistently then things might not be so bad.
One could then try to  replace the space
$\cA$ of all gauge fields on $\bR^3$
appearing in the relevant path integral expressions of the form
$\int_{\cA} \cdots \exp(i S_{CS}(A)) DA$
by the space $\cA_{comp}$ of 1-forms on $\bR^3$
with compact support and hope that the new path integrals
reproduce the  interesting knot invariants
that appeared  in \cite{Wi} for Chern-Simons models on compact manifolds.
 However, if one wants to apply axial gauge fixing
there  is (at least) one argument that makes it necessary to work with the original space $\cA$
 of all gauge fields, cf.
the argument in Subsec. 2.2 in \cite{Ha2}
that the mapping $\tilde{\G} \times \cA^{cax}   \ni (\Omega,A) \mapsto A \cdot \Omega \in
\cA$ is a bijection (here $\cA^{cax}$ is the space of
1-forms which are ``completely axial'' and  $\tilde{\G}:= \{\Omega \in \G \mid \Omega(0)=1\}$).
The analogue of this argument where each of the three spaces
$\cA^{cax}$, $\tilde{\G}$ and $\cA$ is replaced by the corresponding
subspace of elements with compact support does not hold.
In other words: for the approach in \cite{Ha2} it was necessary to ``combine''
results that hold only for spaces with compact support with results
that  hold only for the bigger spaces $\cA^{cax}$, $\tilde{\G}$ and $\cA$.
 It should therefore not be too
 surprising that the axial gauge approach for Chern-Simons models on $\bR^3$
 runs into difficulties (at least when using the implementation of \cite{ASen, Ha1, Ha2}).
  In fact, the loop smearing dependence was not the only
complication/problem in \cite{Ha2}.  There were two other problems:
Firstly, it turned out in \cite{Ha2} that
  the values of the WLOs  differ  from those expected in the standard literature
  even for the few  links for which there was no loop smearing dependence,
    i.e. for loops without double points\footnote{provided that horizontal framing
    (=``strictly vertical framing'' in the terminology of \cite{Ha2}) is used}.
  Secondly, in the approach in \cite{ASen, Ha1, Ha2} it was unclear right from the
    beginning how quantum groups (resp. the corresponding R-matrices)
    could enter the computations.
    Note that a quantum group $U_q(\cG)$, $q \in \bC \backslash \{-1,0,1\}$, is obtained
    from the classical enveloping algebra $U(\cG)$
    by a deformation process that involves a fixed Cartan subalgebra $\ct$.
    But such a Cartan subalgebra never played any role in \cite{ASen, Ha1, Ha2}.\par

By contrast, in the torus gauge  approach to Chern-Simons models on $M=\Sigma \times S^1$
 a Cartan subalgebra $\ct$ plays an important  role right from the beginning.
Moreover, as we have seen in Subsec. \ref{subsec6.3} above,
 in the torus gauge approach to
    Chern-Simons models on $\Sigma \times S^1$  with {\em compact}\footnote{it is interesting to note that
    if one  evaluates Eq. \eqref{eq_WLO_end} for non-compact $\Sigma$,
    for which the set $[\Sigma,G/T]$ consists of only one element
    and the summation $\sum_{\cl \in [\Sigma,G/T]} \cdots$
    is therefore trivial,
one runs into difficulties. In particular,
    the values of the WLOs of links without double points
    are then {\em not} given by the shadow invariant, which
     is an other argument in favor
    of our claim that something  is ``wrong'' with Chern-Simons models on non-compact manifolds}
     $\Sigma$ the values of the WLOs of links without double points
    do agree  with those expressions expected in the standard literature.
    (In \cite{HaHa} it is
     shown that this is also true for general groups $G$ and general link colorings, cf. point (1)
    in Subsec. \ref{subsec7.2} below).
This makes us optimistic that
also the last complication, i.e. the LSD problem
 will not appear in the torus gauge approach.\par

Finally,  we would like to emphasize that even if it turns out that
the  LSD problem does
appear during the  evaluation of the WLOs of general links,
the  torus gauge  approach is still useful:
\begin{itemize}
\item[(a)] By studying the WLOs of links consisting exclusively of
three vertical loops in the torus gauge  approach one obtains
a path integral derivation of the Verlinde formula resp. the fusion rules, cf. \cite{BlTh1}.

\item[(b)] It is shown in \cite{HaHa} (cf. point (1)
in Subsec. \ref{subsec7.2} below) that by studying WLOs of links
  that are obtained by taking a loop without double points like in Subsec. \ref{subsec6.3}
  and adding two  vertical loops
  one can also obtain a path integral derivation of the so-called quantum Racah formula
   (cf. \cite{Sawin03}).

\item[(c)] With the help of the torus gauge approach
one can probably gain a better understanding of Witten's surgery
operations from a path integral point of view. In \cite{Wi}
arguments from Conformal Field Theory are used in order to explain
the appearance of the $S$- and $T$-matrices in the formulas that
relate the values of the WLOs under surgery operations. In
\cite{HaSen} we plan to give an alternative explanation  which only
uses arguments based on the path integral.
\end{itemize}

\subsection{Other Generalizations/Further Directions}
\label{subsec7.2}

The generalization
of the computations of Subsec. \ref{subsec6.3} to links with double points,
 discussed in the previous subsection, is clearly
 the most important open problem that remains to be studied in the torus gauge approach.
 But there are  other directions for a generalization/extension of the results
of the present paper which we also find interesting.
They are given in the following list:

\begin{itemize}
\item[(1)] Generalize the results of Subsec. \ref{subsec6.3}
  and the results that can be expected if the project described in
  Subsec. \ref{subsec7.1} can be completed  successfully
  to arbitrary (simple simply-connected compact) groups $G$
  and arbitrary link colorings. (Recall that in Subsec. \ref{subsec6.3}
 we only considered the special
        situation where $G=SU(2)$ and
        where all the loops are ``colored'' with the fundamental representation $\rho_{SU(2)}$).
        In fact, for the case of links without double points
        this generalization has already been carried out in \cite{HaHa}.
    As a by-product of the computations in \cite{HaHa}
     we obtained a ``path integral derivation''
   of the so-called quantum Racah formula, cf. \cite{Sawin03}.

\item[(2)] In a recent paper, cf. \cite{BlTh4},
 Blau and Thompson study the partition function
 of Chern-Simons models on 3-manifolds $M$
 which are the total spaces of  arbitrary $S^1$-bundles
 (and not only trivial $S^1$-bundles as in \cite{BlTh1,BlTh2,BlTh3}).
 Similarly, one can ask whether the results of the
 present paper can be generalized
 to Chern-Simons models on arbitrary $S^1$-bundles.

\item[(3)] The torus gauge  approach will probably  be useful for
gaining a better understanding of Witten's surgery operations from a path integral point
of view, cf. point (c) in the list in Subsec. \ref{subsec7.1} above.

\item[(4)] It should be  possible  to obtain
 a rigorous realization of the full integral
expression on the right-hand side of Eq. \eqref{eq_WLO_end}
using results/techniques from white noise analysis, cf.  \cite{Ha6}.
However, since this treatment based on white noise analysis  is  rather technical
 it is natural to look for
alternative approaches for making rigorous sense of the right-hand side
of Eq. \eqref{eq_WLO_end}. For example, one can study approaches
which involve a suitable discretization of the base manifold $M = \Sigma \times S^1$.
 For every fixed triangulation $K$ of $\Sigma$
 and every fixed triangulation of $S^1$
  there is a discrete  analogue of the torus gauge fixing procedure
  and it should be possible to ``discretize''  the computations
 in  Secs. \ref{sec5} and \ref{sec6} of the present paper in such a  way
 that the shadow invariant is recovered within this discretized setting
 (possibly after taking a suitable continuum limit).
 Such an approach is currently studied in \cite{Ha7a,Ha7b}.
\end{itemize}

\subsection{Conclusions}
\label{subsec7.3}

In the present paper we have shown how the face models
that were introduced in \cite{Tu2}  arise naturally when  evaluating the
right-hand side of Eq. \eqref{eq_WLO_end},
which generalizes formula (7.1) in \cite{BlTh1}.
Although we have carried out all the details only in some special cases
we think that it is reasonable to expect (cf. the arguments in
 Subsec. \ref{subsec7.1}  above) that
when completing the computations
for general links one will finally arrive at the formula \eqref{Turaevs_general_state_sum}
in Appendix B (or, for $G\neq SU(2)$, at the relevant generalization
of formula \eqref{Turaevs_general_state_sum} described in \cite{Tu3}).
If this turns out to be true then
this would mean that we have solved problem (P1) of the introduction
for manifolds $M$ of the form $M=\Sigma \times S^1$.
Moreover, in view of point (4) in Subsec. \ref{subsec7.2}
this would probably also lead to the solution
of problem (P2)' for such manifolds.

  \setcounter{section}{1}
  \renewcommand{\thesection}{\Alph{section}}
 \setcounter{section}{1}
 \section*{Appendix A: Proof of Eq. \eqref{eq_independence}}

 First we observe that
 for all $j, j' \in \cN$ and $s>0$
 such that
 \begin{equation} \label{eq_supp_cond}
  \pi_{\Sigma}(\supp(j)) \cap \pi_{\Sigma}(\phi_s(\supp(j'))) = \emptyset
  \end{equation}
 holds,  the  functions
 $(\cdot,j)$ and $(\cdot,j')$ on $\cN^*$ are
 independent w.r.t. $\Phi^{\orth}_{B,\phi_s}$.
 This follows  from the $\Phi^{\orth}_{B,\phi_s}$-analogues
 of  Eqs. \eqref{eq_erstes_moment} and \eqref{eq_zweites_moment}
 (with the help of the polarization identity).\par
 Using the general Wick theorem analogue mentioned in Sec. \ref{sec4}
 we see that this statement can be generalized to arbitrary
 (finite) sequences $j_1, j_2, \ldots, j_m \in \cN$, $m \in \bN$,
 such that  condition \eqref{eq_supp_cond}  holds with $j:=j_i$, $j':=j_{i'}$, $i,i' \le m$.
 Thus, for small  $s>0$,
   $m_i \in \bN$,  $u_1 < u_2 < \ldots < u_{m_i}$,  and arbitrary polynomial functions $p_i$
 in $m_i$ variables we have:
 the  $n$-tuple $\psi^{\eps}_1, \ldots, \psi^{\eps}_n$
 given by
  $$ \psi^{\eps}_i = p_i\biggl(D^{l_i^{\eps}}_{u_1}(\cdot + A^{\orth}_c + A^{\orth}_{sing}(\cl) + Bdt ), \ldots,
D^{l_i^{\eps}}_{u_{m_i}}(\cdot + A^{\orth}_c + A^{\orth}_{sing}(\cl) + Bdt )\biggr), \quad \eps>0$$
 is independent w.r.t.  $\Phi^{\orth}_{B,\phi_s}$ if
 $\eps$ is sufficiently small
(that the aforementioned support condition is fulfilled for $\psi^{\eps}_1, \ldots, \psi^{\eps}_n$
and small $\eps>0$
follows from the assumptions that  $\DP(L) = \emptyset$ and that the framing $(\phi_s)_{s>0}$
is horizontal).\par
Eq. \eqref{eq_independence} now follows with the help of a suitable limit argument
(cf. also Proposition 4  in \cite{Ha2} and the paragraph preceding Eq. (6.3) in \cite{Ha2}).

\setcounter{section}{2}

\section*{Appendix B: The shadow invariant for $M=\Sigma \times S^1$}

For the convenience of the reader we will now  recall
some basic notions from  \cite{Tu2},
in particular the definition of the ``shadow invariant'' which was introduced
 there (cf. also  \cite{PoRe}).\par

For an admissible link $L$ in $M= \Sigma \times S^1$
we will set $D(L):=(\DP(L),E(L))$
where $\DP(L)$ denotes, as above, the set of double points of $L$ and
$E(L)$ the set of curves in $\Sigma$
into which the loops $l^1_{\Sigma}, l^2_{\Sigma}, \ldots, l^n_{\Sigma}$
are decomposed when being ``cut''  in the points of $\DP(L)$.
Clearly, $D(L)$ can be considered to be a finite (multi-)graph.
We set
\begin{equation} \Sigma \backslash D(L):=  \Sigma \backslash ( \bigcup_j \arc(l^j_{\Sigma}))
\end{equation}
As $L$ was assumed to be admissible (cf. Subsec. \ref{subsec3.1}) it follows
 that
the set $\cC_{conn}( \Sigma \backslash D(L))$ of connected components
of  $\Sigma \backslash D(L)$
 has only finitely many elements
$X_1, X_2, \ldots, X_{\mu}$, $\mu \in \bN$,  which we will call
the ``faces'' of  $\Sigma \backslash D(L)$.
In Sect. 3 in \cite{Tu2} it was shown how the link $L$ induces naturally
a function $\cC_{conn}( \Sigma \backslash D(L)) \to \bZ$
which associates  to every face  $X_t \in \cC_{conn}( \Sigma \backslash D(L))$ a number $x_t \in \bZ$.
$x_t$ was called the ``gleam'' of $X_t$
and  $x'_t := x_t - z_t/2 \in \tfrac{1}{2} \bZ$
 with $z_t:= \# \{ p \in \DP(L)\mid p \in \partial X_t\}$
 the ``modified gleam'' of $X_t$
(cf. also Remark e) ii in Sec. 1 of \cite{Tu2}).
We will call the pair $sh(L):= (D(L), (x_t)_{t \le \mu})$
  the ``shadow'' of $L$.

Let us now fix an $\bar{r} \in \bN$
and set
$$I:= I_{\bar{r}}:=\{0,1/2,1,3/2, \ldots, (\bar{r}-2)/2\}$$
For each $j \in I$ we set
\begin{align} \label{eqB.2}
 u_j  & := \pi i (j - j(j+1)/\bar{r}) = \pi i j - \tfrac{\pi i}{\bar{r}} j(j+1),\\
 \label{eqB.3}
 v_j & := (-1)^{2j} \frac{ \sin((2j+1) \pi/\bar{r})}{\sin(\pi/\bar{r})}
 \end{align}

A ``coloring'' of $L$ with colors in $I$ is a mapping
$col:\{l_1, l_2 , \ldots, l_n\} \to I$.
An ``area coloring'' of $sh(L)$  with colors in $I$ is a mapping
$\eta:\{X_1, \ldots, X_{\mu} \}  \to I$.
In the sequel let us fix a coloring $col$ of $L$.
Clearly, $col$ induces a mapping
$E(L) \to I$, which will also be denoted by $col$.
For every $e \in E(L)$ let $X_1(e)$ and  $X_2(e)$ denote the two
faces that are ``touched'' by  $e$.
\begin{figure}[h]
\begin{center}
\includegraphics[height=1.5in,width=1.5in]{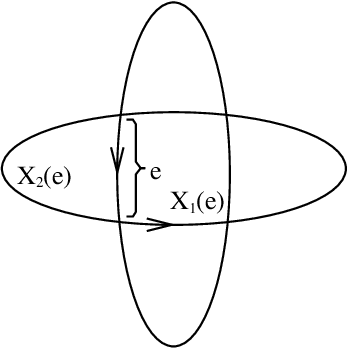}
\caption{} \label{fig1}
\end{center}
\end{figure}
More precisely, $X_1(e)$ (resp. $X_2(e)$) denotes
the unique face $X_t$ such that $e \subset \partial X_t$
and, additionally, the orientation which is induced on $e$ by the orientation
on $\partial X_t$ coincides with (resp. is opposite to)
the orientation which $e$ inherits from the loop on which it lies
(this is illustrated in  Figure \ref{fig1}
for a graph with 4 vertices and 8 edges).\par
An area coloring $\eta$ will be called ``admissible''
w.r.t. $col$ if for all $e \in E(L)$ the triple
$(\bar{i},\bar{j},\bar{k})$ given by
$$\bar{i}=col(e), \quad \bar{j}=\eta(X_1(e)), \quad  \bar{k}=\eta(X_2(e))$$
fulfills the relations
\begin{align} \label{eq_relations_ijk_1}
& \bar{i}+ \bar{j} + \bar{k} \in \bZ\\
\label{eq_relations_ijk_2}
& \bar{i}+ \bar{j} + \bar{k} \le \bar{r} -2 \\
\label{eq_relations_ijk_3}
& \bar{i}\le  \bar{j} +  \bar{k}\\
\label{eq_relations_ijk_4}
& \bar{j} \le \bar{k} +  \bar{i}, \quad  \bar{k} \le  \bar{i}  + \bar{j}
\end{align}
The set of all admissible area colorings $\eta$ of $sh(L)$ w.r.t. $col$
will be denoted by $\ad(sh(L);col )$ or simply by $\ad(sh(L))$.\par
\begin{figure}[h]
\begin{center}
\includegraphics[height=1.5in,width=1.5in]{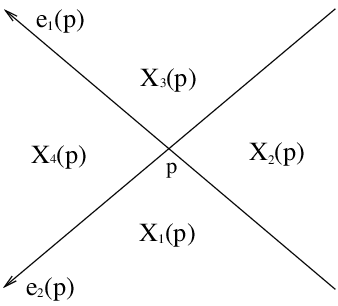}
\caption{}
\label{fig2}
\end{center}
\end{figure}

Note that every pair $(p, \eta) \in \DP(L) \times  \ad(sh(L);col )$
induces a  6-tuple $(\bar{i},\bar{j},\bar{k}, \bar{l},\bar{m},\bar{n}) \in I^6$
given by
$$\bar{i}=col(e_1(p)), \quad \bar{l}=col(e_2(p))$$
and
$$\bar{j}=\eta(X_1(p)), \quad \bar{k}=\eta(X_2(p)), \quad \bar{m}=\eta(X_3(p)), \quad \bar{n}=\eta(X_4(p))$$
where $e_1(p)$ and $e_2(p)$ are the two edges ``starting'' in $p$
and  $X_1(p), X_2(p), X_3(p), X_4(p)$  the four faces
that ``touch'' the point $p$,  cf. Figure \ref{fig2}.
We can now define the ``shadow invariant'' $|\cdot|$  by
\begin{equation} \label{Turaevs_general_state_sum}
| sh(L) |  = \sum_{\eta \in ad(sh(L))} \biggl( \prod_{p \in \DP(L)} \symb_q(\eta,p) \biggr)
 \biggl(  \prod_{t=1}^{\mu} (v_{\eta(X_t)})^{\chi(X_t)} \exp(2 x'_t u_{\eta(X_t)}) \biggr)
\end{equation}
for all $L$ as above.
Here $\symb_q(\eta,p)$ denotes the so-called quantum 6j-symbol which is
associated to the number $q:=\exp(\tfrac{2 \pi i}{\bar{r}}) \in \bC$
and to the 6-tuple $(\bar{i},\bar{j},\bar{k}, \bar{l},\bar{m},\bar{n})$ induced by  $(\eta,p)$
(for more details,  see \cite{ReKi,Tu2}).

\begin{remark} \rm
Actually, $|\cdot|$ was defined in \cite{Tu2} as a function on the
set of {\em framed} shlinks.
What we denote by $| sh(L) |$ is in fact
the (value of the) shadow invariant for the framed shlink $sh(L)$
which is obtained from $L$ when equipping
$L$ with a ``vertical'' framing in the sense of   \cite{Tu2}.
Note that in the special case where the link $L$ has no double points
  ``vertical'' framing in the sense of  \cite{Tu2}
is equivalent to what we called ``horizontal'' framing
 in Subsec. \ref{subsec5.2} above.
\end{remark}

\begin{remark} \rm  \label{rmB.1}
In the special case where the link $L$ has no double points,
i.e.
 $\DP(L) = \emptyset$,
 we have $E(L)=\{l^1_{\Sigma}, \ldots, l^n_{\Sigma}\}$
  and $x'_t = x_t$
so formula \eqref{Turaevs_general_state_sum} reduces to
 \begin{equation}  \label{Turaevs_state_sum}
| sh(L) |  = \sum_{\eta \in ad(sh(L))} \prod_{t=1}^{\mu} (v_{\eta(X_t)})^{\chi(X_t)} \exp(2 x_t u_{\eta(X_t)})
\end{equation}
Moreover, in this special case $x_t$ is simply given by
\begin{equation} \label{eq_formel_xt}
x_t = \sum_{j \text{ with } \arc(l^j_{\Sigma}) \subset \partial X_t}   \eps_j  \cdot \sgn(X_t;l^j_{\Sigma})
\end{equation}
where $\eps_j = \wind(l^j_{S^1})$
and where we have set
$ \sgn(X_t;l^j_{\Sigma}) := 1$ (resp. $ \sgn(X_t;l^j_{\Sigma}) := -1$)
if the orientation on $\partial X_t = \arc(l^j_{\Sigma})$
which is induced by the orientation on $X_t$ coincides with (resp. is opposite to)
the orientation
that is induced by $l^j_{\Sigma}:S^1 \to \arc(l^j_{\Sigma})$.
\end{remark}

\setcounter{section}{3}

\section*{Appendix C: Some comments on the Faddeev-Popov determinant}

\subsection*{A change of variable formula}

 In order to prepare the heuristic derivation of Eqs. \eqref{eq_genFadPop1}--\eqref{eq_genFadPop2} which we will give below   let us first state the following  rigorous
``change of variable formula'' for integrals on (finite dimensional) smooth manifolds:\par
Let $X$ and $Y$ be two diffeomorphic oriented smooth manifolds
and let $f: X \to Y$ be a fixed orientation preserving diffeomorphism.
Moreover, let $\mu_X$  be a
 positive Borel measure  on $X$
 which comes from a  volume form
 $\nu_X$ on $X$.  (Observe that this condition is automatically fulfilled in the special
 case where $X$ is a  Lie group and $\mu_X$ a (right) Haar measure).
 Similarly, let  $\mu_Y$ be a
 positive Borel measure  on  $Y$
 which comes from a  volume form $\nu_Y$ on $Y$. \par

We then have  for every  $\chi \in C^{\infty}_c(Y,\bC)$, i.e. every
smooth function $\chi: Y \to \bC$  with compact support
\begin{multline} \label{eq_ch_of_var0} \int \chi(y) \ d\mu_Y(y) =
\int \chi \ \nu_Y = \int f^*( \chi \ \nu_Y) = \int  (\chi \circ f) \ f^*(\nu_Y) \\
=  \int (\chi \circ f) \ \det(\theta^{-1} \circ df) \ \nu_X = \int \chi(f(x)) \ \det(\theta^{-1}_x \circ df_x)  \ d\mu_X(x)
\end{multline}
where $df: TX \to TY$ is the differential of $f$ and
where $\theta = (\theta_x)_{x \in X}$ is an arbitrary fixed smooth\footnote{i.e. $\theta$ is smooth
when considered as a map $\theta: TX \to TY$}
 family of linear isomorphisms  $\theta_x: T_x X \to T_{f(x)} Y $
such that
\begin{equation}\label{eq_property} \theta_x^* ((\nu_Y)_{f(x)}) =  (\nu_X)_{x} \text{ for all $x \in X$ }
\end{equation}
Eq. \eqref{eq_ch_of_var0} can easily
 be generalized to all smooth functions $\chi:Y \to \bC$
 which are integrable w.r.t. $\mu_Y$ (but do not necessarily have compact support).
 Also for such functions (the right-hand side of the following equation will exist and) we have
\begin{equation} \label{eq_ch_of_var0b} \int \chi(y) \ d\mu_Y(y)  = \int \chi(f(x)) \ \det(\theta^{-1}_x \circ df_x)  \ d\mu_X(x)
\end{equation}
Observe that in the special case where $X$ and $Y$ are (oriented) Lie groups
and $Lie(X)$ and $Lie(Y)$ are the corresponding Lie algebras
we can make the identifications $T_x X \cong Lie(X)$, $x \in X$, and  $T_y Y \cong Lie(Y)$, $y \in Y$,
induced by the  right-translations on $X$ and $Y$.
After doing so
 every linear isomorphism  $\phi:Lie(Y) \to Lie(X)$  induces such a family $\theta = (\theta_x)_{x \in X}$ in the obvious way.
Moreover, if $\mu_X$ and $\mu_Y$ are (right) Haar measures and $\nu_X$ and $\nu_Y$
the corresponding (right-invariant) volume forms
then  the family $\theta = (\theta_x)_{x \in X}$ associated to $\phi$ will automatically have the property \eqref{eq_property}  above
up to a multiplicative constant. In this case Eq. \eqref{eq_ch_of_var0b} implies
\begin{equation} \label{eq_ch_of_var}
\int_Y \chi(y) \ d\mu_Y(y) \sim \int_X \chi(f(x)) \ \det(\phi \circ df_x)  \ d\mu_X(x)
\end{equation}
where $\sim$ denotes equality up to a multiplicative constant independent of $\chi$.

\medskip

\begin{remark} \label{rm_appC}
Observe that  if $U$ is an open subset of a Lie group $X$
we have  $T_x U \cong Lie(X)$ for all $x \in U$. Moreover,
the restriction $(\mu_X)_{|U}$ of the (right) Haar measure $\mu_X$
is a well-defined positive Borel measure on $U$.
From these two observations it follows easily that Eq. \eqref{eq_ch_of_var}
above can be generalized to the situation where $X$ (and therefore $Y$) is not a Lie group itself
 but only an open subset (of an oriented Lie group).
\end{remark}

\medskip

\subsection*{Derivation of Eqs. \eqref{eq_genFadPop1}--\eqref{eq_genFadPop2} in Sec. \ref{subsec2.3}}

Let $M$ be a smooth manifold
and  $G$ be a compact connected Lie group with Lie algebra $\cG$.
Let $\cA:= \Omega^1(M,\cG)$  be the space of smooth $\cG$-valued 1-forms on $M$
and set
$$\G:= C^{\infty}(M,G)$$
The group $\G$ operates on $\cA$ from the right by\footnote{the notation which we use here
is a bit sloppy. If we want we can assume without loss of generality that
$G$ is a matrix Lie group (cf. Sec. \ref{subsec3.1} above)
 and in this case  we can rewrite Eq. \eqref{eq_gaugetransf} as
$ A \cdot \Omega = \Omega^{-1} \cdot  A  \cdot \Omega
+ \Omega^{-1} \cdot d\Omega$ where the two ``$\cdot$'' on the right-hand side
are the obvious multiplications induced
by the corresponding matrix multiplication}
\begin{equation} \label{eq_gaugetransf} A \cdot \Omega := \Omega^{-1}  A  \Omega
+ \Omega^{-1} d\Omega
\end{equation}

\medskip

Let $\cA_{gf}$ be a linear\footnote{the assumption that the subspace $\cA_{gf}$ is {linear}
 can  be weakened, cf. Remark \ref{rm_final_appC} below} subspace of $\cA$, which we assume to be
 ``gauge fixing'' in the sense that the map
$$q : \cA_{gf} \times \G \ni (A,\Omega) \mapsto A \cdot \Omega  \in \cA$$
is a bijection.
The Faddeev-Popov determinant is essentially the ``Jacobian'' of the map $q$.
 More precisely, for each $A_0 \in \cA_{gf}$ and $\Omega_0 \in \G$
we can consider  the differential
 $$dq_{| (A_0,\Omega_0)} : T_{A_0} \cA_{gf} \times T_{\Omega_0} \G \to T_{q(A_0,\Omega_0)} \cA $$
 as a linear map
  $dq_{| (A_0,\Omega_0)}: \cA_{gf} \oplus C^{\infty}(M,\cG) \to \cA$
 provided that we have made the identifications
 \begin{subequations} \label{eq_identifications}
 \begin{align}
 T_{A_0} \cA_{gf} & \cong \cA_{gf},\\
  T_{q(A_0,\Omega_0)} \cA & \cong  \cA\\
 T_{\Omega_0} \G  & \cong C^{\infty}(M,\cG)
 \end{align}
 \end{subequations}
 (Here the first two identifications are obvious; the last identification is the one
 via the (informal) linear isomorphism  $T_{\Omega_0} \G \to  T_{1} \G \cong C^{\infty}(M,\cG)$
induced by the right-translation $R_{\Omega_0^{-1}}: \G \ni \Omega \mapsto   \Omega \cdot \Omega_0^{-1}\in \G$).
  If we now  fix a linear isomorphism
 $$\Psi: \cA \to \cA_{gf} \oplus C^{\infty}(M,\cG)$$
  we can define, informally,\footnote{clearly, $\triangle_{FP}(A_0,\Omega_0)$ depends on $\Psi$
  via a multiplicative constant but this is irrelevant for our purposes}
 \begin{equation} \label{eq_def_triangle}
 \triangle_{FP}(A_0,\Omega_0) :=  \det(\Psi \circ dq_{|(A_0,\Omega_0)})
 \end{equation}

\begin{observation}  $\triangle_{FP}(A_0,\Omega_0)$ is independent
 of $\Omega_0$, i.e. we have $\triangle_{FP}(A_0,\Omega_0) = \triangle_{FP}(A_0,1) $
 where $1 \in \G$ is the unit element of $\G$.
   \end{observation}

 \noindent {\em ``Proof'':}   First observe that
 after making the identifications above we have\footnote{we emphasize that
 the translation part $\Omega_0^{-1} d\Omega_0$
 in $q(A_0,\Omega_0) = A_0 \cdot \Omega_0 = \Omega^{-1}_0  A_0  \Omega_0
+ \Omega^{-1}_0 d\Omega_0$  does not make an appearance in $dq_{|(A_0,\Omega_0)}$
if the identifications \eqref{eq_identifications} above are used}
  $dq_{|(A_0,\Omega_0)} = \Omega_0^{-1} dq_{|(A_0,1)} \Omega_0 = \Ad(\Omega_0) \circ dq_{|(A_0,1)}$.
 Since $G$ is compact we have $\det(\Ad(\Omega_0(x)))=1$ for all $x \in M$ so formally
 we obtain   $1=\det(\Ad(\Omega_0)) =  \det(\Psi \circ \Ad(\Omega_0) \circ \Psi^{-1})$
and therefore
$\det(\Psi \circ dq_{|(A_0,\Omega_0)}) = \det((\Psi  \circ \Ad(\Omega_0) \circ \Psi^{-1}) \circ (\Psi \circ dq_{|(A_0,1)})) = \det(\Psi \circ dq_{|(A_0,1)})$.

\bigskip

 Let $D\Omega$ be the informal normalized (right) Haar measure on $\G$ and let $DA$ and $DA_{gf}$ denote  the informal Lebesgue measures on $\cA$ and $\cA_{gf}$. (Here we have equipped $\cA$ with a fixed scalar product).
Observe that also  the product measure $DA_{gf} \otimes D\Omega$ is then
an informal (right) Haar measure  on $\cA_{gf} \times \G$.
 For  every $\G$-invariant function $\chi: \cA \to \bC$ we therefore obtain
by applying  the heuristic infinite dimensional analogue of Eq. \eqref{eq_ch_of_var} above
\begin{align} \label{eq_crucial_appC}
 \int_{\cA} \chi(A) DA  & \sim
   \int_{\cA_{gf} \times \G} \chi(q(A_{gf},\Omega))   \det(\Psi \circ dq_{|(A_{gf},\Omega)}) \ (DA_{gf} \otimes D\Omega) \nonumber \\
   & =   \int_{\cA_{gf}} \int_{\G} \chi(A_{gf} \cdot \Omega) \triangle_{FP}(A_{gf},\Omega) D\Omega DA_{gf}
     \nonumber \\
 & = \int_{\cA_{gf}} \chi(A_{gf}) \triangle_{FP}[A_{gf}] DA_{gf}
 \end{align}
 where we have set
 \begin{equation} \label{eq_def_triangle_FP}
 \triangle_{FP}[A_0] := \int_{\G} \triangle_{FP}(A_0,\Omega) D\Omega =  \int_{\G} \triangle_{FP}(A_0,1) D\Omega
 = \triangle_{FP}(A_0,1)
 \end{equation}
 for every $A_0 \in \cA_{gf}$.

\begin{remark} \label{rm_final_appC} In view of Remark \ref{rm_appC} above it makes sense to assume
that  Eq.  \eqref{eq_crucial_appC} can be generalized to suitable ``open''\footnote{since we are arguing on a heuristic level we will not bother to specify a topology here} subsets of $\cA$ and $\cA_{gf}$ (we referred to this
observation in the last footnote before Eq. \eqref{eq_FadPop0} in Subsec. \ref{subsec2.3} above).
\end{remark}

\noindent In order to evaluate $\triangle_{FP}(A_0,1)$ explicitly it will be convenient to
consider the function
$$H: \cA_{gf} \oplus C^{\infty}(M,\cG)  \to \cA_{gf} \oplus C^{\infty}(M,\cG)$$
given by
$$H = \Psi \circ q \circ (\id_{\cA_{gf}}, \exp)$$
where $\exp:= \exp_{C^{\infty}(M,G)} : C^{\infty}(M,\cG) \to C^{\infty}(M,G)$ is the ``pointwise'' exponential map.
From the chain rule and the relation $(d \exp_G)_{|0} = \id_{\cG}$
which, informally, implies $d \exp_{|0} = \id_{C^{\infty}(M,\cG)}$
we obtain for every $A_0 \in \cA_{gf}$
\begin{equation} \label{eq_C15}
 \triangle_{FP}[A_0] = \triangle_{FP}(A_0,1) = \det(\Psi \circ dq_{|(A_0,1)}) = \det(dH_{|(A_0,0)})
\end{equation}
where $dH$ is now the ``usual'' total differential of a smooth map between (infinite-dimensional)
vector spaces.
 \begin{observation} In the special case where
 $\Psi$ was chosen such that  $\Psi(A_{gf})= A_{gf}$ for all $A_{gf} \in \cA_{gf}$
 we have\footnote{as  a preparation for the final formula
 Eq. \eqref{eq_final_appC} below, which  uses the physicist notation
 $\frac{\delta F(A_0 \cdot \exp(\eta))}{\delta \eta}_{| \eta = 0 }$
 we now begin to use a similar notation  on the right-hand side of Eq. \eqref{eq_C16}}
  \begin{equation} \label{eq_C16}
 \det(dH_{|(A_0,0)}) = \det\left( \begin{matrix} \frac{\delta H_1(A_{gf},0)}{\delta A_{gf}}_{| A_{gf} = A_0} && \frac{\delta H_1(A_0,\eta)}{\delta \eta}_{| \eta = 0 } \\
 \frac{\delta H_2(A_{gf},0)}{\delta A_{gf}}_{| A_{gf} = A_0} && \frac{\delta H_2(A_0,\eta)}{\delta \eta}_{| \eta = 0 }
 \end{matrix}  \right) = \det\left( \frac{\delta H_2(A_0,\eta)}{\delta \eta}_{| \eta = 0} \right) \end{equation}
since  $H_1(A_{gf},0)=A_{gf}$ and $H_2(A_{gf},0)=0$
 and therefore
 $\frac{\delta H_1(A_{gf},0)}{\delta A_{gf}}_{| A_{gf} = A_0} = \id_{\cA_{gf}}$
and  $\frac{\delta H_2(A_{gf},0)}{\delta A_{gf}}_{| A_{gf} = A_0} = 0$
 \end{observation}

\medskip

 Clearly, we can choose the linear isomorphism $\Psi$ above such that
$\Psi_2 = F$  where $F$ is as in Sec. \ref{subsec2.3} above and
Eqs. \eqref{eq_C15} and \eqref{eq_C16} then imply
\begin{equation} \label{eq_final_appC}
\triangle_{FP}[A_0] =  \det\bigl(\frac{\delta H_2(A_0, \eta)}{\delta \eta}_{| \eta = 0 }\bigr) =
\det\bigl(\frac{\delta F(A_0 \cdot \exp(\eta))}{\delta \eta}_{| \eta = 0 }\bigr)
\end{equation}
which coincides with Eq. \eqref{eq_genFadPop2} in Sec. \ref{subsec2.3} above.

\bigskip

\bigskip

{\em Acknowledgements:} It is a pleasure for me to thank Sebastian de Haro
for several  interesting and useful discussions
on  q-deformed Yang-Mills theory and quantum topology,
and for drawing my attention to the paper \cite{Tu2}.
I also want to thank the referee for several useful comments.
Financial support from the SFB 611, the Max Planck Gesellschaft, and, above all,
the Alexander von Humboldt-Stiftung is
gratefully acknowledged.


\bigskip

\begin{thebibliography}{10}

\bibitem{Ad1}
D.~H.~Adams.
\newblock {A doubled discretization of abelian Chern-Simons theory}.
\newblock {\em Phys. Rev. Lett.}, 78(22):4155--4158, 1997.

\bibitem{Ad2}
D.~H.~Adams.
\newblock {R-Torsion and Linking Numbers from Simplicial Abelian Gauge Theories}.
\newblock {Preprint, arXiv:hep-th/9612009}

\bibitem{AS}
S.~Albeverio and J.~Sch{\"a}fer.
\newblock {Abelian Chern-Simons theory and linking numbers via oscillatory
  integrals}.
\newblock {\em J. Math. Phys.}, 36(5):2135--2169, 1994.

\bibitem{ASen}
S.~Albeverio and A.N. Sengupta.
\newblock {A Mathematical Construction of the Non-Abelian Chern-Simons
  Functional Integral}.
\newblock {\em Commun. Math. Phys.}, 186:563--579, 1997.

\bibitem{AlFr}
D.~Altschuler and L.~Freidel.
\newblock {Vassiliev knot invariants and Chern-Simons perturbation theory to
  all orders}.
\newblock {\em Comm. Math. Phys.}, 187:261--287, 1997.

\bibitem{AxSi1}
S.~Axelrod and I.M. Singer.
\newblock {Chern-Simons perturbation theory.}
\newblock In {Catto, Sultan et al.}, editor, {\em {Differential geometric
  methods in theoretical physics. Proceedings of the 20th international
  conference, June 3-7, 1991, New York City, NY, USA}}, volume 1-2, pages
  3--45. World Scientific, Singapore, 1992.

\bibitem{AxSi2}
S.~Axelrod and I.M. Singer.
\newblock {Chern-Simons perturbation theory. II.}
\newblock {\em J. Differ. Geom.}, 39(1):173--213, 1994.

\bibitem{Bar2}
D.~Bar-Natan.
\newblock {On the Vassiliev knot invariants}.
\newblock {\em Topology}, 34:423--472, 1995.

\bibitem{Bar}
D.~Bar-Natan.
\newblock {Perturbative Chern-Simons theory}.
\newblock {\em J. Knot Theory and its Ramifications}, 4:503--547, 1995.

\bibitem{BlTh1}
M.~Blau and G.~Thompson.
\newblock {Derivation of the Verlinde Formula from Chern-Simons Theory and the
  G/G model}.
\newblock {\em Nucl. Phys.}, B408(1):345--390, 1993.

\bibitem{BlTh2}
M.~Blau and G.~Thompson.
\newblock {Lectures on 2d Gauge Theories: Topological Aspects and Path Integral
  Techniques}.
\newblock In E.~Gava et~al., editor, {\em Proceedings of the 1993 Trieste
  Summer School on High Energy Physics and Cosmology}, pages 175--244. World
  Scientific, Singapore, 1994.

\bibitem{BlTh3}
M.~Blau and G.~Thompson.
\newblock {On Diagonalization in $Map(M,G)$}.
\newblock {\em Commun. Math. Phys.}, 171:639--660, 1995.

\bibitem{BlTh4}
M.~Blau and G.~Thompson.
\newblock {Chern-Simons Theory on $S^1$-Bundles: Abelianisation and q-deformed Yang-Mills
Theory}.
\newblock {\em J. High Energy Phys.}, 5:3--37, 2006 (electronic).


\bibitem{BoTa}
R.~Bott and C.~Taubes.
\newblock {On the self-linking of knots}.
\newblock {\em J. Math. Phys.}, 35(10):5247--5287, 1994.

\bibitem{Br_tD}
Th. Br{\"o}cker and T.~tom Dieck.
\newblock {\em Representations of compact {L}ie groups}, volume~98 of {\em
  Graduate Texts in Mathematics}.
\newblock Springer-Verlag, New York, 1985.

\bibitem{deHa1}
S.~de~Haro.
\newblock{A Note on Knot Invariants and $q$-Deformed $2d$ Yang-Mills},
\newblock {\em Phys. Lett.}, B634:78--83, 2006

\bibitem{HaHa}
S.~de~Haro and A. Hahn.
\newblock {Chern-Simons theory and the quantum Racah formula},
\newblock {Preprint, arXiv:math-phys/0611084}


\bibitem{Freed1}
D.~S. Freed.
\newblock Quantum groups from path integrals.
\newblock q-alg/9501025 Preprint.

\bibitem{FK}
J.~Fr{\"o}hlich and C.~King.
\newblock {The Chern-Simons Theory and Knot Polynomials}.
\newblock {\em Commun. Math. Phys.}, 126:167--199, 1989.

\bibitem{GMM}
E.~Guadagnini, M.~Martellini, and M.~Mintchev.
\newblock {Wilson Lines in Chern-Simons theory and Link invariants}.
\newblock {\em Nucl. Phys. B}, 330:575--607, 1990.


\bibitem{Ha1}
A.~Hahn.
\newblock {Chern-Simons theory on ${\mathbb R}^3$ in axial gauge: a rigorous
  approach}.
\newblock {\em J. Funct. Anal.}, 211(2):483--507, 2004.

\bibitem{Ha2}
A.~Hahn.
\newblock {The Wilson loop observables of Chern-Simons theory on ${\mathbb
  R}^3$ in axial gauge}.
\newblock {\em {Commun. Math. Phys.}}, 248(3):467--499, 2004.

\bibitem{Ha3b}
A.~Hahn.
\newblock {Chern-Simons models on $S^2 \times S^1$, torus gauge fixing, and
  link invariants I}.
\newblock {\em {J. Geom. Phys.}}, 53(3):275--314, 2005.


\bibitem{Ha3c}
A.~Hahn.
\newblock {Chern-Simons models on $S^2 \times S^1$, torus gauge fixing, and
  link invariants II}.
\newblock {\em {J. Geom. Phys.}}, 58:1124--1136, 2008.


\bibitem{Ha6}
A.~Hahn.
\newblock {White noise analysis in the theory of three-manifold quantum
  invariants}.
\newblock In A.N. Sengupta and P.~Sundar, editors, {\em Infinite Dimensional
  Stochastic Analysis}, volume XXII of {\em Quantum Probability and White Noise
  Analysis}, pages 201--225. World Scientific, Singapore, 2008.

\bibitem{Ha7a}
A.~Hahn.
\newblock {From simplicial Chern-Simons theory to the shadow invariant I},
\newblock {in preparation}.

\bibitem{Ha7b}
A.~Hahn.
\newblock {From simplicial Chern-Simons theory to the shadow invariant II},
\newblock {in preparation}.


\bibitem{HaSen}
A.~Hahn.
\newblock {Surgery operations on the Chern-Simons path integral from a conditional expectations point of view},
\newblock {in preparation}.



\bibitem{Hu}
S.-T. Hu.
\newblock {\em Homotopy Theory}.
\newblock Academic Press, New York and London, 1959.

\bibitem{Ka}
L.~Kauffman.
\newblock {\em {Knots}}.
\newblock Singapore: World Scientific, 1993.

\bibitem{Ka3}
L.~Kauffman.
\newblock {Functional integration, Kontsevich integral and formal integration.}
\newblock {\em J. Korean Math. Soc.}, 38(2):437--468, 2001.


\bibitem{ReKi} A. N. Kirillov and N.Y. Reshetikhin.
Representations of the algebra $U_q(sl_2)$,
 $q$-orthogonal polynomials and invariants of links.
In V.G. Kac et al., editor, {\em Infinite Dimensional Lie Algebras and Groups},
Vol. 7 of {\em Advanced Ser. in Math. Phys.},   pages 285--339, 1988.


\bibitem{Kup}
G.~Kuperberg.
\newblock {Quantum invariants of knots and 3-manifolds (book review)}.
\newblock {\em Bull. Amer. Math. Soc.}, 33(1):107--110, 1996.

\bibitem{LS}
P.~Leukert and J.~Sch{\"a}fer.
\newblock {A Rigorous Construction of Abelian Chern-Simons Path Integrals using
  White Noise Analysis}.
\newblock {\em Rev. Math. Phys.}, 8(3):445--456, 1996.

\bibitem{Pok}
S. Pokorski.
\newblock {\em Gauge field theories}.
\newblock Cambridge University Press, 2000.

\bibitem{PoRe}
M.~Polyak and N.~Reshetikhin.
\newblock {On 2D Yang-Mills Theory and Invariants of Links}.
\newblock In Sternheimer~D. et~al, editor, {\em Deformation Theory and
  Symplectic Geometry}, pages 223--246. Kluwer Academic Publishers, 1997.

\bibitem{ReTu2}
N.Y. Reshetikhin and V.G. Turaev.
\newblock {Ribbon graphs and their invariants derived from quantum groups.}
\newblock {\em Commun. Math. Phys.}, 127:1--26, 1990.

\bibitem{ReTu1}
N.Y. Reshetikhin and V.G. Turaev.
\newblock {Invariants of three manifolds via link polynomials and quantum
  groups.}
\newblock {\em Invent. Math.}, 103:547--597, 1991.


\bibitem{Sawin03}
S.~Sawin.
\newblock {Quantum groups at roots of unity and modularity}.
\newblock{\em {J. Knot Theory Ramifications}}, 15(10):1245--1277,
2006.

\bibitem{Tu3}
V.~G. Turaev.
\newblock {\em Quantum invariants of knots and 3-manifolds}.
\newblock de Gruyter, 1994.

\bibitem{Tu2}
V.G. Turaev.
\newblock {Shadow links and face models of statistical mechanics}.
\newblock {\em J. Diff. Geom.}, 36:35--74, 1992.

\bibitem{TuVi}
V.G. Turaev and O.~G. Viro.
\newblock {State sum invariants of 3-manifolds and quantum 6j-symbols}.
\newblock {\em Topology}, 31(4):865--902, 1992.

\bibitem{Wi89_3}
E.~Witten.
\newblock Gauge theories and integrable lattice models.
\newblock {\em Nuclear Phys. B}, 322(3):629--697, 1989.

\bibitem{Wi}
E.~Witten.
\newblock {Quantum Field Theory and the Jones Polynomial}.
\newblock {\em Commun. Math. Phys.}, 121:351--399, 1989.

\bibitem{Wi90}
E.~Witten.
\newblock Gauge theories, vertex models, and quantum groups.
\newblock {\em Nuclear Phys. B}, 330(2-3):285--346, 1990.

\end{thebibliography}
\end{document}